\documentclass[acmsmall]{acmart}

\setcopyright{rightsretained}
\acmJournal{TECS}
\acmYear{2022} \acmVolume{1} \acmNumber{1} \acmArticle{1} \acmMonth{1} \acmPrice{}
\acmDOI{10.1145/3529257}

\acmJournal{TECS}
\acmVolume{37}
\acmNumber{4}
\acmArticle{111}
\acmMonth{8}

\usepackage{graphicx}
\graphicspath{
    {biography_photos/}
    {figures/}
}
\usepackage{svg}

\usepackage[utf8]{inputenc} %

\usepackage{xcolor}
\usepackage{ulem} %
\usepackage{soul} %

\usepackage{listings}

\usepackage{caption}
\usepackage{subfigure}

\usepackage{tabularx}
\usepackage{multirow}

\usepackage{comment}

\usepackage{pifont}

\include{ushyphex.tex}

\usepackage{listings}
\usepackage{xcolor}

\colorlet{punct}{red!60!black}
\definecolor{jsonbackground}{HTML}{EFEFEF}
\definecolor{delim}{RGB}{20,105,176}
\colorlet{numb}{magenta!60!black}

\lstdefinelanguage{json}{
    basicstyle=\normalfont\ttfamily,
    stepnumber=1,
    numbersep=6pt,
    numbers=left,
    showstringspaces=false,
    breaklines=true,
    backgroundcolor=\color{jsonbackground},
    framerule=0pt,
    frameround=tttf,
    frame=tlBR,
    literate=
     *{0}{{{\color{numb}0}}}{1}
      {1}{{{\color{numb}1}}}{1}
      {2}{{{\color{numb}2}}}{1}
      {3}{{{\color{numb}3}}}{1}
      {4}{{{\color{numb}4}}}{1}
      {5}{{{\color{numb}5}}}{1}
      {6}{{{\color{numb}6}}}{1}
      {7}{{{\color{numb}7}}}{1}
      {8}{{{\color{numb}8}}}{1}
      {9}{{{\color{numb}9}}}{1}
      {:}{{{\color{punct}{:}}}}{1}
      {,}{{{\color{punct}{,}}}}{1}
      {\{}{{{\color{delim}{\{}}}}{1}
      {\}}{{{\color{delim}{\}}}}}{1}
      {[}{{{\color{delim}{[}}}}{1}
      {]}{{{\color{delim}{]}}}}{1},
}

\definecolor{darkgreen}{rgb}{0.13, 0.5, 0.13}

\begin{document}

\title[]{CEDR - A Compiler-integrated, Extensible DSSoC Runtime}

\author{Joshua Mack}
\authornote{Both authors contributed equally to this research.}
\email{jmack2545@email.arizona.edu}
\orcid{0000-0003-1066-5578}
\author{Sahil Hassan}
\authornotemark[1]
\email{sahilhassan@email.arizona.edu}
\orcid{0000-0002-4574-9555}

\author{Nirmal Kumbhare}
\orcid{0000-0002-1578-5504}
\email{nirmalk@catmail.arizona.edu}

\author{Miguel Castro Gonzalez}
\orcid{0000-0002-3575-2320}
\email{migor2d2@email.arizona.edu}

\author{Ali Akoglu}
\affiliation{%
	\institution{University of Arizona Department of Electrical \& Computer Engineering}
	\city{Tucson}
	\state{Arizona}
	\country{USA}
	\postcode{85719}
}
\orcid{0000-0001-7982-8991}
\email{akoglu@ece.arizona.edu}

\renewcommand{\shortauthors}{Mack and Hassan, et al.}

\begin{abstract}
In this work, we present CEDR, a \emph{C}ompiler-integrated, \emph{E}xtensible \emph{D}omain Specific System on Chip \emph{R}untime ecosystem to facilitate research towards addressing the challenges of architecture, system software and application development with distinct plug-and-play integration points in a unified compile time and run time workflow. 
We demonstrate the utility of CEDR on the Xilinx Zynq MPSoC-ZCU102 for evaluating performance of pre-silicon hardware in the trade space of SoC configuration, scheduling policy and workload complexity based on dynamically arriving workload scenarios composed of real-life signal processing applications scaling to thousands of application instances with FFT and matrix multiply accelerators. 
We provide insights into the trade-offs present in this design space through a number of distinct case studies.
CEDR is portable and has been deployed and validated on Odroid-XU3, X86 and Nvidia Jetson Xavier based SoC platforms. 
Taken together, CEDR is a capable environment for enabling research in exploring the boundaries of productive application development, resource management heuristic development, and hardware configuration analysis for heterogeneous architectures.

\end{abstract}



\maketitle

\section{Introduction}\label{sec:introduction}
Heterogeneous computing systems, while offering a large potential for performance gains relative to homogeneous counterparts, traditionally pair efficiency gains with reductions in ease of use and programmer productivity.
Platforms such as Domain-specific System on Chip (DSSoC) devices have been proposed as one such solution for addressing this divergence with the hope that the focus on a smaller domain of applications will enable more productive software and programming abstractions.
However, despite the advantages gained through reducing the problem size, many of the core challenges of utilizing and programming heterogeneous systems still apply.
In traditional heterogeneous programming paradigms, massive amounts of effort are put into offline performance analysis by domain experts to determine the portions of an application that must be accelerated, the type of accelerators needed, and effective implementation strategy for the target hardware configuration. 
Low-performance serial implementations are then replaced with their optimized heterogeneous implementations, and a static binary that represents a single, expertly-tuned instance of that application is produced.
Such static and offline resource allocation decisions result in a greedily optimized implementation that assumes it does not share the heterogeneous accelerators with any other applications. 
However, in a computing environment where heterogeneity is widespread, this assumption has the potential to lead to drastic mismanagement or under-utilization of the target hardware.
In homogeneous computing, this kind of dilemma has been thoroughly addressed through the integration of intermediate layers of resource-management software -- like operating systems -- that work to ensure that all applications can share the underlying hardware effectively despite being unaware the others exist.
We find that these same requirements apply to heterogeneous computing, and in particular DSSoCs, as well. 
Hence, to meet these requirements, there is a need for an intelligent runtime system and programming framework to enable effective utilization of DSSoC platforms and take full advantage of their underlying hardware without requiring users to become hardware experts in the process.
Furthermore, we envision that the DSSoC system should also enable a productive programming and deployment experience in such a way that multiple users can coexist and share the hardware as a service by supporting execution of any combination of dynamically arriving applications. 

This goal, while providing an ambitious target to aim for, tends to hide some of the complexity that becomes apparent when working in the design and study of DSSoCs. 
In the process of constructing such a system, there are a number of research questions and avenues that become readily apparent, and foundational across all of these questions is the underlying definition of a DSSoC itself.
While it may seem that this definition is broadly apparent -- a DSSoC is an SoC that is tailored for a particular domain of computation -- there are challenges and subtleties in how exactly to define the scope of a domain.
Trivially, we know that a DSSoC needs to support more than one application, as otherwise it is an application-specific integrated circuit (ASIC), but at the same time, we know that a DSSoC cannot be suitable for all applications, as otherwise it loses the specialization of being domain specific.
With this in mind, one of the biggest questions in the area of domain specific architecture research is how precisely to determine which heterogeneous processing elements (PEs) to use for which domain as well as how both power and data need to be managed between all of these PEs.
We believe that one of the most promising avenues for exploring these questions is through direct empirical experimentation with frameworks that enable users to rapidly sweep and explore the design space itself.
However, while in the literature, various frameworks have been proposed that enable exploration of certain aspects of this design space -- such as SoC and application design without a focus on scheduling~\cite{mantovani_openesp_2020, shao_2016_CodesigningAccelerators, nazarian_2020_S4oCSelfOptimizing, chen_2016_ARAPrototyperEnabling} or standalone application programming interfaces that are independent of hardware~\cite{sujeeth_2014_DeliteCompiler, huang_cpp-taskflow_2019} -- to the best of our knowledge, no frameworks thus far have been presented that bring together application development, resource management, and accelerator design capabilities into a single unified compilation and runtime toolchain that targets DSSoC hardware.
Towards this end, in this work, we introduce CEDR: a \emph{C}ompiler-integrated, \emph{E}xtensible, \emph{D}SSoC \emph{R}untime.

As the name implies, CEDR is a novel open-source\footnote{Available at: \url{https://github.com/UA-RCL/CEDR}} ecosystem that integrates compile-time application analysis with a Linux-based runtime system and holistically targets the aforementioned requirements and capabilities.
By coupling these components, it enables compilation and development of user applications, evaluation of resource management strategies, and validation of hardware configurations in one unified framework.
Importantly, this framework is itself portable across a wide number of Linux-based systems, ensuring that effort to migrate across systems is minimal for all developers involved.
Using the CEDR compilation environment, application developers can develop and validate large, non-trivial applications to serve as workloads for scheduling heuristic developers and hardware architects; scheduling heuristic developers can easily implement their policies in a common environment for cross-validation and evaluation; and hardware architects can design new DSSoC architectures with the knowledge that they will be able to build on an existing library of validated applications and schedulers rather than rely on simple micro-benchmarks executed in unrealistic simulation environments.
While there are a large number of existing works that provide software and runtime environments for heterogeneous architectures~\cite{moazzemi_2019_HESSLEFREEHeterogeneous, bolchini_opencl_2018, christodoulis_2018_FPGATarget, tan_picos++_2019, boutellier_prune_2018, auerbach_2012_liquidmetal, hsieh_surf_2019}, CEDR is unique in the way it brings together all of these aspects of DSSoC development and couples them with unique features like task-level measurement of performance counters or support for software-based pipelining (``streaming'') of application tasks.
We believe that the CEDR ecosystem, with its integrated compile-time and runtime workflows, will empower researchers to conduct design space explorations, and consequently, it will help the research community move towards answering the aforementioned questions and establishing a more general understanding of DSSoCs and their broader role in an era of increasingly heterogeneous computing systems.

A preliminary version of this work appeared in the \textit{``Heterogeneity in Computing Workshop (HCW 2020)"}~\cite{MackUserspace2020} where we presented a baseline FPGA-based emulation framework.
In this paper, we expand the preliminary work with the following contributions:

\begin{itemize}
    \item We rearchitect the runtime to allow for launching a workloads with a user-friendly, daemon-based job submission process akin to those used in high performance computing (HPC) systems. Through this interface, users can easily construct arbitrarily complex, highly interleaved application scenarios.
    \item Enabled by this new runtime architecture, we scale our analysis to 3480 experimental configurations that collectively result in scheduling and analysis of over 10 million total tasks.
    \item We expand and further validate CEDR's ability to schedule and dispatch to arbitrary heterogeneous hardware through integration of new FPGA accelerators (matrix multiplication) and new hardware platforms (CUDA-based accelerators on Nvidia Jetson AGX Xavier).
    \item We expand the list of built-in schedulers by incorporating a variant of Heterogeneous Earliest Finish Time~\cite{Mack_Performant_2022} (HEFT-RT) and Earliest Task First (ETF).  
    \item We add support for PE-level work queues that optimize runtime overhead, enable integration of more complex resource management heuristics that rely on reservation-based policies, and help lower the inter-task overhead on heavily-utilized PEs.
    \item We enable management and scheduling of concurrent, pipelined execution of tasks in an application to vastly improve performance of highly recurring, stream-based application graphs. 
    \item We implement the ability to access performance counters and dynamically collect a rich, configurable set of fine-grained execution characteristics that allow us to extract insights about our application domain, enable rich workload characterization and support future scheduler design. 
    \item We incorporate the ability to cache scheduling decisions to reduce overhead associated with scheduling via complex heuristics.
\end{itemize}

 The rest of the paper is organized as follows: Section~\ref{sec:cedr_overview} presents an overview of the CEDR framework. 
 Section~\ref{sec:experimental_setup} presents the experimental setup and discusses the applications, workloads, and hardware configurations studied in the experiments.
 Section~\ref{sec:experimental_evaluation} presents validation-based sweeping experiments that demonstrate the utility of CEDR for extracting insights from large-scale workload scenarios.
 Section~\ref{sec:case_studies} presents series of case studies to demonstrate the utility of CEDR in design space exploration from application engineer, hardware designer and scheduling heuristic developer perspectives.
 Section~\ref{sec:related_work} discusses other work in this area and contextualizes CEDR within it.
 Finally, Section~\ref{sec:conclusion} concludes by summarizing this work and discussing avenues for future work.

\section{CEDR Overview} \label{sec:cedr_overview}

CEDR is composed of two components: a compilation workflow and a runtime workflow.
The compilation workflow is used to convert C/C++ applications into CEDR-compatible binaries, and the runtime workflow is leveraged to then parse, schedule, dispatch, and execute those applications across a heterogeneous pool of resources on a given compute platform.
CEDR provides these capabilities while remaining independent of any one scheduling heuristic or hardware platform, ensuring that it can be ported across any number of execution environments without requiring major development effort to port any given scheduler to a new SoC platform or vice versa.
To achieve this, CEDR is built to schedule and execute applications that can be expressed as Directed Acyclic Graphs (DAGs), where each node in the graph represents a given task that requires scheduling to a given resource, and edges between the nodes represent temporal dependencies.
In this abstraction, the structure of the user application itself is captured by the graph, and the actual implementation of each node can be decoupled from this graph through the use of shared libraries and function pointers.
With applications structured in this fashion, the role of the CEDR runtime is then to dynamically bind function implementations to their corresponding DAG nodes and execute them on the system's compute resources.
For instance, a single DAG node may represent a 256 point FFT, and such a node is represented with a flexible binary structure that may have both a CPU and accelerator implementations associated with it.
The goal of the runtime is to then ask the user-selected scheduling heuristic to choose an implementation based on the current state of the compute resources and dynamically dispatch to the chosen implementation. 
In the following subsections, we will explore each of these aspects of CEDR (compilation and runtime) in detail.
As understanding the mechanics of the runtime helps to motivate the goals of the compiler, we begin by discussing the runtime in Section~\ref{subsec:overview_runtime_workflow} before discussing the compilation approaches in Section~\ref{subsec:overview_compilation_workflow}.

\subsection{Runtime Workflow} \label{subsec:overview_runtime_workflow}

\begin{figure}
    \centering
    \includegraphics[width=0.5\linewidth]{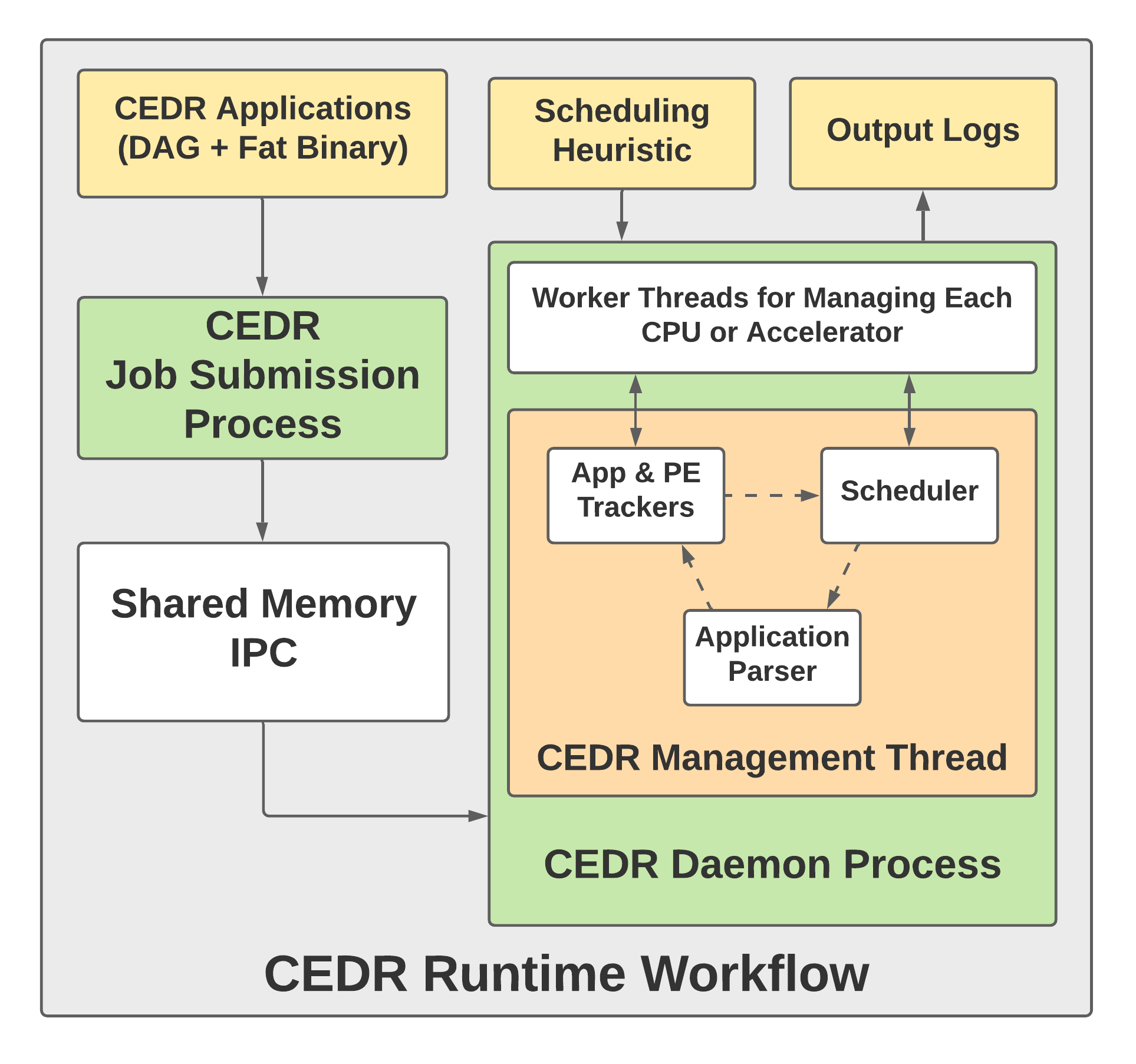}
    \caption{Overview of the components present in the CEDR Runtime}
    \label{fig:runtime_workflow}
\end{figure}

The architecture of the CEDR runtime is presented in Fig.~\ref{fig:runtime_workflow}.
The runtime itself operates as a background \textit{Daemon Process}, shown in the right half of the figure, and the user submits jobs for execution via inter-process communication (IPC) using the \textit{Job Submission Process}.
Then, the daemon process consists of two key components: the \textit{Worker Threads} and the \textit{CEDR Management Thread}.
For each resource in the system -- whether it is a CPU core or an accelerator -- we spawn one worker thread that is tasked with receiving, executing, and reporting back on work assigned to that particular resource.
As an example, suppose we are running on a system with one CPU core (CPU 1) and one FFT accelerator (FFT 1).
In this case, the worker thread for CPU 1 is, itself, assigned via its processor affinity to run on CPU 1, and as it receives tasks that are scheduled to CPU 1, it executes them and reports back to the management thread on their completion. 
Meanwhile, the worker thread for FFT 1 is tasked with running on one of the CPU cores and facilitating data transfers to and from the underlying accelerator.
While it is configurable, by default, each of these worker threads utilize Linux's real-time \texttt{SCHED\_RR} policy with a static priority of 99 to minimize the amount of time spent executing non-CEDR-related tasks on their managed resources.
One of the advantages of this architecture is that we can easily scale to systems with any number of resources simply by changing the number of worker threads spawned to manage them.
These worker threads are managed using the widely utilized POSIX thread library~\cite{mueller1993library} by the main CEDR management thread, and when coupled with the fact that all of the components shown here operate in Linux userspace, we can see that CEDR is trivially portable across a wide range of Linux-based SoC platforms.

\begin{lstlisting}[
    language=json,
    firstnumber=1,
    basicstyle=\scriptsize\ttfamily,
    label=lst:test_application,
    caption=Sample Application JSON,
    float=tbp,
    escapechar=\%
]
"AppName": "sample_application",
"SharedObject": "sample_application.so",
"Variables": {
    "var_0": {
        "bytes": 4,
        "is_ptr": false,
        "ptr_alloc_bytes": 0,
        "val": []
    },
    "var_1": {
        "bytes": 4,
        "is_ptr": false,
        "ptr_alloc_bytes": 0,
        "val": [0, 1, 0, 0]
    },
    "var_2": {
        "bytes": 8,
        "is_ptr": true,
        "ptr_alloc_bytes": 2048,
        "val": []
    }
},
"DAG": {
  "Node 0": {
    "arguments": ["var_0", "var_1"]
    "predecessors": [],
    "successors": [{"name": "Node 1", "edgecost": 1.0}],
    "platforms": 
    [{"name": "cpu", "runfunc": "Node_0_CPU", "nodecost": 1.0}]},
  "Node 1": {
    "arguments": ["var_1", "var_2"]
    "predecessors": [{"name": "Node 0", "edgecost": 1.0}],
    "successors": [{"name": "Node 2", "edgecost": 1.0}],
    "platforms": 
    [{"name":"cpu", "runfunc": "Node_1_CPU", "nodecost": 1.0},
     {"name":"fft", "runfunc": "FFT_Accel_Dispatch", "shared_object": "fft_accel.so", 
                    "nodecost": 1.0}]},
  "Node 2": {
    "arguments": ["var_0", "var_1", "var_2"]
    "predecessors": [{"name": "Node 1", "edgecost": 1.0}],
    "successors": [],
    "platforms": 
    [{"name":"cpu", "runfunc": "Node_2_CPU", "nodecost": 1.0}]}
}
\end{lstlisting}

Switching focus to the management thread, we can see that it operates in a continuous loop of application parsing, application \& PE tracking, and task scheduling.
The application parser forms the entry-point by which applications are received by the runtime.
Each CEDR application is submitted in the form of a JSON-based DAG file and a flexible binary format (also known as a ``Fat Binary'') that contains the various invocations needed by each node for each heterogeneous PE.
As applications are received, the application parser reads the provided JSON and binary objects and initializes CEDR's internal application representation.
These parsed applications are themselves cached and stored as ``application prototypes'' such that, if they are to be submitted again in the future, the runtime doesn't need to re-parse, instead just instantiating another copy.
An example of a CEDR application JSON is provided in Listing~\ref{lst:test_application}.
We can see that this JSON has four main top-level keys: \texttt{AppName}, \texttt{SharedObject}, \texttt{Variables}, and \texttt{DAG}.
The \texttt{AppName} string captures the name of the application to be used internally for logging related to this application as well as the key used to identify if an application has been parsed and cached previously, the \texttt{SharedObject} string defines the binary that contains the functions utilized by each DAG node, the \texttt{Variables} object gives CEDR insight into the memory requirements of this application, and the \texttt{DAG} object captures the application DAG's structure.
Each variable defined in the \texttt{Variables} object is allocated and managed by CEDR upon every instantiation of a given application with the understanding that the application can use this memory to share data between DAG nodes.
A given variable is identified throughout the JSON by its key, and the fields inside -- \texttt{bytes}, \texttt{is\_ptr}, \texttt{ptr\_alloc\_bytes}, and \texttt{val} -- respectively refer to the bytes required for that variable's type; whether that variable is itself a pointer; if it is a pointer, how many bytes of storage it requires; and a list of bytes that can serve as a variable's initial value.
Finally, each key in the \texttt{DAG} object represents one node in the application's DAG, with each node specified via four fields: \texttt{arguments}, \texttt{predecessors}, \texttt{successors}, and \texttt{platforms}.
The \texttt{predecessors} and \texttt{successors} lists capture the set of predecessor and successor tasks for a given node, respectively, along with their communication costs as \texttt{edgecost}. With this, we can infer the structure of a given DAG.
Meanwhile, the \texttt{arguments} list captures the set of variables (defined in the \texttt{Variables} object) that this node requires when it is invoked, and the \texttt{platforms} list captures the set of heterogeneous-resource specific implementations that can be used to execute a given task on the desired resources.
Each platform entry specifies a \texttt{runfunc} that, by default, is searched for in the top level \texttt{SharedObject}, and a \texttt{nodecost} that specifies the expected execution time of the \texttt{runfunc} on the desired resource in microseconds. The user can also optionally override this behavior by specifying a different, platform-specific \texttt{shared\_object} key within the entry itself.
For instance, as shown on line 29 of the listing, \texttt{Node 0} can only be invoked on the CPU. 
Meanwhile, on lines 35-36, we can see that \texttt{Node 1} can be invoked on both CPU and FFT resource types, with the FFT resource type overriding the top-level \texttt{SharedObject} binary.
Later, we will discuss how ``accelerator supported'' nodes are specified during the compilation process in Section~\ref{subsec:overview_compilation_workflow}.

As parsing completes, applications are handed off to the application \& PE tracker, which begins by pushing the head nodes from each application -- the nodes with empty \texttt{predecessors} lists -- into the runtime's \textit{ready queue} for scheduling and dispatch.
From there, tasks are allocated by the user's specified scheduling heuristic to run on particular PEs by passing them to their corresponding worker threads and choosing the appropriate function that was previously parsed from their \texttt{platforms} list.
By default, CEDR provides five scheduling policies: round robin (RR), minimum execution time (MET), a custom HEFT~\cite{Mack_Performant_2022} inspired scheduler, earliest finish time (EFT), and earliest task first (ETF) schedulers.
Together, these policies provide a useful foundation for scheduler experimentation, covering a broad range of heuristics whose characteristics range from low complexity and low runtime overhead up through those with high complexity and high runtime overhead.
For users interested in integrating their own scheduling heuristic, any policy can be integrated trivially so long as it can receive and schedule tasks from the runtime's ready queue as shown in the \texttt{Scheduler} block in Fig.~\ref{fig:runtime_workflow}.
As tasks are received and executed by the various worker threads, they signal their completion back to the application \& PE tracker, which responds by checking the dependency resolution of their successor nodes and pushing them into the ready queue as necessary, after which the process repeats.
If an application runs out of successors to enqueue, it is marked as completed by the runtime, timing logs are generated representing its execution, and the memory associated with the application instance is released.
These timing logs capture all of the relevant scheduling and timing information about when each task in a given application ran, on which PE it ran, and so on.
This cycle of application parsing, application dispatch, and log generation repeats indefinitely until an IPC command is received that signals for the runtime to terminate.

\subsection{Compilation Workflow} \label{subsec:overview_compilation_workflow}

\begin{figure}
    \centering
    \includegraphics[width=\linewidth]{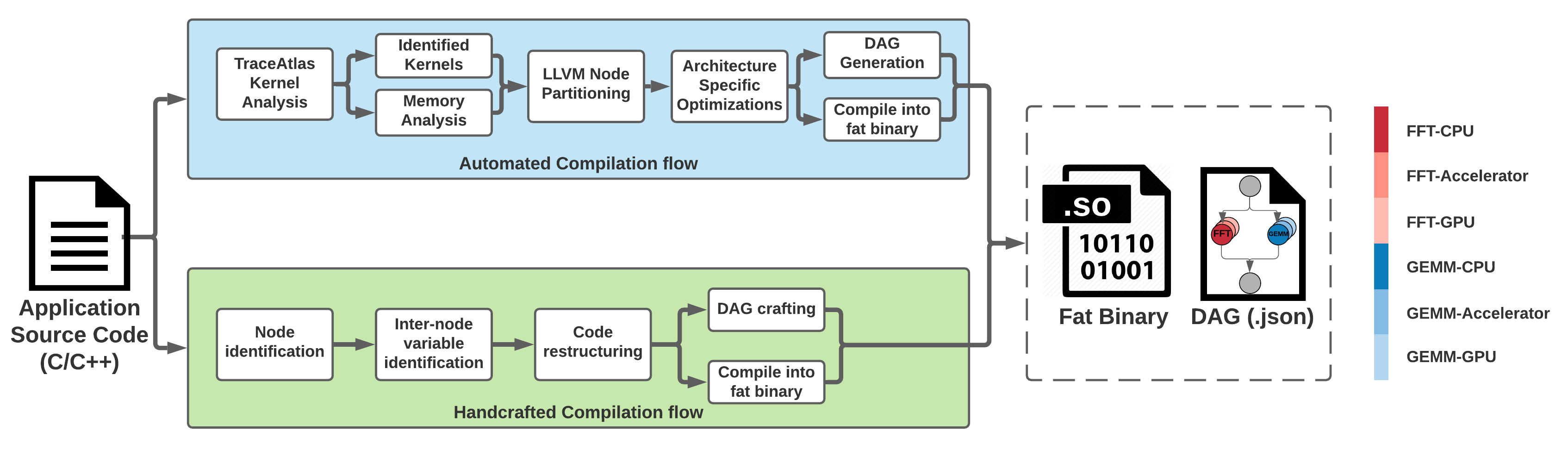}
    \caption{Two distinct methodologies for preparing applications for execution in CEDR}
    \label{fig:compiler_workflow}
\end{figure}

To prepare applications for CEDR, there are two methodologies available as shown in Fig.~\ref{fig:compiler_workflow}.
The first path, shown in the top half of the figure, involves passing an off-the-shelf user application through an automated toolchain that can produce functionally correct DAG-based applications for CEDR.
Meanwhile, the second path, shown in the bottom half of the figure, involves transforming a user application through a hand-crafted process that, despite being more time consuming, can produce binaries that better exploit the available opportunities for parallelism and heterogeneous execution.
In the following paragraphs, each of these two workflows will be explored in detail, starting with the automated compilation workflow.

\subsubsection{Automated Compilation Process}
In converting an off-the-shelf application for use in CEDR, the broad goal is to take an arbitrary C/C++ application and pass it through analysis tooling to (i) determine the optimal boundaries to use in segmenting the baseline code into a sequence of DAG nodes and (ii) determine where heterogeneous execution opportunities are present and ensure that those opportunities can be maximally exploited by the scheduling heuristic in CEDR.
To do this, we begin by converting the user's application into the LLVM~\cite{lattner2004llvm} intermediate representation (IR).
The user's LLVM IR is then passed through TraceAtlas~\cite{uhrie_automated_2020}, an open source\footnote{Source available at \url{https://github.com/ruhrie/TraceAtlas}} toolchain for collecting and analyzing dynamic application traces from arbitrary application code.
Using TraceAtlas, we modify the user's LLVM IR to include tracing instrumentation, compile a \textit{tracing executable}, and execute it.
As the \textit{tracing executable} runs, it dumps a runtime trace to disk that captures temporal aspects of the code's execution such as the full sequence of basic block transitions that occurred during execution.
Next, we analyze this trace using TraceAtlas, and it suggests the regions in the code that should be labeled as ``kernels'' (roughly analogous to ``hot'' sections of the original program) and ``non-kernels'' (analogous to the ``cold'' sections of the program).
With this information extracted, we proceed to the \textit{LLVM Node Partitioning} phase.
Using the provided information about suggested kernel regions, we launch a second phase of analysis to determine whether those regions can, themselves, be easily \textit{outlined} from their original source code into standalone functions that can serve as nodes in the application's DAG.
Adjustments to these node boundaries are necessary 
if extracting a kernel would cause issues with the program’s control flow structures. 
For example, the two branches of an if/else statement must lie within the same function after outlining as otherwise, directing execution down at least one of the two paths would require unconditionally jumping from one function body to the other without respecting function call semantics. 
We resolve these issues by expanding the provided kernel boundaries until they contain only a single entry and single exit point for their respective sets of basic blocks. 
After this stage, all of the user’s application code is partitioned into standalone functions that, when called in sequence, recreate the behavior of the original user application.
With that, we proceed to the \textit{Architecture Specific Optimizations} phase.
In this phase, we determine whether any DAG nodes that have been \textit{outlined} are supported with acceleration via heterogeneous execution on our target architecture.
To do this, we apply a combination of (i) pattern matching of the outlined DAG nodes against a known library of kernels for the target architecture and (ii) optional ``hints'' from the user in the form of \textit{kernel labels} added to the original source code.
Based on the results of this analysis, each DAG node is annotated with the set of supported execution platforms, and we proceed onto the final phase: \textit{DAG Generation} and \textit{Fat Binary Compilation}.
In the final phase, we perform a high level memory analysis to identify const variable allocations as well as variables that are assigned const-inferable allocations of heap memory (such as those that utilize const arguments to \texttt{malloc}).
These variables along with their memory allocation sizes are all populated into a \textit{variables} dictionary for use in the remaining \textit{DAG Generation} steps.
With those memory allocations in place, we proceed to generating the JSON that represents our DAG.
We iterate through each of the nodes extracted in the \textit{LLVM Node Partitioning} phase, and we create a serial chain of DAG nodes that preserves the linear execution flow of the original program.
As a result of the outlining process, each DAG node's function receives, as arguments, pointers to all of the externally-defined variables that it requires.
These variables are linked with their corresponding entries in the variables dictionary defined earlier, and with this, we have a DAG structure that captures the linear execution flow of the program as well as most of the const-inferable memory behaviors.
Additionally, for each DAG node that supports heterogeneous execution, we append entries to their implementation function lists that correspond to the given pre-defined heterogeneous kernel.
Finally, to generate the \textit{Fat Binary}, we simply compile the modified LLVM IR into a shared object, and by combining this with a library of other shared objects that handle the heterogeneous kernels in CEDR, the effective output is akin to a fat binary format, with each node having multiple implementations depending on the hardware platform to be used.

\subsubsection{Hand-crafted Compilation Process}\label{subsubsec:handcraft_compile}

Compared to the automated compilation workflow, the process of hand-crafting an arbitrary C/C++ application to a CEDR-compatible form has a larger degree of flexibility, providing more freedom to utilize accelerators and exploit parallelism within an application.
This is due to the fact that, to a certain extent, CEDR is capable of executing anything that can be compiled to a shared object and coupled with an appropriate JSON file.
Given this large degree of flexibility, the path for hand-crafting an application consists less of a fixed set of procedures and more of a general set of guidelines.
Towards this end, in the bottom half of Fig.~\ref{fig:compiler_workflow}, we present a workflow that captures these guidelines.
Starting with a baseline serial C/C++ application, the first stage, \textit{Node Identification}, involves identifying the boundaries that should be used to delineate the node boundaries for the application's underlying DAG.
These boundaries should be informed by (i) the desired level of task granularity (fine/coarse grain), (ii) desired accelerator support, and (iii) the degree of parallelism.
For instance, any code that will be accelerated needs to be placed in a node by itself with no other extraneous logic or side effects as the DAG structure of CEDR itself relies on the ability to swap out one function-level implementation for another to implement its model of heterogeneity: all implementations of a given DAG node should accept the same set of arguments, and they should have the same functional behavior \& side effects with the only difference being their method of computation.
Similarly, applications that exhibit a higher degree of inter-task parallelism or a finer-grained level of task granularity have the potential to outperform coarser, less parallel options, but this parallelism needs to be balanced against the extra stress it creates on the scheduling heuristic to avoid large increases in scheduling overhead.
Regardless, with these steps in mind, by the end of this phase, the developer has determined how to segment their code into standalone DAG nodes, and they can proceed to the next phase: \textit{Inter-node Variable Identification}.
In this phase, the goal is to identify the variables from the original source program that are shared among tasks in the task flow graph.
These variables will need to either be (i) extracted to global variables in the original source code such that all of the functions involved have access to them or (ii) passed in to each of the DAG nodes as arguments and added to the \texttt{Variables} dictionary of the application's JSON.
If the variables are extracted to a global scope in the original source code rather than moved to the application's JSON, then it remains the responsibility of the user application to manage them and ensure that they can be passed to the heterogeneous accelerator.
This is due to the fact that, as the memory holding data to be processed is within the user application's control, the code to send it elsewhere must remain in their control as well.
However, there are certain cases where this enables a higher degree of flexibility than what CEDR's memory initialization methods can accommodate.
Meanwhile, if the variables are re-defined using the JSON, they will be managed by CEDR, and less work is required on the application's side to pass them to an accelerator.
With these characteristics in mind, in the next phase, \textit{Code Restructuring}, the code is rewritten to have the new DAG-based structure and memory layout that was determined in the preceding stages.
Finally, we end with the \textit{DAG Generation} and \textit{Fat Binary Compilation} phases, in which the user constructs the DAG that corresponds to their restructured application and compiles their application binary, where this last phase is typically done by compiling and linking the restructured application code into a shared object.

\subsection{Research Questions} \label{subsec:overview_research_questions}

Establishing an understanding of the DSSoC in general will require
addressing a number of smaller questions across the spectrum of compilation and runtime characteristics.
First, on the compilation front: how do we compile applications that take full advantage of the heterogeneous nature of the underlying DSSoC, and when porting from serial application code, how do we leverage memory analysis results to enable parallel execution of nodes in the resulting DAG?
On the runtime front: what kind of runtime management practices enable integration of advanced resource management techniques while minimizing overhead of the runtime itself?
One investigation into this is done via schedule caching in Section~\ref{subsec:schedule_caching}, and another will be done via PE-level work queues in Section~\ref{subsec:pe_work_queues}.
Additionally, for applications that are highly throughput-oriented, how can a runtime such as CEDR be optimized for these "stream-based" applications while still allowing for full flexibility on the part of the scheduler?
We shall investigate optimizations that can be enabled for stream-based applications in Section~\ref{subsec:dag_streaming}.
Perhaps the most important role of CEDR is to enable application developers and hardware architects to interrogate design decisions in the trade space of scheduler heuristics, hardware configuration and workload characteristics. 
In Section~\ref{sec:experimental_evaluation}, we will demonstrate this multi-dimensional search space analysis capability with a case study on identifying the most suitable scheduler while sweeping through dynamically arriving workload scenarios and hardware compositions with execution time, scheduling overhead, and resource utilization metrics. Section~\ref{sec:experimental_evaluation} will allow us to answer research questions such as “Is accelerator always the best choice?” and “Is the scheduler with best cumulative execution time performance always the best choice?”. Answering these questions will lead to better understanding of why task to PE mapping decisions should be made dynamically by the runtime system rather than statically by the application developers, and why a complex and sophisticated scheduler isn’t always the best choice for dynamically arriving applications with high degree of concurrency.

Taken together, we believe that, in the context of the broader research efforts on domain-specific architectures, CEDR provides a unique set of features that are particularly tailored for investigating some of the open questions in this area.
Towards this end, CEDR has been designed with the following users in mind:
\begin{itemize}
    \item \textbf{Application programmers}: application programmers are domain experts who have the knowledge of the underlying domain, and they are willing to write their applications subject to any number of constraints or requirements as determined by the compiler, runtime, and hardware developers. They serve the critical role of enabling each of the three other roles to meaningfully test their work with workloads that extend beyond otherwise trivial benchmarks.
    \item \textbf{Compiler developers}: for compiler developers who wish to compile applications to DAG-based representations, CEDR provides a flexible runtime for which they can evaluate the output of their compiler across heterogeneous platforms.
    \item \textbf{Hardware architects}: for hardware architects who wish to determine the optimal accelerators to include in their DSSoC, CEDR's ability to assist in rapidly porting applications to novel architectures allows them to quantitatively evaluate their pre-silicon hardware designs across workloads that extend beyond simple micro-benchmarks.
    \item \textbf{Scheduling heuristic developer}: for scheduling heuristic developers, CEDR provides a means by which policies can be easily implemented, integrated and compared across applications and architectures without any changes required in the policies themselves.
\end{itemize}
We believe that in its current state CEDR has the ability to address, at some level, the needs of each of these users respectively.
In the subsequent sections, we will explore the ways in which this is achieved, starting first with characterization of the target applications and experimental methodology used in the remainder of the work.

\section{Experimental Setup} \label{sec:experimental_setup}

In Section~\ref{sec:experimental_evaluation}, we will demonstrate the capability of the CEDR framework through a number of experiments.
In this section, we provide background on the applications and hardware platforms that are used in these studies.
All experiments are performed on a Xilinx Zynq Ultrascale+ ZCU102 MPSoC development board~\cite{ZCU102}.
This MPSoC combines general purpose CPUs (4x Arm Cortex A53 processors) with programmable FPGA fabric. 
To demonstrate the portability of CEDR, we also conduct experiments on the Nvidia Jetson AGX SoC platform that couples 8 Arm cores with a Volta GPU. 
We vary the heterogeneous hardware configurations by adjusting the number and types of processing elements (PEs) used by CEDR, such as the number of CPUs as well as the accelerators present in the FPGA or GPU fabric.  
On the FPGA fabric, we add the Xilinx FFT IP accelerator for Fast Fourier Transform (FFT) and a custom designed accelerator for Matrix Multiply (MMULT) computation, with the FFT IP able to support all the sizes of radix-2 between 8 point and 2048 point FFT computations.
To facilitate data transfer to and from accelerators, we use direct memory access (DMA) blocks to move data between the host CPUs and the hardware accelerators via the AXI4-Stream protocol~\cite{AXI4}.
On the host side, we utilize \textit{udmabuf}~\cite{udmabuf} to enable contiguous userspace-accessible buffers for transferring data to and from the hardware accelerators.
A user application communicates with the accelerators by writing the data into a udmabuf buffer and a DMA engine is then configured to move data from this buffer into an accelerator for processing.
Once the accelerator completes processing the task, it then writes the data, through the DMA engine, back into the udmabuf buffer.
All-in-all, the ZCU102 platform has 3 Arm CPU cores, 1 FFT accelerator, and 1 MMULT accelerator that can be used as PEs by the CEDR runtime. One of the 4 present Arm cores is used to execute the runtime itself.

\begin{table*}[]
    \centering
    \begin{tabular}{|c|c|c|c|c|c|}
        \hline
        Application         &  
            \begin{tabular}{c}Avg. Exec. Time\\CPU (ms)\end{tabular} & 
            \begin{tabular}{c}Task\\Count\end{tabular} & 
            \begin{tabular}{c} FFT\\Support \end{tabular} & 
            \begin{tabular}{c} MMULT\\Support \end{tabular} \\
        \hline \hline
        \begin{tabular}{c}Radar\\Correlator\end{tabular}    &  0.82   & 7    &  \checkmark  &      \\
        \hline
        \begin{tabular}{c}Temporal Mitigation\end{tabular} &  4.39   & 11    &     &  \checkmark   \\
        \hline
        WiFi TX             &  16.12   & 93    &  \checkmark   &      \\
        \hline
        Pulse Doppler      &  95.83   & 1027    &  \checkmark  &      \\
        \hline
    \end{tabular}
    \caption{Basic characteristics of the applications used in experimental studies}
    \label{tab:application_characteristics}
\end{table*}
On the Jetson platform, in order to run the accelerator functions (FFT and MMULT) on the GPU, we use the \texttt{cuFFT} and the \texttt{cuBLAS} APIs provided by CUDA respectively. Within \texttt{cuFFT}, we use the \texttt{cufftExecZ2Z} function that performs a complex to complex transform on double precision data. For the MMULT, we use the \texttt{cublasCgemm} function that multiplies two matrices with complex numbers of single precision. 
Both functions use the standard \texttt{cudaMemcpy} functions for copying data between the GPU device memory and the host memory, over the PCIe interface. 

For our analysis we use four representative real-world applications from the domain of software defined radio: Radar Correlator (RC), Temporal Interference Mitigation (TM), Pulse Doppler (PD), and WiFi TX (TX).
Radar correlator models the use of a radar pulse to determine distance to an object by looking at the time delay in the received pulse compared to the input pulse. 
This application involves calculating the time shift in the received signal with respect to the transmitted signal, using two 256 point FFT computations. Hence it is a suitable candidate for studying moderate use of FFT accelerator.
Temporal Interference Mitigation is a computational kernel that receives a signal consisting of low-energy radar signals combined with high-energy communications data and applies a technique known as \textit{successive interference cancellation} to cancel out the communications data and extract the radar signals for further processing. 
This process relies heavily on matrix multiplication to cancel out the incoming communications data, and as such, it is an excellent candidate for the matrix multiplication accelerator.
In Pulse Doppler, a series of short radar pulses are emitted, and the user application observes the shift in the frequencies of the return pulses with respect to the input pulse, to determine both the distance of an object and its velocity.
This application has up to 128 parallel 256 point FFT task nodes, which makes it an ideal candidate for studying heavy use of FFT accelerator.
Finally, WiFi TX works by implementing a WiFi transmit chain, generating a single packet with 64 bits of input data and scrambling, encoding, modulating, and adding forward error correction to it for transmission over an arbitrary communications channel.
This application involves one inverse FFT operation of size 128 point per packet transmission. Hence it can be considered as a good scenario of moderate FFT utilization in a high latency job. The serial chain of processing seen in the WiFi TX also makes this application a suitable application to couple   with Pulse Doppler type of application and generate a workload that mixes serial execution with concurrency.  
In Table~\ref{tab:application_characteristics}, we summarize execution time and complexity in terms of number of tasks in the DAG based representation of each application. 

For the purposes of our studies in this paper, we aim to use these four applications according to their ability to stress the CEDR runtime differently, with the low latency applications helping to expose runtime overhead in various aspects of the system and high latency applications stressing the ability of the scheduler to effectively manage resources without becoming over-encumbered.
With this aim in mind, from Table~\ref{tab:application_characteristics}, we can see that the applications clearly fall into two distinct categorizations based on average observed latencies: Radar Correlator and Temporal Mitigation fall into the category of \textit{low} latency applications, whereas WiFi TX and Pulse Doppler fall relatively into the category of \textit{high} latency applications. 
Therefore, we classify these two sets of applications with different scale of latencies into two workloads as shown in Table~\ref{tab:workload_characteristics}. 
Each workload consists of even mixture of its constituent applications in terms of the number of application instances. 
Mixing applications of different latencies in the same workload would result in high latency applications dominating PE utilization and, in turn, the runtime execution, while correspondingly delaying and diluting the impact of low latency applications in the broader scope of the experiment. 
Furthermore, Table~\ref{tab:workload_characteristics} shows that the input data sizes for low and high latency workloads are smaller and larger respectively. 
We can see that the adopted workload categorization allows us to simultaneously observe the effects of application latency and input data rate on the runtime execution and utilization of PEs.
These workloads will be referred to as \textit{Low Latency} Workload and \textit{High Latency} Workload respectively.
\begin{table*}[]
    \centering
    \begin{tabular}{|c|c|c|c|c|}
        \hline
        Workload & Applications & \begin{tabular}{c}Application\\instances\end{tabular} & \begin{tabular}{c}Total Input\\data size (Kb)\end{tabular} & Accelerator\\
        \hline \hline
        Low latency & \begin{tabular}{c}Radar Correlator,\\Temporal Mitigation\end{tabular} & 10, 10 & 640 & FFT, MMULT \\
        \hline
        High latency & \begin{tabular}{c}Pulse Doppler,\\WiFi TX\end{tabular} &  5, 5 & 5185 & FFT\\
        \hline
    \end{tabular}
    \caption{Characterization of workloads used for CEDR experiments.}
    \label{tab:workload_characteristics}
\end{table*}
In Table~\ref{tab:exp_configurations}, we summarize the configuration parameters used in our experiments. On the ZCU102 platform with 3 Arm Cores along with FFT and MMULT accelerators, we compose 12 hardware configurations. 
We use 29 injection rates, where each injection rate defines a periodic rate of job arrival for its given workload in microseconds. 
We sweep all input configurations, repeat each experiment five times, and collect performance metrics listed in the table by averaging across the five runs. 
The scheduling overhead metric captures the time spent by the runtime in making scheduling decisions. This time is proportional to the number of scheduling rounds made by the runtime as well as the complexity of the scheduling algorithm.
The cumulative execution time of an application is the sum of execution times of its individual tasks, ignoring the overhead associated with scheduling them. Lower cumulative execution time indicates better scheduling decisions made and better exploitation of the available heterogeneity at hand.
The application execution time is the difference between the end of the last task and the start of the first task of an application, including the overhead of all scheduling decisions in between. Lower execution times indicate the scheduler's capability to manage the workload efficiently. 
The resource utilization ratio is the ratio between the total time a PE is active and the overall end-to-end execution time. A higher resource utilization on each type of PE indicates scheduler's capability of better exploiting heterogeneity.
To make the first three metrics comparable across different runtime configurations, we normalize them with the number of applications (per application). 
Then we average these three normalized metrics along with the resource utilization ratio, across five repetitions. 
For brevity, throughout the paper, we will take each metric (for instance ``cumulative execution time'') to refer to its corresponding averaged-per-application version (``average cumulative execution time / application'').
Overall, we conduct an exhaustive sweeping experiment that covers 3480 configuration scenarios for our analysis on an off-the-shelf SoC platform under three hours.
While three hours may seem fairly long, we note that this is orders of magnitude faster than an equivalent sweep relative in many cycle-accurate and discrete-event simulators.
As one point of reference, Fig. 19(c) in DS3~\cite{arda_2020_DS3} quantifies the simulation overhead versus the number of tasks present in the simulation.
In our experiments, each design point in our high latency workload experiments consists of over 5,000 application tasks, and accordingly, we can estimate that this workload would incur a simulation overhead of approximately 10 ms/$\mu$s, a 10,000x slowdown.
Assuming that this simulation overhead can be reduced through means such as rescaling the shortest task in the workload to execute for a handful of simulation timesteps, it is perhaps reasonable to optimistically assume that this overhead can be reduced to 50 - 100x.
Even then, that would imply that the 3 hour configuration sweep presented here would require on the order of 150 - 300 hours, and this disregards the fact that there will undoubtedly be differences in the simulation output given that executing in a framework like CEDR is inherently ``cycle-accurate'' while high level simulators are not.
If the goal is to avoid these differences, then a cycle-accurate methodology must be applied.
Again referring to DS3, they find that their relative simulation speedup against gem5 is 600x, and as such, by leveraging this number, we can estimate that an equivalent sweep would incur a 30,000 - 60,000x simulation overhead, requiring a simulation time on the order of months to execute.
In the following subsections, we present these metrics with respect to varying runtime configurations and analyze their implications.

\begin{table*}[]
    \centering
    \begin{tabular}{|c||c|}
        \hline
        \multirow{8}{*}{Input Configurations }& \begin{tabular}{c}12 Hardware configurations\\3 CPUs (C1-C3), 1 FFT (F0-F1), 1 MMULT (M0-M1)\end{tabular}\\\cline{2-2}
        & 5 Schedulers (SIMPLE, MET, EFT, ETF, HEFT\textsubscript{RT}) \\
        \cline{2-2}
        & 2 Workloads (High latency, Low latency) (Table~\ref{tab:workload_characteristics})\\
        \cline{2-2}
        & \begin{tabular}{c}29 Injection rates\\High latency (29 points between 10-2000 Mbps)\\Low latency (29 points between 1-1000 Mbps)\end{tabular}\\
        \hline \hline
        \multirow{4}{*}{Output Metrics} & Average cumulative execution time/ application\\
        \cline{2-2}
        & Average execution time / application \\
        \cline{2-2}
        & Average scheduling overhead / application\\
        \cline{2-2}
        & Average resource utilization ratio\\
        \hline
    \end{tabular}
    \caption{Runtime configurations to sweep across, and performance metrics to capture in large scale CEDR validation. CPU core configuration is indexed as 1 to 3 since minimal resource pool is composed of a single CPU core, but notation for accelerators use 0 or 1 indicating whether a specific accelerator is used or not. }
    \label{tab:exp_configurations}
\end{table*}

\section{Experimental Evaluation} \label{sec:experimental_evaluation}

In this section, we will demonstrate the utility of CEDR through validation-based sweeping experiments that show its ability to yield insights in application design, scheduler integration, and accelerator verification before moving on to showing advanced features that are possible through the use of schedule caching, queuing, streaming, and performance counters. The experiments presented in this section involve sweeping the search space of runtime configurations based on hardware composition, scheduling heuristics, and application workloads for various data rates as defined in Table~\ref{tab:exp_configurations}.
In the following subsections, we present the four performance metrics captured from the sweeping experiments, and analyze the trends in these results.

\subsection{Runtime Configuration Sweep}\label{subsec:runtime_config_sweep}

\subsubsection{Average Cumulative Execution Time Analysis}\label{subsubsec:avg_cum_exec_analysis}

\begin{figure}[t]
    \centering
    \subfigure[]{\includegraphics[width=0.48\textwidth, trim=10cm 1cm 9cm 3.3cm, clip=true]{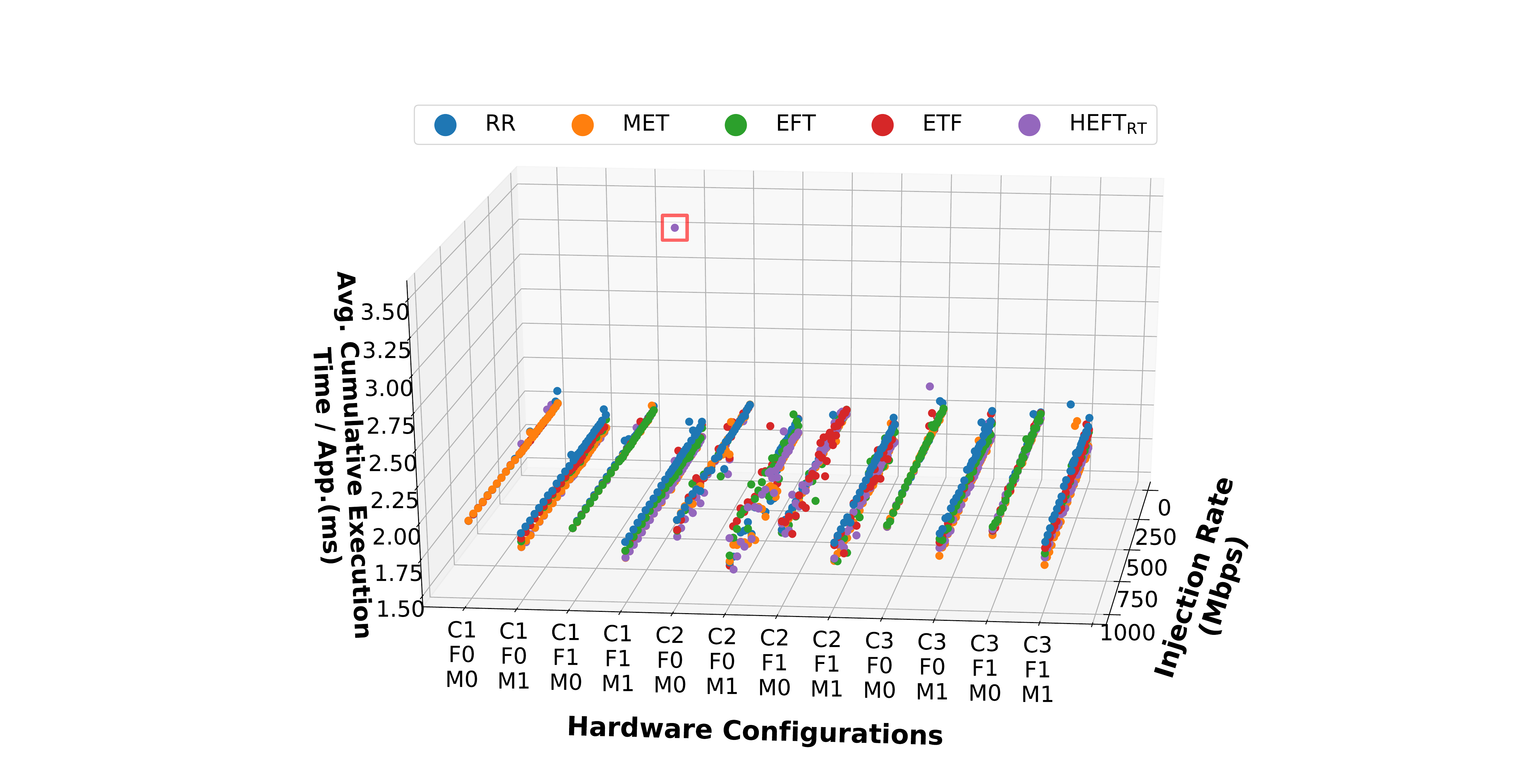}}
    \hspace{0.02\textwidth}
    \subfigure[]{\includegraphics[width=0.48\textwidth, trim=10cm 1cm 9cm 3.3cm, clip=true]{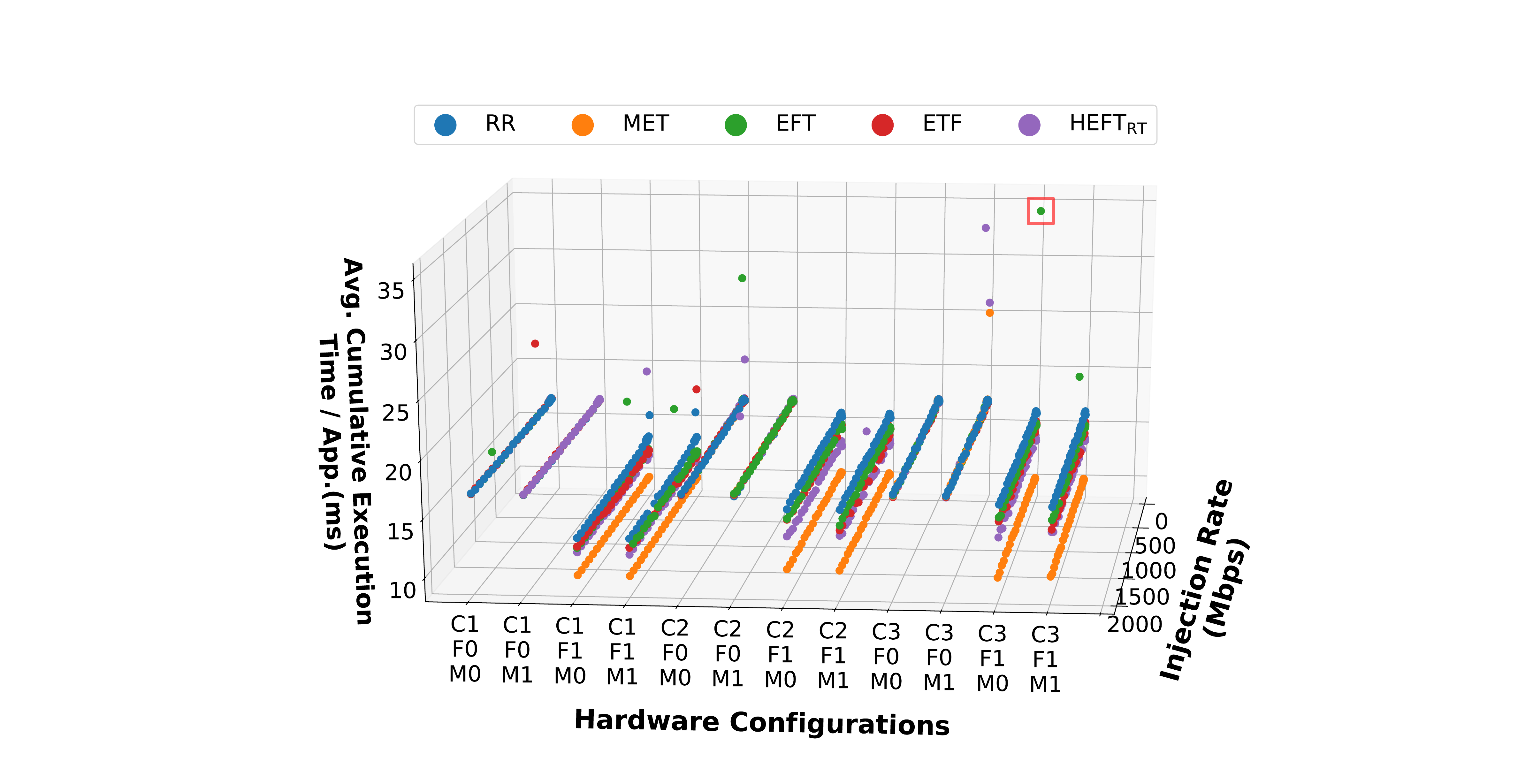}}
    \caption{Average cumulative execution time per application for 12 resource pool configurations, 5 schedulers and 29 injection rates, using (a) \textit{low} latency, (b) \textit{high} latency workloads.}
    \label{fig:cum_exec}
\end{figure}

Figure~\ref{fig:cum_exec} (a) and (b) present the average cumulative execution time per application for \textit{low} latency and \textit{high} latency workloads respectively, using the input configurations specified in Table~\ref{tab:exp_configurations}. Here, X-axis shows the distinct hardware configurations, Y-axis shows the injection rates, and Z-axis presents the average cumulative execution time per application. The unique colors of the data points correspond to different schedulers. 

We observe that the average cumulative execution time per application remains about the same for CPU-only hardware configurations (C1-F0-M0, C2-F0-M0 and C3-F0-M0) among both \textit{low} and \textit{high} workloads.
As expected, we observe a reduction in the cumulative execution time with the addition of accelerators listed in Table~\ref{tab:workload_characteristics} to the pool of PEs. 
For example, the \textit{low} latency workload benefits from both FFT and MMULT accelerators, while the \textit{high} latency workload benefits from the FFT accelerator.
Besides the data points that follow the above described execution trends, we also notice some outlier points with significantly higher cumulative execution times in Figure~\ref{fig:cum_exec} (a) and (b) as marked by a selected red square on each plot. These rarely occurring outliers are caused by one or more tasks within a workload running for up to an order longer than their expected execution time. Although these points may seem erroneous, we shall functionally verify them in Section~\ref{subsec:cedr_verification} and further explain the reason behind these long running tasks with the help of performance counters.

\subsubsection{Average Execution Time Analysis}\label{subsubsec:avg_exec_analysis}

\begin{figure}[t]
    \centering
    \subfigure[]{\includegraphics[width=0.48\textwidth, trim=8.8cm 1cm 8.5cm 3.3cm, clip=true]{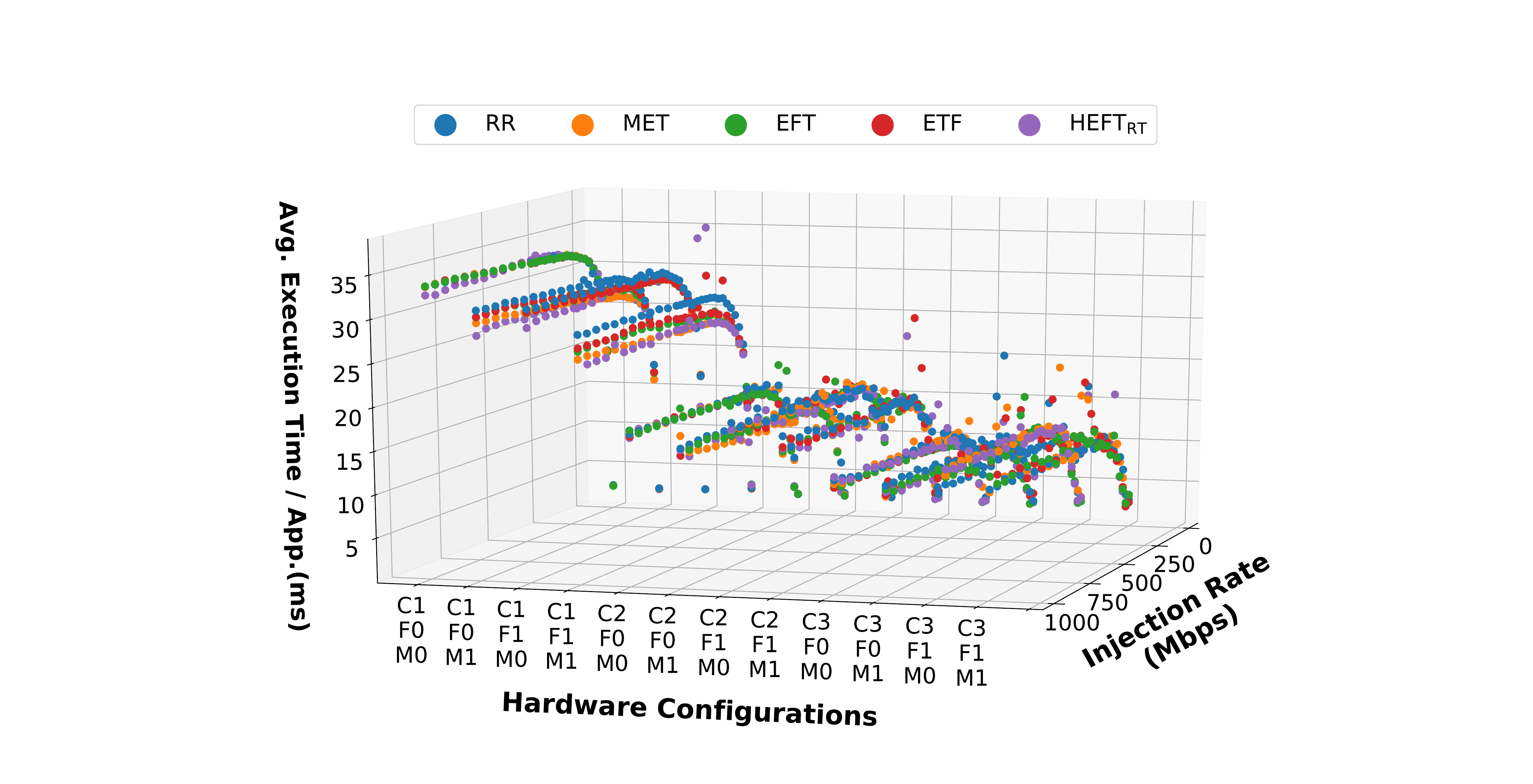}}
    \hspace{0.02\textwidth}
    \subfigure[]{\includegraphics[width=0.48\textwidth, trim=8.7cm 1cm 8.5cm 3.3cm, clip=true]{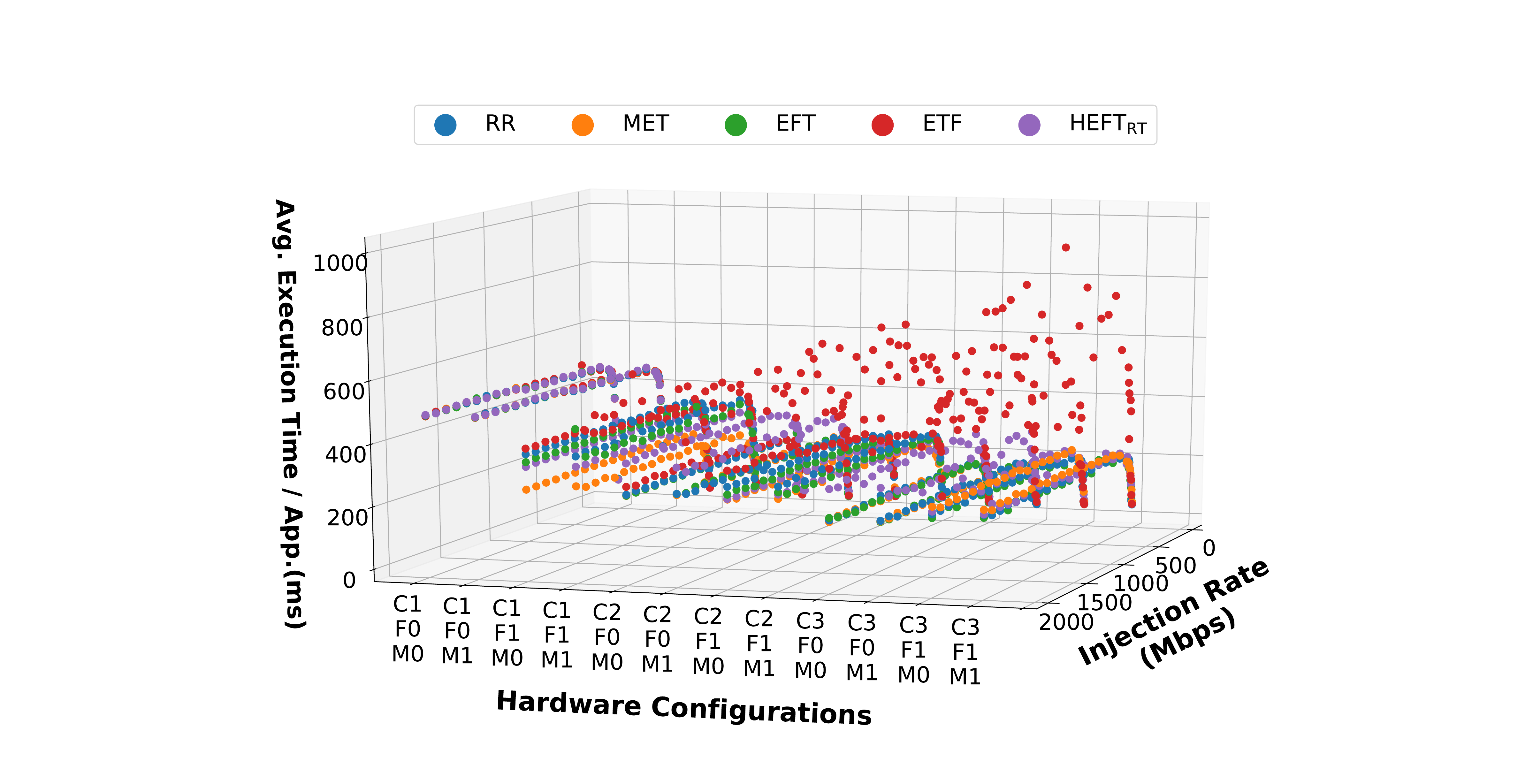}}
    \caption{Average execution time per application for 12 resource pool configurations, 5 schedulers and 29 injection rates, using (a) \textit{low} latency, (b) \textit{high} latency workloads.}
    \label{fig:exec}
\end{figure}

Figure~\ref{fig:exec}(a) and (b) show the average execution time per application (Z-axis) for \textit{low} and \textit{high} workloads respectively across the same configuration sweep used for generating Figure~\ref{fig:cum_exec}. 

Looking at these figures from the perspective of injection rate shows that, as we increase the injection rate, a saturation trend for average execution time is observed across all hardware configurations and schedulers for both \textit{low} and \textit{high} latency workloads.
This saturation trend is expected across all the injection rates, as we run the experiments for a fixed number of application instances rather than a fixed amount of time. 
Therefore, beyond a certain injection rate, the different application arrivals are frequent enough such that CEDR effectively receives them all simultaneously, and any increase in injection rate stops causing a meaningful difference in the workload as observed by CEDR.
At this point, CEDR becomes oversubscribed, and for fair/insightful evaluations, it is desired that our experimental evaluations capture these oversubscribed saturation regions.
From the hardware configuration perspective, we observe a decreasing/downward trend in execution time as the number of PEs and PE types increase, excluding the scenario where MMULT accelerator is added to the resource pool for the \textit{high} latency workload.

\begin{figure}
    \centering
     \subfigure[]{\includegraphics[width=0.49\textwidth, trim=0cm 0cm 0cm 0cm, clip=true]{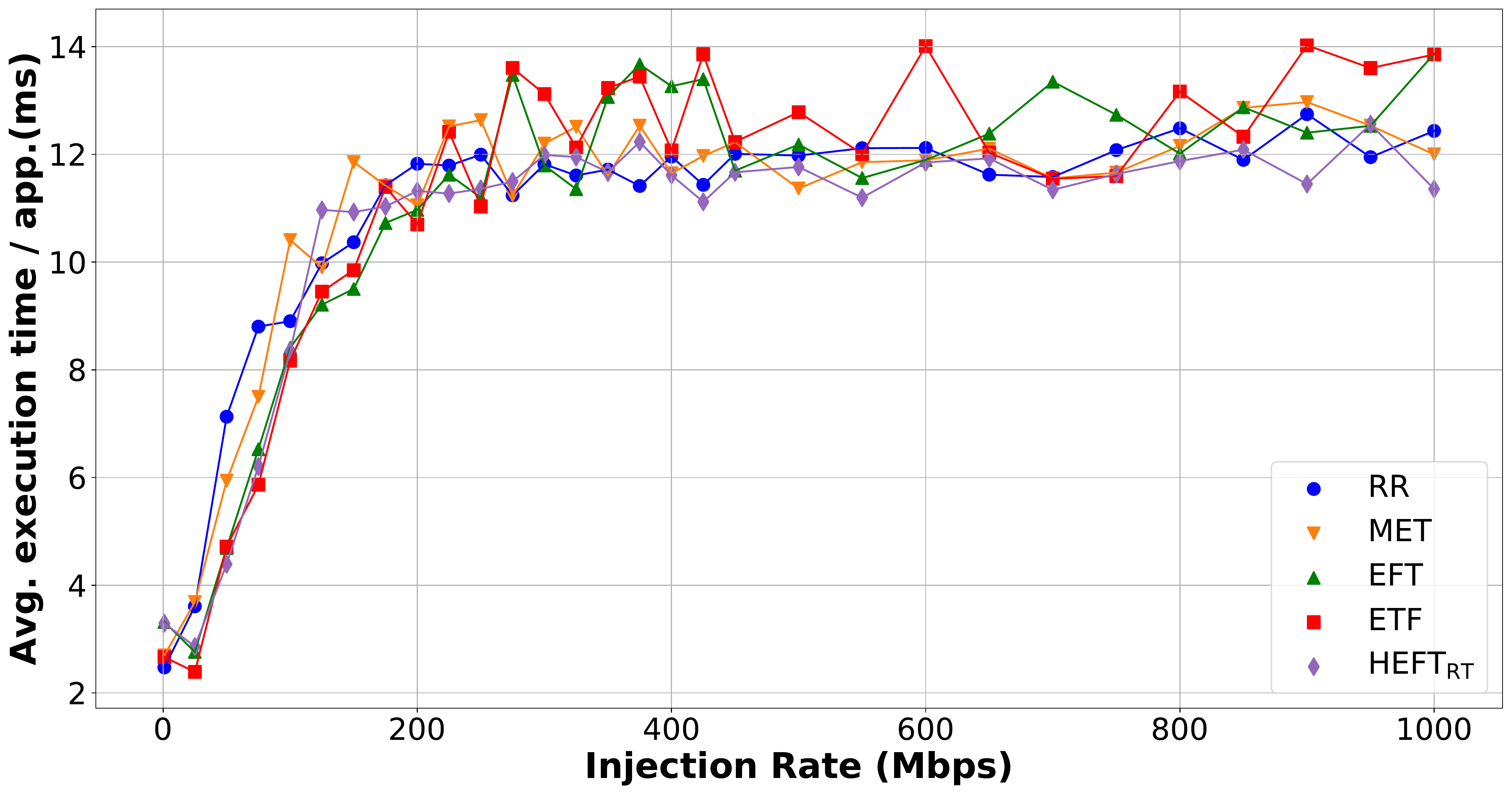}}
    \subfigure[]{\includegraphics[width=0.49\textwidth, trim=0cm 0cm 0cm 0cm, clip=true]{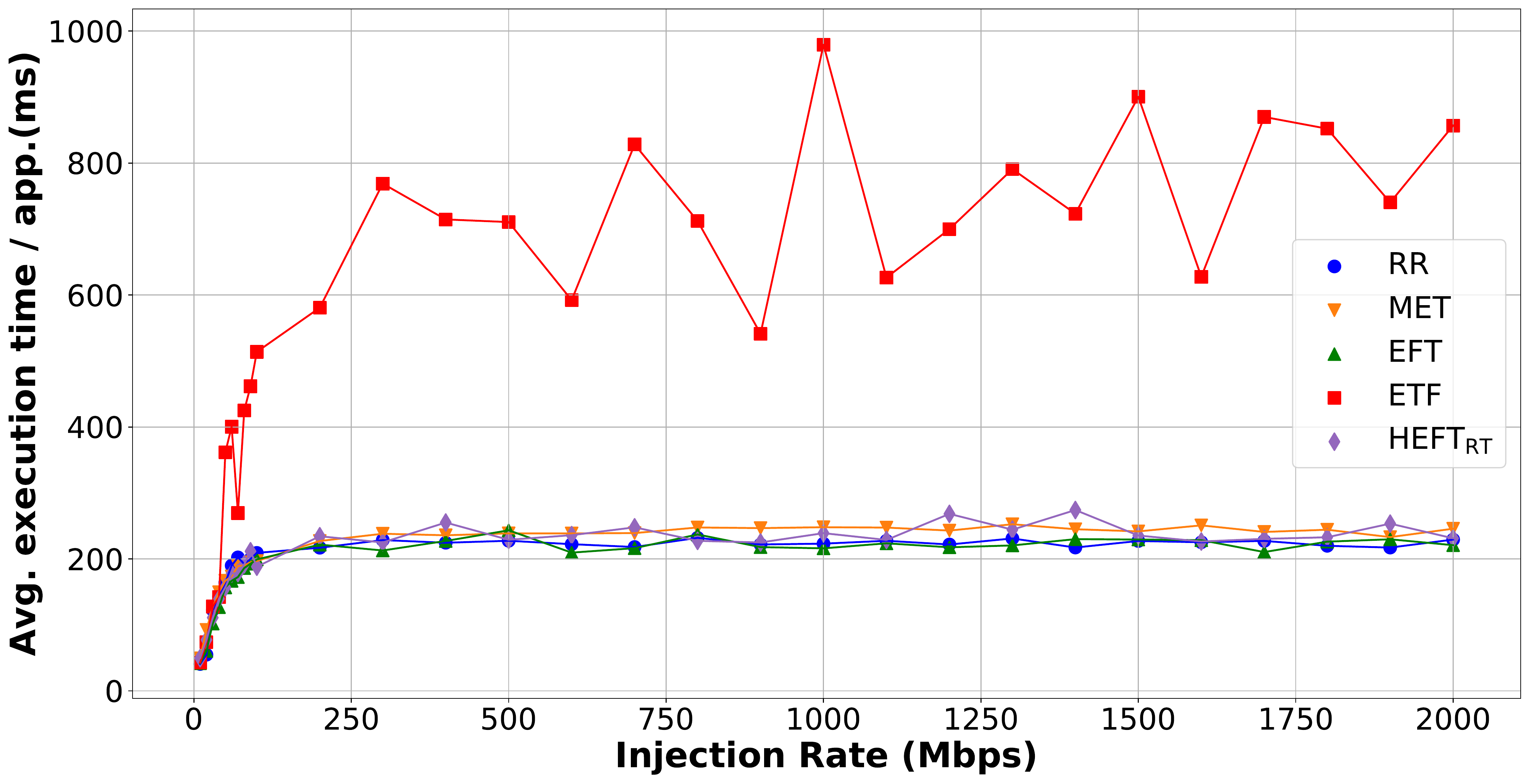}}
    \caption{Average execution time per application with respect to varying injection rates, for different schedulers with 3 CPUs, 1 FFT and 1 MMULT; using (a) \textit{low} latency workload, (b) \textit{high} latency workload.}
    \label{fig:exec_2d}
\end{figure}

Now, we analyze the downward trend in average execution time from scheduling heuristic point of view. 
We observe that for \textit{low} latency workload all the schedulers follow this downward trend. 
For \textit{high} latency workload, all schedulers follow the downward trend except for the ETF scheduler. 
Instead, ETF shows an upward trend in execution time with increasing number of PEs in the hardware configuration. 
To better view this trend, we fix the hardware configuration to the most heterogeneous one (3 CPUs, 1 FFT and 1 MMULT) and present the average execution time per application with varying injection rates in Figure~\ref{fig:exec_2d}(a) and (b) for \textit{low} and \textit{high} latency workloads respectively. For the \textit{low} latency workload, we notice that all of the schedulers show similar performance in Figure~\ref{fig:exec_2d}(a), confirming our observation from Figure~\ref{fig:exec}(a). For the \textit{high} latency workload, the ETF is the only divergent case with significantly higher execution time performance, while showing a saturation trend similar to the other schedulers. We attribute the higher execution time performance of ETF to the increase in its scheduling overhead with respect to increase in the number of PEs and types of PEs. The ETF scheduler traverses over the tasks ready to be scheduled, as well as the list of available PEs. Therefore, the complexity and overhead of ETF scheduler increases with the number of PEs and number of tasks ready to be scheduled. We will take a closer look at the scheduling overhead analysis in Subsection~\ref{subsubsec:avg_sched_overhead_analysis}.

\subsubsection{Average Scheduling Overhead Analysis}\label{subsubsec:avg_sched_overhead_analysis}

\begin{figure}[t]
    \centering
    \subfigure[]{\includegraphics[width=0.48\textwidth, trim=11cm 1.5cm 8cm 3cm, clip=true]{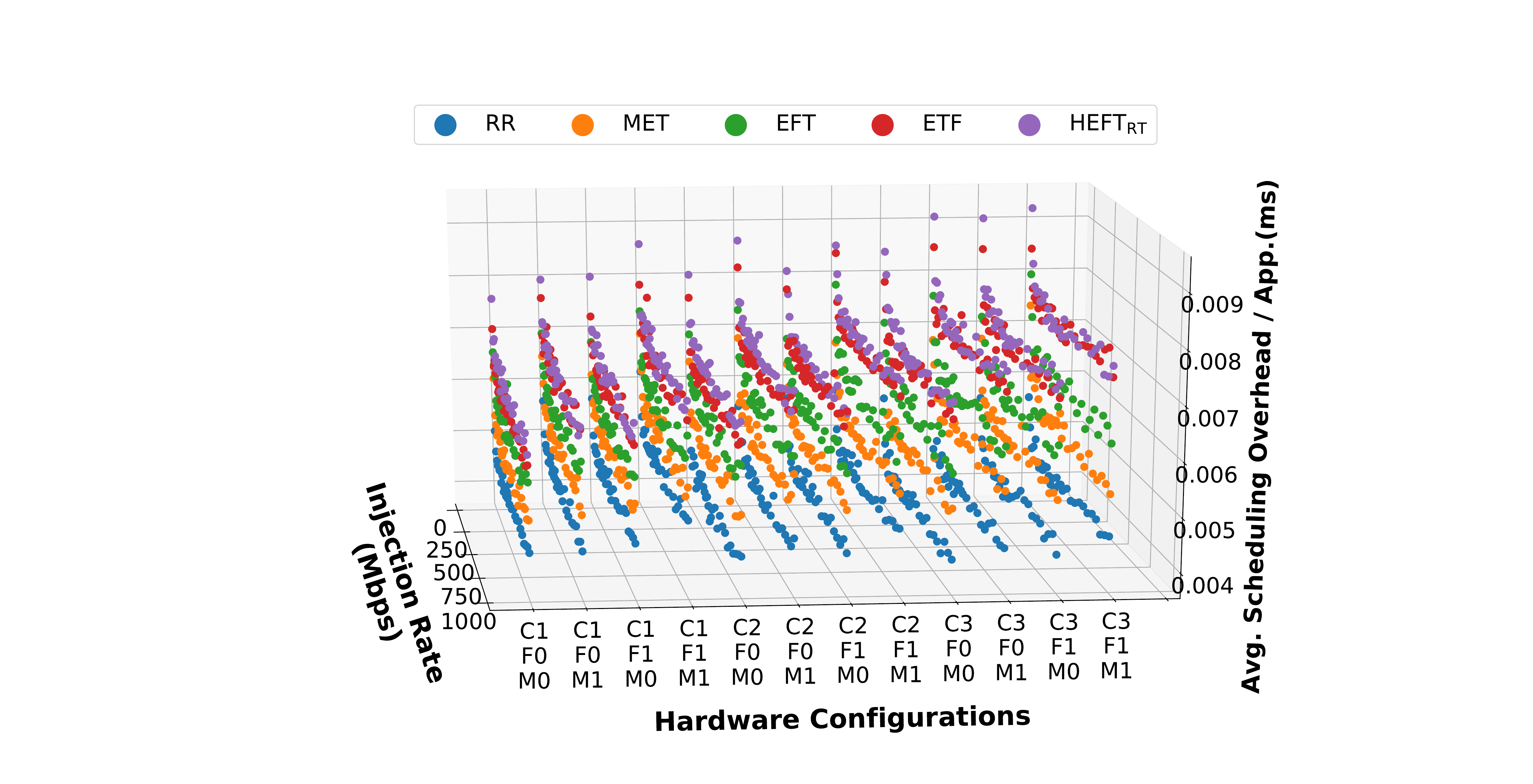}}
    \hspace{0.02\textwidth}
    \subfigure[]{\includegraphics[width=0.48\textwidth, trim=11cm 1.5cm 8cm 3cm, clip=true]{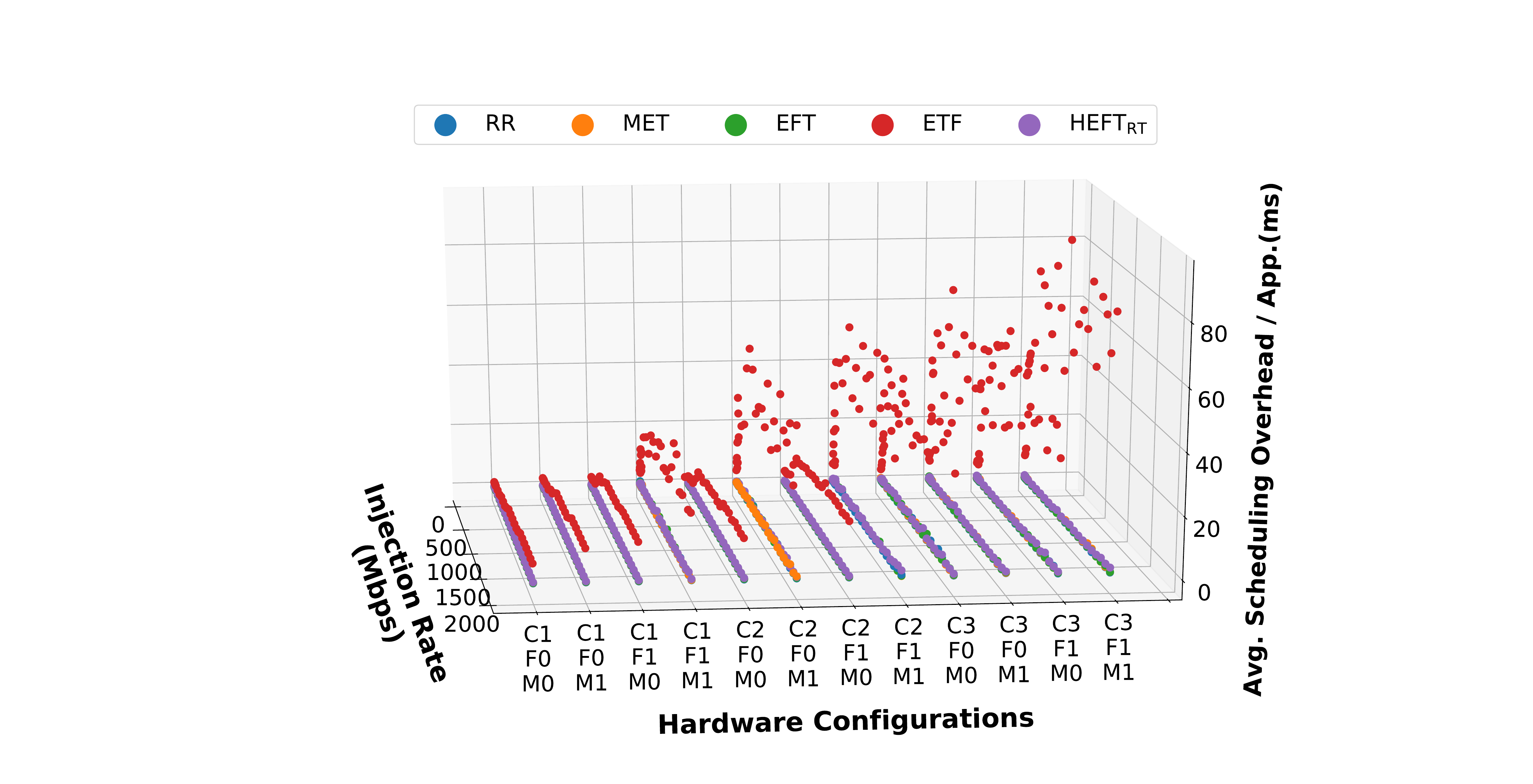}}
    \caption{Average scheduling overhead per application for 12 resource pool configurations , 5 schedulers and 29 injection rates, using (a) \textit{low} latency, (b) \textit{high} latency workloads.}
    \label{fig:sched_overhead}
\end{figure}

We show the average scheduling overhead per application (Z-axis) across the  sweeping input configurations in Figure~\ref{fig:sched_overhead}(a) and (b) for \textit{low} and \textit{high} latency workloads respectively.

We notice that the average scheduling overhead per application does not vary by any significant margin for the \textit{low} latency workload as shown in Figure~\ref{fig:sched_overhead}(a). However, organizing the five schedulers in an increasing order of average scheduling overhead gives us the order of- RR, MET, EFT, ETF and HEFT\textsubscript{RT}, where ETF and HEFT\textsubscript{RT} have similar average scheduling overheads. The ordering of schedulers aligns with the increasing complexity of these heuristics. %

\begin{figure}
    \centering
     \subfigure[]{\includegraphics[width=0.49\textwidth, trim=0cm 0cm 0cm 0cm, clip=true]{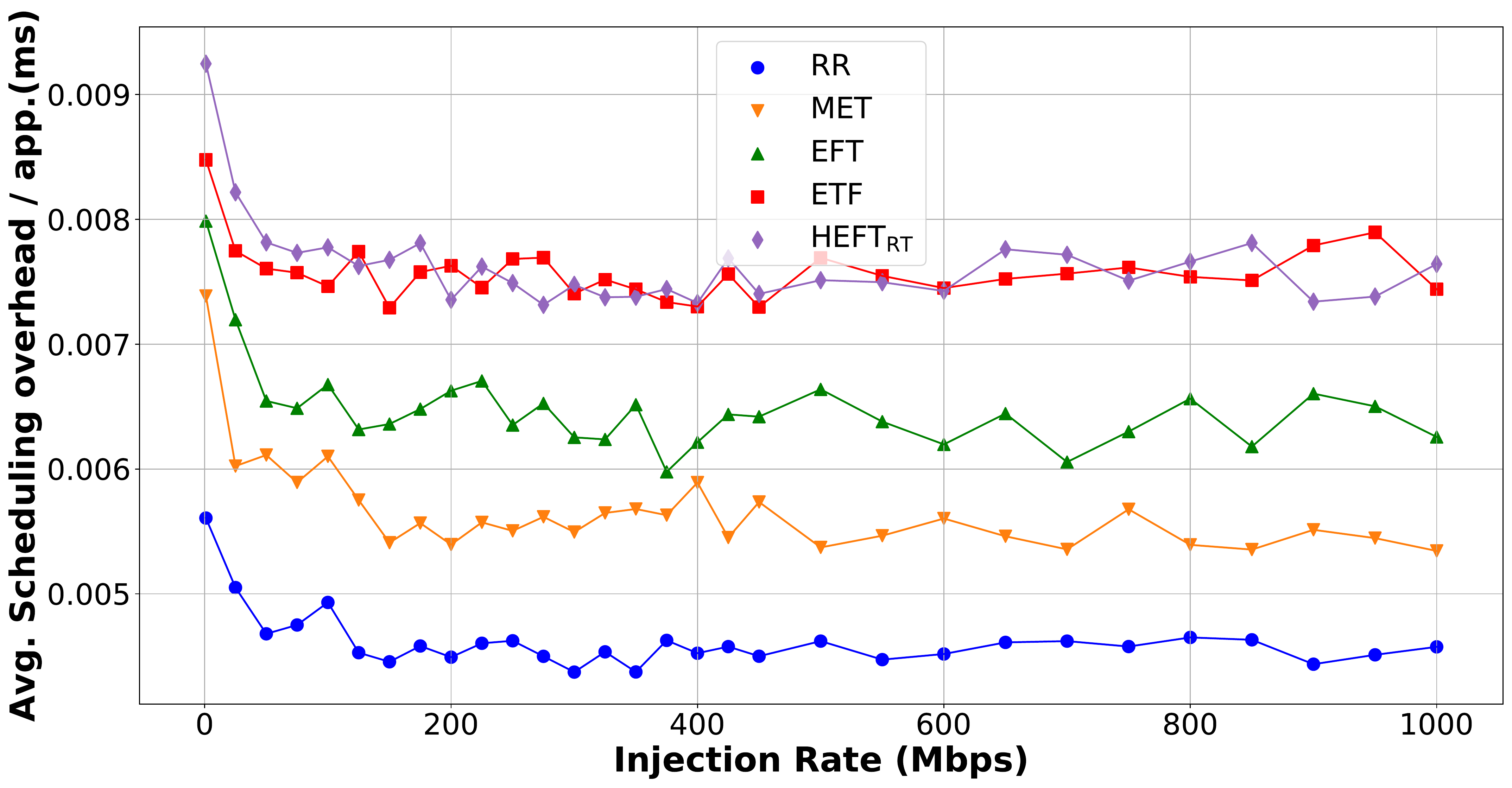}}
    \subfigure[]{\includegraphics[width=0.49\textwidth, trim=0cm 0cm 0cm 0cm, clip=true]{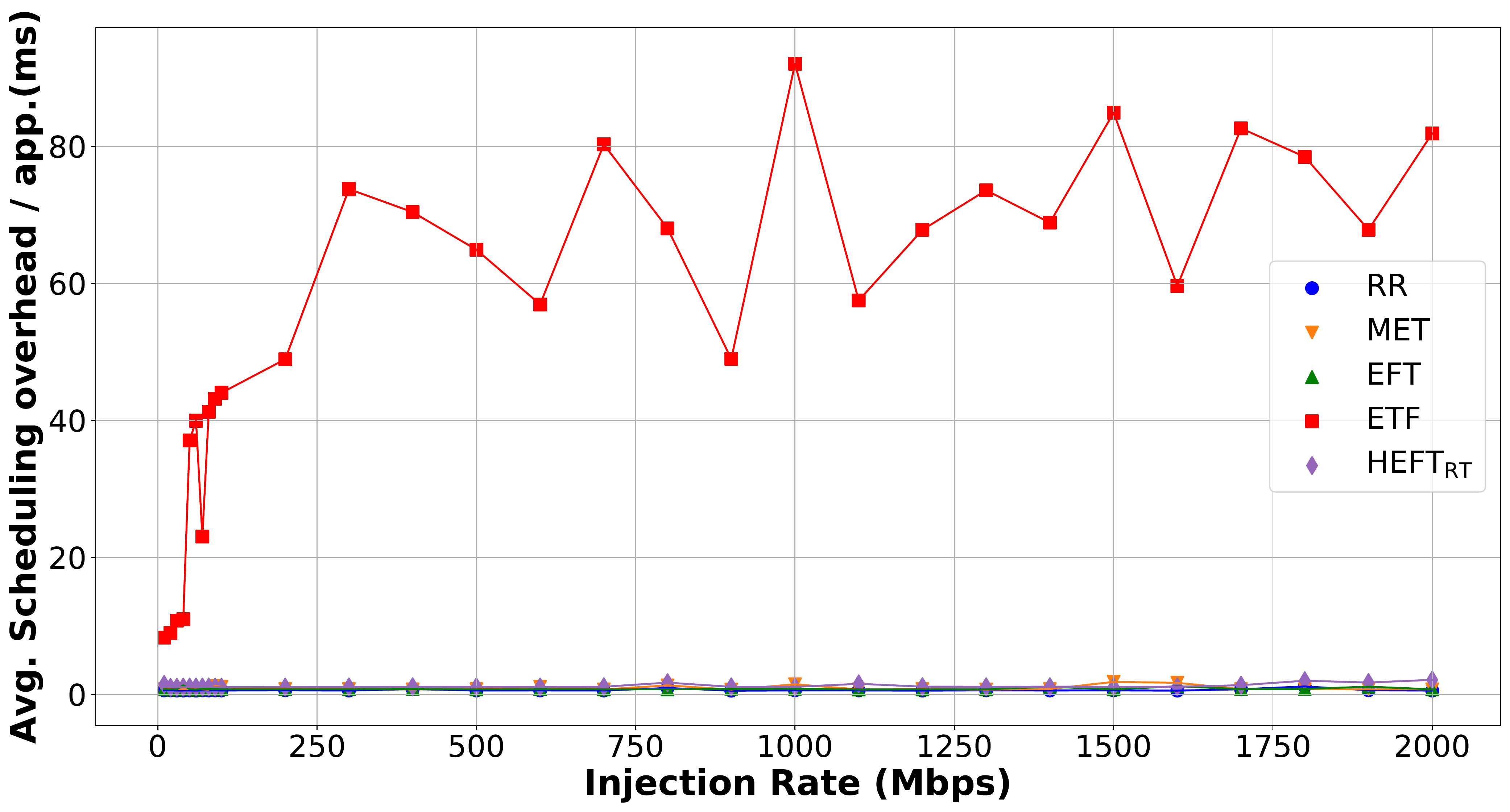}}
    \caption{Average scheduling overhead per application for the (a) \textit{low} latency and (b) \textit{high} latency workloads on 3 CPU, 1 FFT and 1 MMULT.}
    \label{fig:sched_2d}
\end{figure}

For the \textit{high} latency workload, as shown in Figure~\ref{fig:sched_overhead}(b), we observe that except for ETF, the remaining schedulers perform very similar across all injection rates and hardware configurations. Instead, the scheduling overhead of ETF increases with respect to increasing PEs (similar to our previous observation in Section~\ref{subsubsec:avg_exec_analysis}). 
Next we take a slice from the 3D plots by fixing the hardware configuration to most heterogeneous setup (3 CPUs, 1 FFT and 1 MMULT), and compare scheduling overhead trend with respect to injection rate in Figure~\ref{fig:sched_2d}(a) and (b) for \textit{low} and \textit{high} latency workloads respectively. In the \textit{low} latency workload scenario, the average scheduling overhead is slightly higher at lowest injection rates, but saturates at a lower value for increasing injection rates. The increased scheduling overhead at lower injection rates is caused by the applications arriving to the runtime in a less overlapped manner. Hence the scheduler runs more number of scheduling rounds for the same amount of workload. 
In the case of \textit{high} latency workload, the overhead of running ETF grows substantially for higher injection rates, compared to the remaining schedulers as seen in Figure~\ref{fig:sched_2d}(b). This increase can be attributed to complexity of ETF growing with the number of tasks ready to be scheduled, as well as the number of PEs. The \textit{high} workload consists of Pulse Doppler, which contains up to 512 parallel task nodes- making the cost of running ETF significantly higher on this workload. %
However, for the \textit{low} latency workload, the average scheduling overhead of ETF follows the trend of remaining schedulers as seen in Figure~\ref{fig:sched_2d}(a). We attribute this to the low number of tasks present in this workload, which is not enough to stress ETF, as opposed to the \textit{high} latency workload.

\subsubsection{Average Resource Utilization Ratio}

\begin{figure}
    \centering
     \subfigure[]{\includegraphics[width=0.48\textwidth, trim=0cm 0cm 2cm 1.5cm, clip=true]{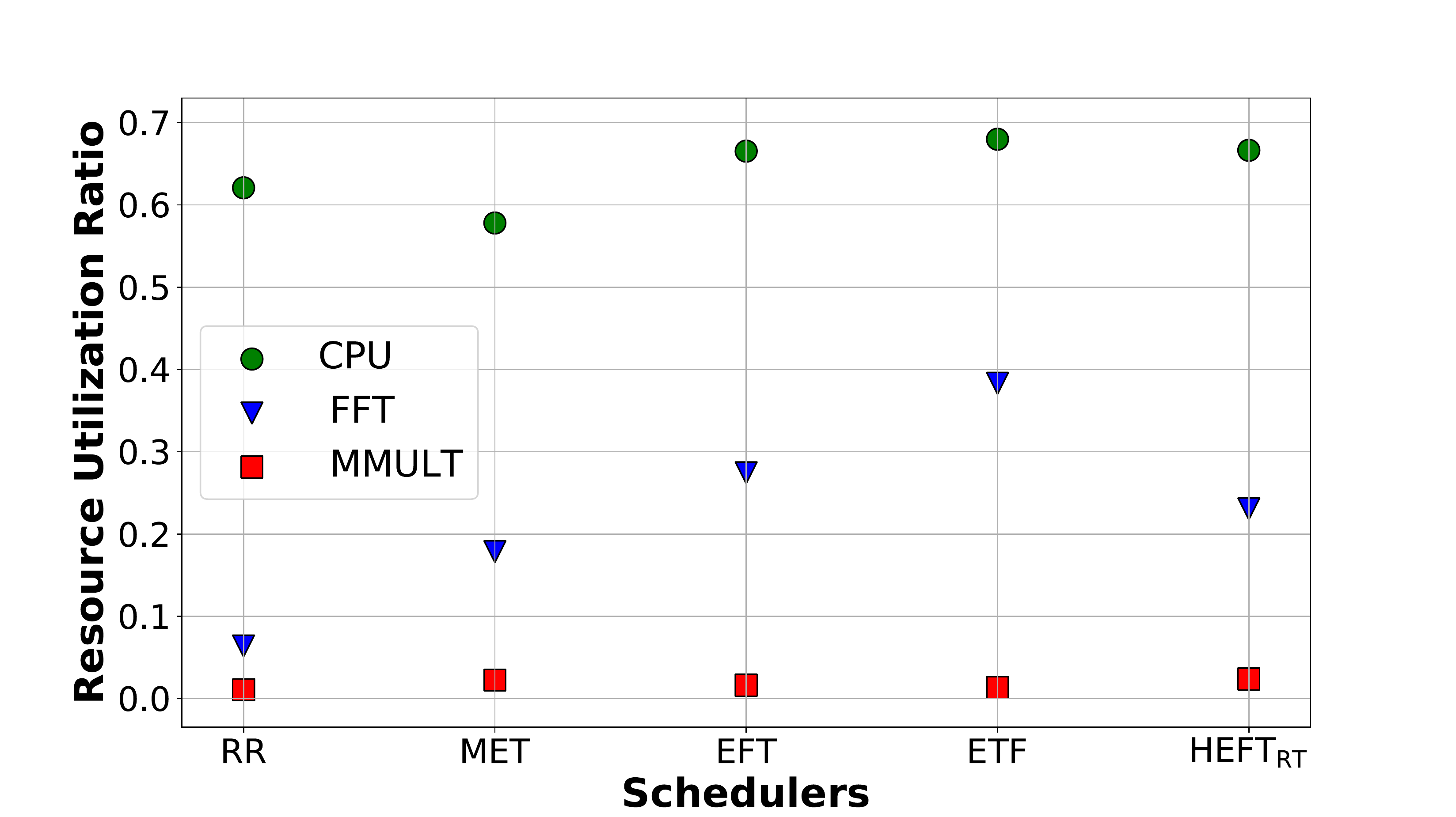}}
    \hspace{0.02\textwidth}
    \subfigure[]{\includegraphics[width=0.48\textwidth, trim=0cm 0cm 2cm 1.3cm, clip=true]{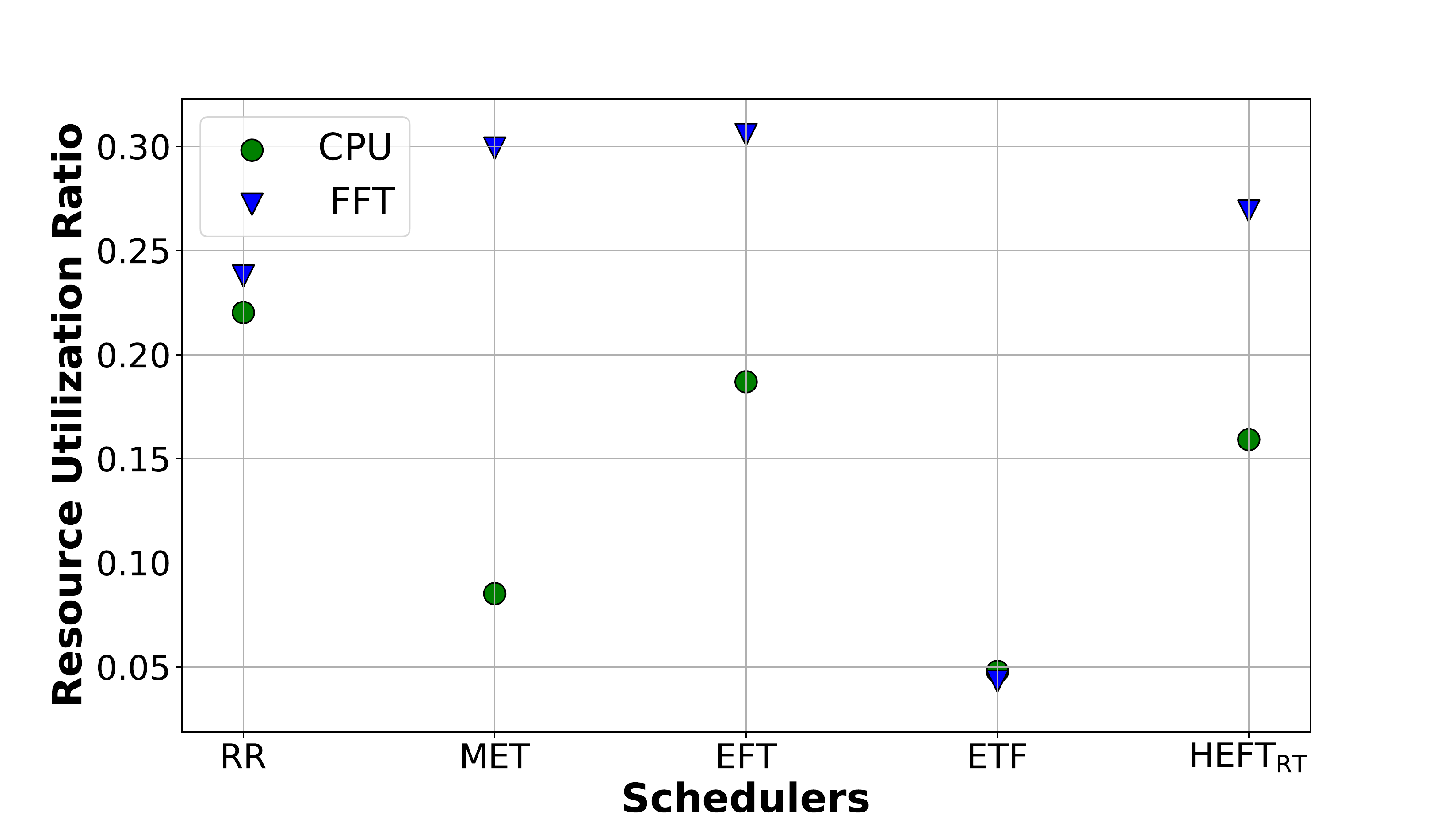}}
    \caption{Average resource utilization ratio for different types of PEs (for hardware configuration having 3 CPUs, 1 FFT, 1 MMULT) using (a) \textit{low} latency (injection rate 1000 Mbps) and (b) \textit{high} latency (injection rate 2000 Mbps) workload.}
    \label{fig:resutil_2D}
\end{figure}

Another metric for scheduler performance analysis is the measurement on how well the system resources are utilized. Improving the utilization of the resources enables realizing better parallel execution of tasks in the applications, thereby improving the end-to-end execution time. A heuristic that prioritizes minimum execution time such as MET, or a scheduler with high penalty in terms of time spent on scheduling decisions such as ETF are expected to play critical role on determining the scheduling objective function for the interleaved workload scenarios. In this experiment, we evaluate resource utilization performance of the five schedulers. For this, we choose the 3 CPU core, 1 FFT and 1 MMULT based hardware configuration for the \textit{high} latency and \textit{low} latency workloads. We set the injection rate to a high enough value (2000 Mbps) to ensure that the system is oversubscribed. 
Figure~\ref{fig:resutil_2D}(a) and (b) present the five schedulers along X-axis, and the corresponding average resource utilization ratio for each resource type that is color coded for CPU, MMULT and FFT along Y-axis for \textit{low} and \textit{high} latency workloads respectively. 

For the \textit{low} latency workload, Figure~\ref{fig:resutil_2D}(a) shows that, in terms of FFT accelerator utilization, ETF performs the best at $38.3\%$, followed by EFT and HEFT\textsubscript{RT} with $27.4\%$ and $23.1\%$ utilization respectively. The MET although assigns all of the FFT tasks to FFT accelerator, interestingly achieves a lower FFT accelerator utilization at $17.9\%$. We attribute this to MET's inability to fully take advantage of the remaining 3 CPUs to parallelize execution, causing the overall execution time of the workload to grow and in turn reducing the resource utilization ratio of FFT accelerator. Finally, as expected, the simple RR scheduler utilizes the FFT accelerator the least at $6.4\%$. Here we note that we don't observe insightful trend for the MMULT utilization since the amount of time spent on matrix multiplication in the low latency workload is very small. 

For the \textit{high} latency workload, Figure~\ref{fig:resutil_2D}(b) shows that EFT, MET and HEFT\textsubscript{RT} perform relatively well in terms of FFT accelerator utilization, with average resource utilization of 30.6\%, 29.9\% and 26.9\% respectively. The accelerator utilization drops to 23.7\% for the RR scheduler. For the ETF scheduler however, opposite to the trend observed for \textit{low} latency workload, here we notice poor average resource utilization of 4.3\% and 4.8\% for FFT accelerator and CPU. This behavior can be attributed to the large scheduling overhead of ETF for \textit{high} latency workload as shown in Section~\ref{subsubsec:avg_sched_overhead_analysis}. Due to increased overhead of scheduling, the PEs have to remain idle for longer period of time, causing the resource utilization ratio to drop significantly.

\subsubsection{Research Questions: Runtime Configuration Perspective}\label{subsec:runtime_rq}

The capability of CEDR to conduct large scale experiments as shown in Section~\ref{subsec:runtime_config_sweep} allows us to explore the answers to some crucial research questions related to the field of domain specific computing. In this section we explore the answer to the following two questions:

\begin{itemize}
    \item [RQ$_1$] Is accelerator always the best choice?
    \item [RQ$_2$] Is the scheduler with best cumulative execution time performance always the best choice?
\end{itemize}

In order to answer RQ$_1$, we utilize the end-to-end execution time as the key indicator of the efficient execution of a given workload on the runtime framework. This metric captures the total latency of the execution of the workload, along with any overhead encountered by the runtime to manage the workload. Furthermore, the effect of various task to PE type mapping decisions is reflected in the end-to-end execution time.

\begin{figure}[t]
    \centering
    \subfigure[]{\includegraphics[width=0.49\textwidth, trim=1.8cm 0.5cm 3cm 2.5cm, clip=true]{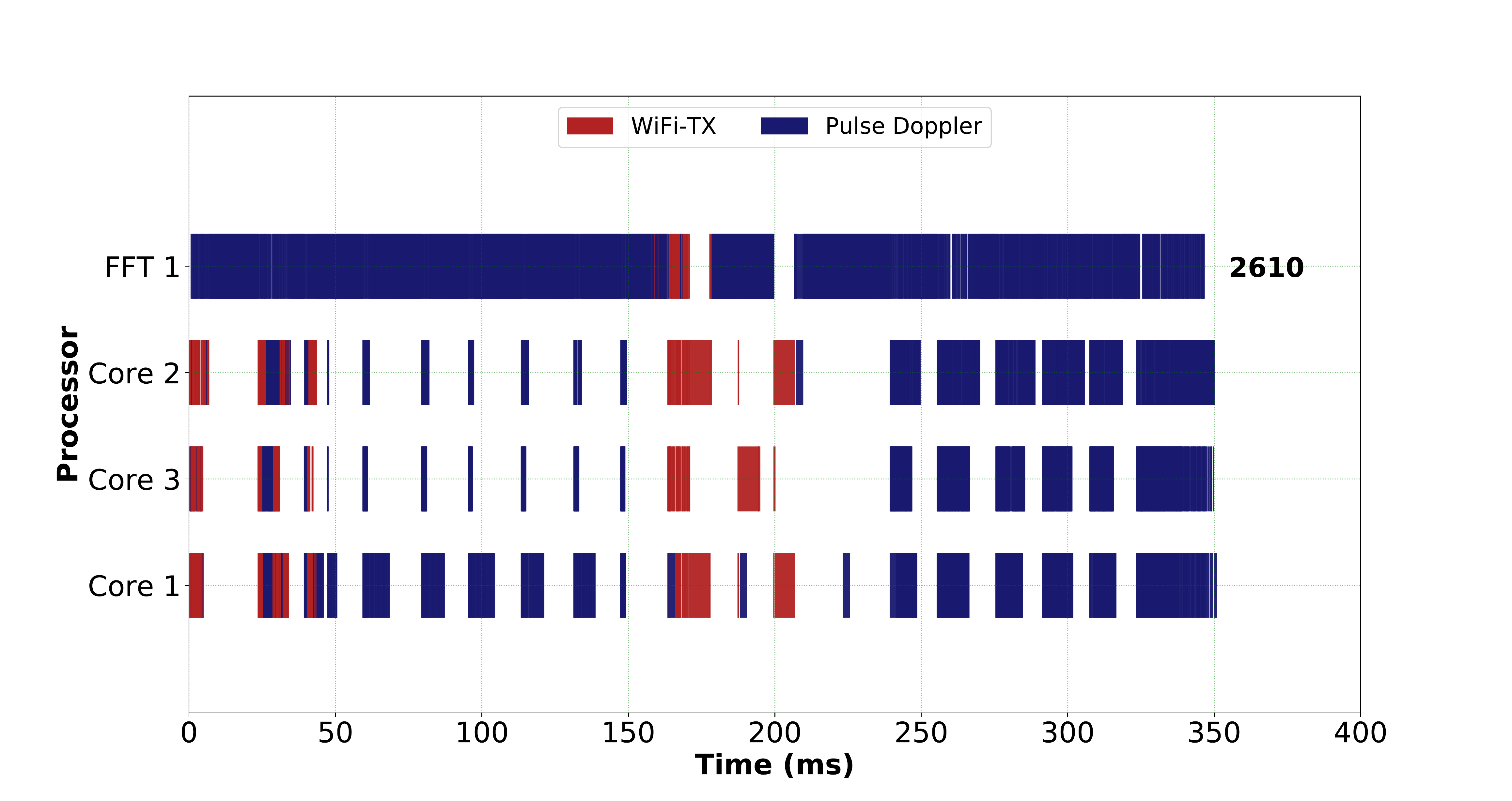}}
    \subfigure[]{\includegraphics[width=0.49\textwidth, trim=1.8cm 0.5cm 3cm 2.5cm, clip=true]{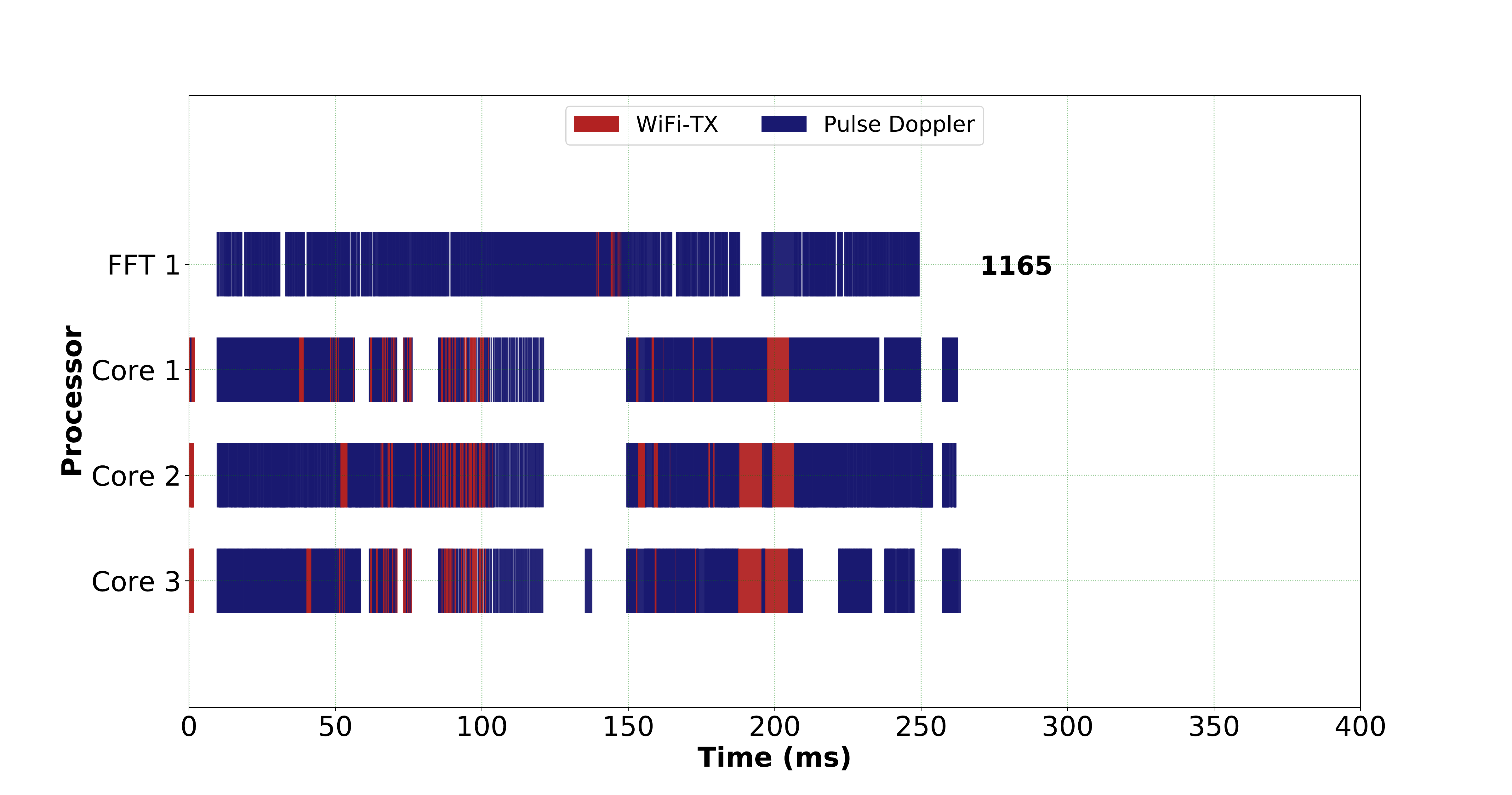}}
    \caption{Gantt chart showing execution of \textit{high} latency workload on 3 CPUs and 1 FFT using (a) accelerator only policy (\texttt{ACC\_only}), (b) both accelerator and CPU selection policy (\texttt{ACC+CPU}).}
    \label{fig:acc_vs_mixedPEs}
\end{figure}

We create two execution scenarios in CEDR, using the \textit{high} latency workload with the injection rate of 2000 Mbps, which belongs to the saturated region of average execution time per application as shown in Figure~\ref{fig:exec_2d}(b). This injection rate ensures that the system is oversubscribed. In the first execution scenario, MET scheduler is used. MET, due to its objective function, favors scheduling and executing FFT tasks on the accelerator (\texttt{ACC\_only}). In the second execution scenario, we use EFT scheduler that provides the flexibility of mapping FFT task to either accelerator or CPU (\texttt{ACC+CPU}). The resulting timing details are plotted as Gantt charts in Figure~\ref{fig:acc_vs_mixedPEs}(a) and (b), for \texttt{ACC\_only} and \texttt{ACC+CPU} cases respectively.

The \textit{high} latency workload consists of 2610 FFT task nodes. As denoted in Figure~\ref{fig:acc_vs_mixedPEs}(a), when all of these task nodes are executed on the FFT accelerator, the end-to-end execution time is approximately 350 ms. 
On the other hand, the \texttt{ACC+CPU} policy uses the FFT accelerator for executing only 1165 of the FFT task nodes, but completes execution in approximately 260 ms, a 25\% reduction. 
Furthermore, we observe that the CPU cores remain comparatively underutilized for \texttt{ACC\_only} case, compared to \texttt{ACC+CPU} case. 
Due to the freedom of running FFT tasks on either type of resources, the \texttt{ACC+CPU} policy better exploits the parallelism within the workload, and achieves improved end-to-end execution time. 
This result indicates that during the application design process, naively mapping/scheduling the kernels to accelerator at compile time can lead to poor execution performance of the application at runtime, especially for workloads with dynamically arriving applications in an interleaved manner. 
This implies that a DSSoC should provide users with a development environment where application programmers can design their applications in a hardware-agnostic manner.

\begin{figure}[t]
    \centering
    \subfigure[]{\includegraphics[width=0.49\textwidth, trim=0cm 0cm 0cm 0cm, clip=true]{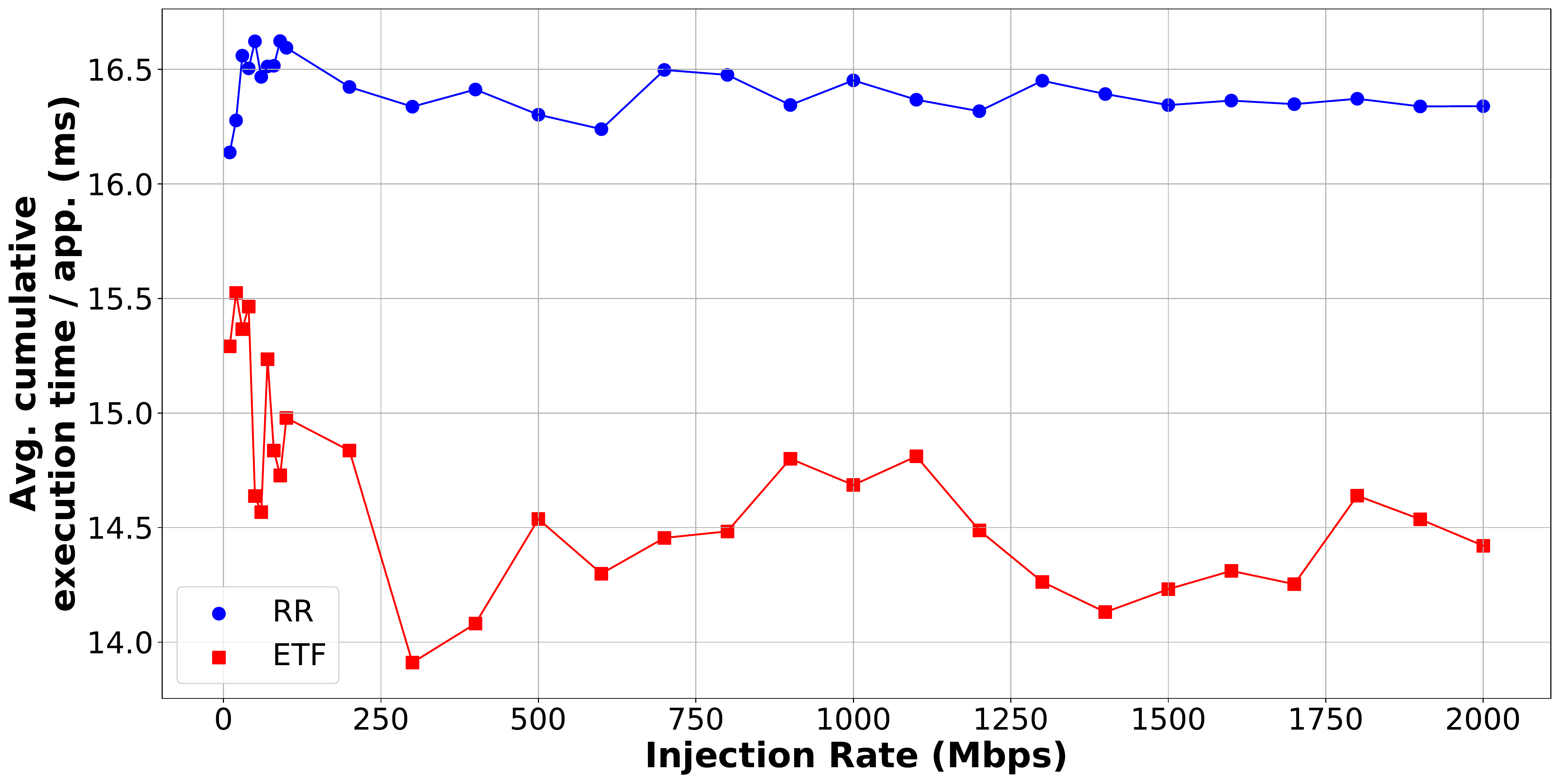}}
    \subfigure[]{\includegraphics[width=0.49\textwidth, trim=0cm 0cm 0cm 0cm, clip=true]{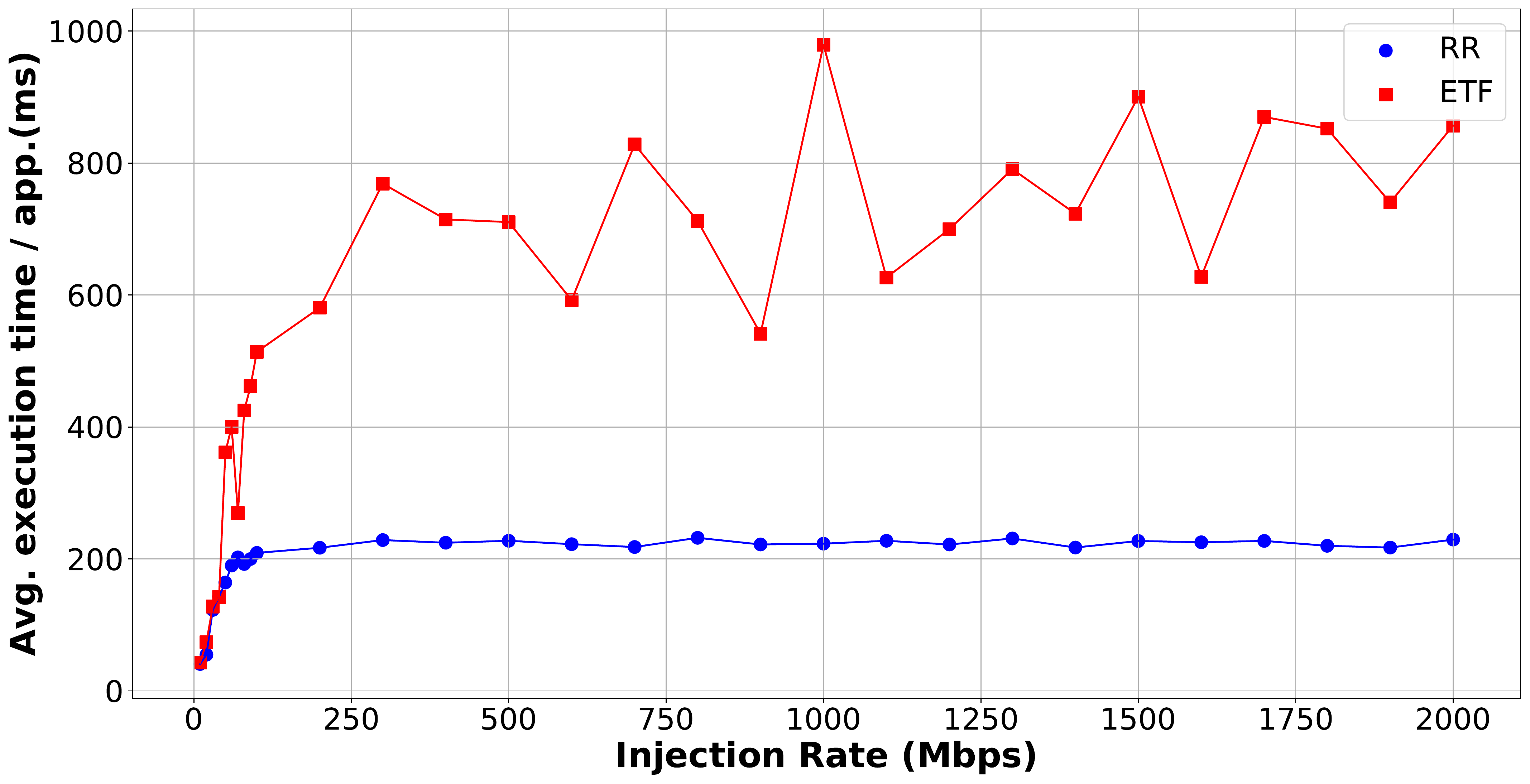}}
    \caption{(a) Average cumulative execution time per application and (b) average execution time per application, for \textit{high} latency workload using RR and ETF schedulers.}
    \label{fig:rq2}
\end{figure}

In order to answer RQ$_2$, we 
focus on the RR and ETF schedulers as these schedulers exhibit a large variation in typical cumulative execution time performance. 
We present the average cumulative execution time per application and average execution time per application in Figure~\ref{fig:rq2}(a) and (b) respectively with respect to various injection rates for C3-M1-F1 hardware configuration based on the \textit{high} latency workload. In Figure~\ref{fig:rq2}(a), we observe that the cumulative execution time for ETF across all injection rates are significantly lower than RR. At the highest injection rate (2000 Mbps), ETF shows 1.9 ms lower cumulative execution time compared to RR. This implies that ETF makes better scheduling decisions by better exploiting the available accelerators. However, Figure~\ref{fig:rq2}(b) suggests that as the system becomes oversubscribed with higher injection rates, the average execution time per application suffers severely for ETF compared to RR. At the highest injection rate, ETF takes 468 ms longer on average than RR to complete each application instance. This extra overhead of ETF is caused by its lengthy scheduling process as demonstrated earlier in Section~\ref{subsubsec:avg_sched_overhead_analysis}. With this, we conclude that despite making better task-to-PE mapping decisions, the complexity of a sophisticated scheduling heuristic can cause it to perform worse than a simpler scheduling heuristic that makes less informed decisions. Therefore, new class of scheduling heuristics are needed for heterogeneous platforms to balance the trade-off between the quality and complexity of scheduling decisions.

\subsection{CEDR Verification} \label{subsec:cedr_verification}

\subsubsection{Performance Counter Collection} \label{subsec:papi_counters}

Having access to the state of task execution in each of the concurrently running applications in terms of performance counters is a key capability for debugging, hot spot analysis and performance optimization. 
Performance Application Programming Interface (PAPI)~\cite{terpstra2010CollectingPerformance} is a well known library for interfacing with performance counters and it is integrated within the runtime workflow of the CEDR. 
This enables low-level performance profiling and workload characterization of applications at the granularity of individual kernels or DAG nodes, and it is worth noting that collection of these counters is possible without needing changes in the user code itself.

\begin{table*}[]
    \centering
    \begin{tabular}{|c|c|c|c|c|c|}
        \hline
        Applications & Instructions & Branches & Branch Misses & L1 Cache Loads & L1 Cache Misses\\
        \hline \hline
        Radar Correlator & 158341 & 6273 & 958 & 69348 & 1435 \\
        \hline
        \begin{tabular}{c}Temporal\\Mitigation\end{tabular} & 3543527 & 349478 & 11944 & 1351507 & 4063 \\
        \hline
        Pulse Doppler & 15016980 & 686875 & 80525 & 6484258 & 192936\\
        \hline
        WiFi-TX & 9861806 & 1102819 & 60703 & 3339442 & 11475\\
        \hline
    \end{tabular}
    \caption{Application characterization in CEDR using PAPI counters on ZCU102.}
    \label{tab:app_papicounters}
\end{table*}

In Table~\ref{tab:app_papicounters}, we provide five frequently studied low-level performance values collected through CEDR using PAPI, for the four applications used in this paper. 
These values include number of instructions, branches, branch misses, L1 cache loads and L1 cache misses. 
The instruction counts for Pulse Doppler and WiFi-TX are significantly higher compared to Radar Correlator and Temporal Mitigation -- reaffirming our knowledge about these application characteristics from Table~\ref{tab:application_characteristics}. 
To see this, we can focus on the high latency applications and see that, for instance, WiFi-TX contains around 1.6 times more branches compared to pulse doppler, whereas Pulse Doppler performs around 1.9 times higher L1 cache loads compared to WiFi-TX. 
This suggests that WiFi-TX is more compute intensive whereas Pulse Doppler is more communication intensive.

\begin{table*}[]
    \centering
    \begin{tabular}{|c|c|c|c|c|c|}
        \hline
        Task Name & Instructions & Branches & Branch Misses & L1 Cache Loads & \begin{tabular}{c}L1 Cache\\Misses\end{tabular}\\
        \hline \hline
        Head Node & 728 & 65 & 43 & 476 & 38\\
        \hline
        \begin{tabular}{c}Linear Frequency\\Modulation\end{tabular} & 13417 & 875 & 110 & 6146 & 189\\
        \hline
        FFT\_0 & 33411 & 1299 & 204 & 14781 & 384\\
        \hline
        FFT\_1 & 47703 & 1398 & 126 & 21029 & 317\\
        \hline
        Multiplication &  23607 & 382 & 54 & 10499 & 176\\
        \hline
        IFFT & 23556 & 667 & 64 & 10010 & 195\\
        \hline
        Find maximum & 15919 & 1587 & 357 & 6407 & 136\\
        \hline
    \end{tabular}
    \caption{Task-level characterization of Radar Correlator in CEDR using PAPI counters on ZCU102.}
    \label{tab:rc_app_papicounters}
\end{table*}
We demonstrate CEDR's capability to collect finer grained performance counters by presenting task-level counters for Radar Correlator application in Table~\ref{tab:rc_app_papicounters}. 
This table helps identify the FFT\_1 task containing largest number of instructions, and requiring largest number of L1 cache loads among the tasks of Radar Correlator. This kind of deep workload characterization can enable application users and scheduling heuristic designer to truly craft workloads that can be built to stress specific aspects of the scheduler and conduct root cause analysis as we will demonstrate in the following subsection.

\subsubsection{Application Verification \& Outlier Analysis}

In Section~\ref{subsec:runtime_config_sweep}, we extensively demonstrated CEDR's capability to execute different workloads with a wide variety of runtime configurations. 
To ensure the validity of these results, we cross check the outputs from these experiments with the outputs collected from the corresponding standalone execution of each application. The standalone versions refer to the application code prior to being modified into a CEDR-executable DAG and shared object. All outputs from CEDR-based execution match with the output of standalone execution for each application. 

Now, to further validate the results, we will investigate the two outlier points marked by red squares in Figures~\ref{fig:cum_exec}(a) and (b).
We monitor the performance counters for the unusually high average cumulative execution times for these two data points.
For the \textit{low} latency workload, the outlier point corresponds to hardware configuration with 1 CPU core, 1 FFT and 1 MMULT (C1-F1-M1), injection rate of 325 Mbps, and the HEFT\textsubscript{RT} scheduler. 
In this workload there are ten Temporal Mitigation instances, and only one of those instances show the outlier behavior because of the last task node associated with file I/O taking up to 40x longer time than the same task in remaining application instances. 
For all ten instances of this Temporal Mitigation task the performance counters such as the number of instructions, cycles, and cache or branch misses associated with the last node are identical showing no abnormal behavior in task execution, and all outputs match their serial standalone counterparts.
The file I/O operation relies on the OS and lies beyond the control of CEDR despite configuring each worker thread to utilize the highest possible static priority. 
Therefore, we believe that for the outlier instance, the last task node is interrupted by a kernel-level operating system job in order to use the CPU that the task was executing on.  
Similar trend is noticed in the outlier point for \textit{high} workload as shown in Figure~\ref{fig:cum_exec}(b), which belongs to hardware configuration of C3-F1-M0, injection rate of 40 Mbps with EFT scheduler. 
The outlier task in  this case is also a file I/O process. 
This particular task takes up to 20x longer to execute compared to the other instances. 
However, no significant variation is observed among the performance counter values of each instance of the outlier task. Apart from these minor number of outliers that are reliant on the OS and beyond the control of the CEDR runtime, CEDR is well validated and produces functionally correct output in 100\% of the experiments, thus enabling an efficient runtime framework for DSSoCs.

\section{Case studies} \label{sec:case_studies}

With the functional validation of CEDR established, in this section, we will explore a selection of unique features of CEDR and provide examples for the types of analysis these features enable.
We begin, in Section~\ref{subsec:schedule_caching}, with an investigation of schedule caching, a simple and yet effective feature in enabling scheduling heuristic developers to explore the trade space of scheduling quality versus scheduling overhead.
We continue with PE-level work queues in Section~\ref{subsec:pe_work_queues} and discuss how they are another avenue by which scheduler developers can reduce scheduling overhead while trading off against the delay between task scheduling and task execution.
For application developers interested in heavily optimizing the performance of a single application, in Section~\ref{subsec:dag_streaming}, we describe CEDR's support for application graphs that exploit their concurrency through software pipelining rather than DAG concurrency.
Finally, in Section~\ref{subsec:cedr_portability}, we illustrate the portable nature of CEDR through its deployment and expanded analysis on the Nvidia Jetson AGX Xavier development board.

\subsection{Schedule Caching} \label{subsec:schedule_caching}

Earlier in Section~\ref{subsubsec:avg_sched_overhead_analysis}, we showed that, with the increase in number of PEs and heterogeneity, ETF scheduler suffered from time spent on scheduling decisions, while providing high quality scheduling decisions. 
In this case a research problem is investigation of opportunities for addressing the scheduling overhead of ETF like algorithms by leveraging the design principles of the DSSoC architectures. 
In the context of a DSSoC, the target architecture is designed for a specific application domain and certain applications will likely be executed repeatedly on periodically arriving input data. From the scheduler's point of view, the repetition in an arriving workload will likely result into similar list of tasks that require scheduling. Therefore, instead of making task to PE mapping decisions through complex scheduling heuristics on reoccurring lists of task nodes over and over again; caching those decisions into memory, and simply looking up those decisions can significantly reduce the overhead of a scheduler like ETF. 
CEDR runtime has the ability save prior scheduling decisions, which we refer to as schedule-caching. 
The CEDR runtime uses the prior scheduling decision made for that task if it exists in the schedule cache. If such a decision is not present, %
then CEDR calls the scheduling heuristic, dispatches the task to its assigned PE and caches the decision made by the scheduler to the schedule cache. 
However, even in the context of ETF, this simplistic approach shouldn't be sufficient to increase overall performance of the runtime, as with dynamically arriving workloads, the ideal task to PE mapping decision for a specific task may change over time depending on the current state of the PEs and the workload characteristics as illustrated in Subsection~\ref{subsec:runtime_rq}. We analyze this trade-off over ETF and naive RR schedulers, while evaluating the benefit of schedule-caching.

\begin{figure}
    \centering
    \subfigure[]{\includegraphics[width=0.48\textwidth]{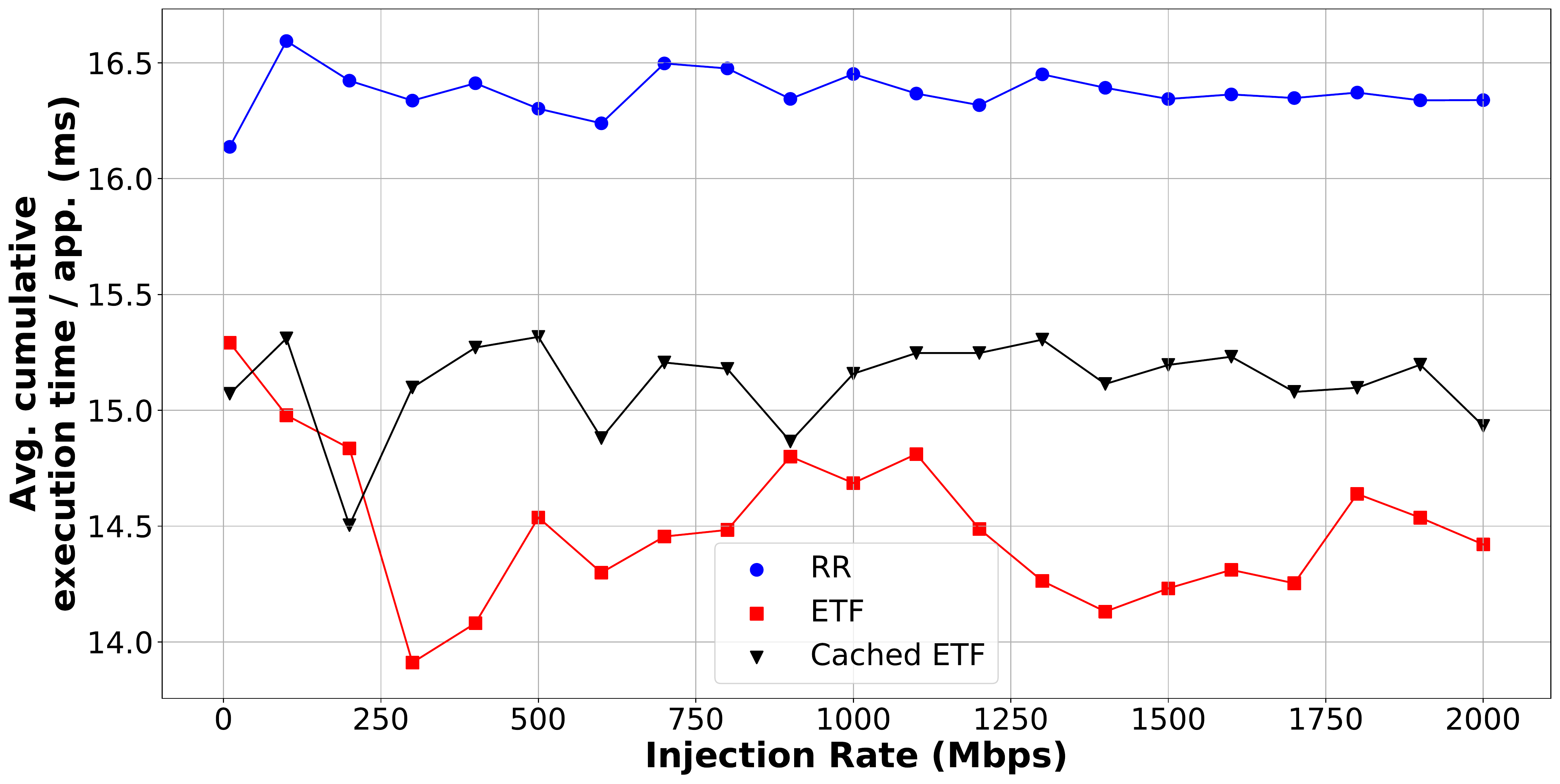}}
    \hspace{0.02\textwidth}
    \subfigure[]{\includegraphics[width=0.48\textwidth]{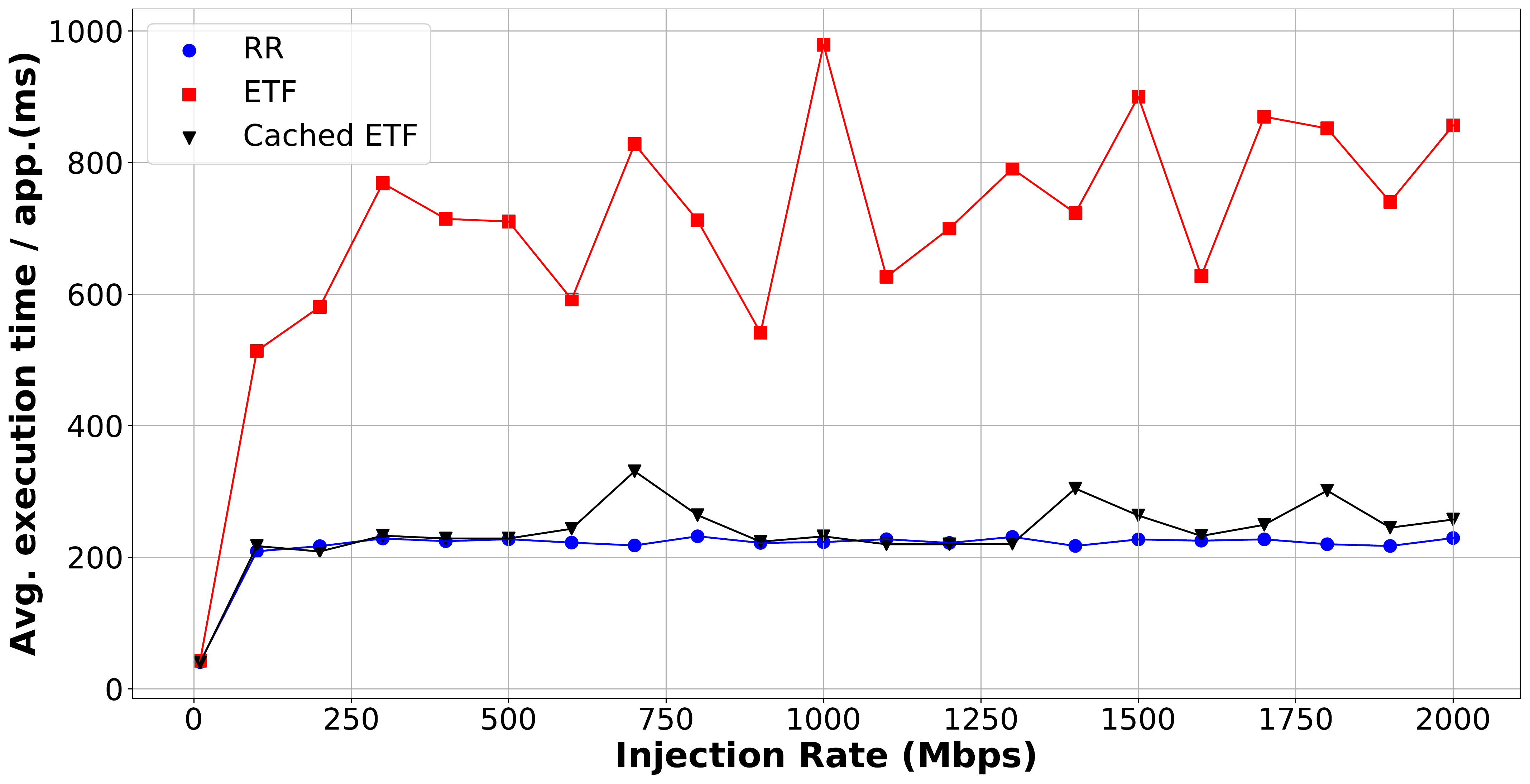}}\\
    \subfigure[]{\includegraphics[width=0.48\textwidth]{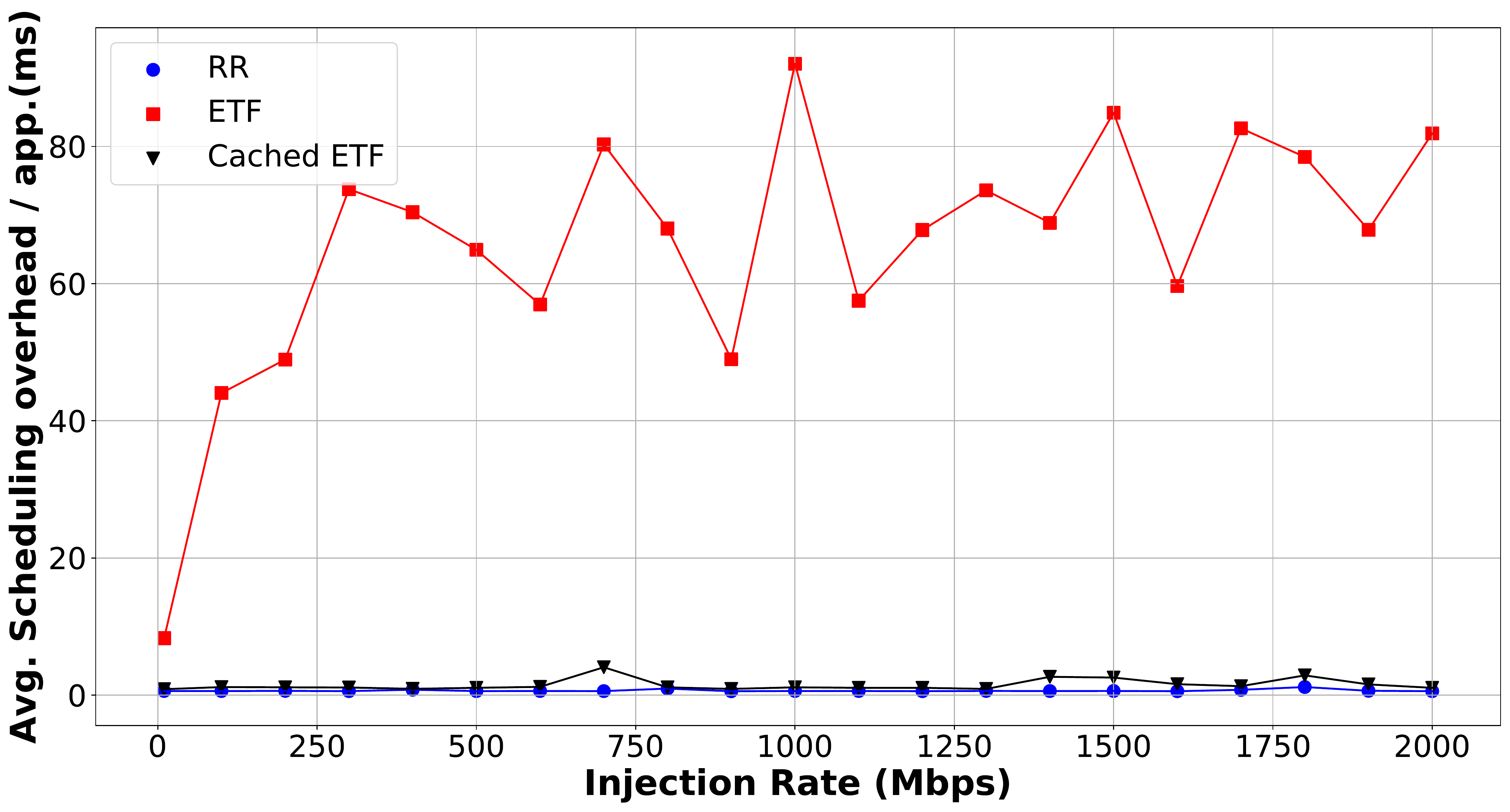}}
    \caption{Comparative analysis between RR, ETF and Cached ETF scheduling in terms of (a) average cumulative execution time, (b) average execution time, and (c) average scheduling overhead.}
    \label{fig:etf_vs_cache}
\end{figure}

Figures~\ref{fig:etf_vs_cache}(a), (b) and (c) present the average cumulative execution time, average execution time and average scheduling overhead (per application) respectively with respect to change in injection rate for ETF, RR, and  ETF with schedule-caching (Cached ETF) on the C3-F1-M1 hardware configuration.
In Figure~\ref{fig:etf_vs_cache}(a), we observe that the quality of decisions made by Cached ETF is better than RR but worse than ETF across all injection rates. On average across all injection rates Cached ETF results with 4.3\% higher cumulative execution time compared to the ETF. 
This observation is in agreement with the fact that in dynamically arriving workload scenarios, the ideal task to PE mapping decisions may vary.

However, focusing on Figure~\ref{fig:etf_vs_cache}(b), we notice that caching scheduling decisions clearly solves the scheduling overhead bottleneck for ETF as Cached ETF performs almost identical to RR in terms of average execution time per application. Looking closely into the time spent on scheduling decisions in Figure~\ref{fig:etf_vs_cache}(c), we attribute this execution time benefit observed in Cached ETF to its low scheduling overhead achieved by schedule-caching. 

These results demonstrate that reusing the historical scheduling decision made by a sophisticated scheduler for repeatedly arriving workloads can reduce the scheduling overhead significantly while maintaining an acceptable level of scheduling quality. Exploring the trade-off between complex schedulers and schedule caching can lead to low-overhead high-quality schedulers, which has been shown by the recently proposed Dynamic Adaptive Scheduler (DAS) by Goksoy et al.~\cite{goksoy_2021_DynamicAdaptive}.

\subsection{PE-level Work Queues} \label{subsec:pe_work_queues}

\begin{figure}
    \centering
    \includegraphics[width=0.6\linewidth]{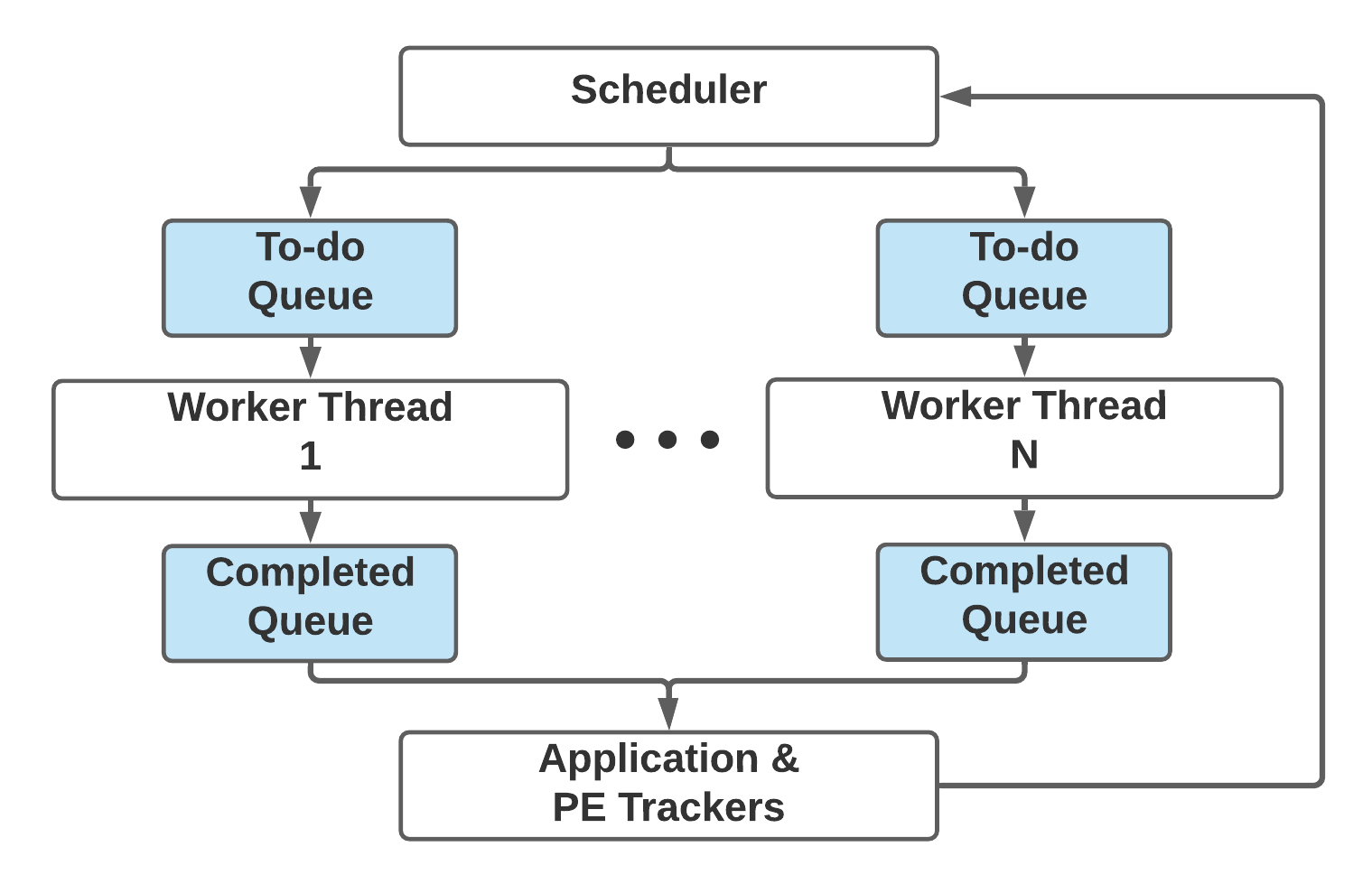}
    \caption{Architectural overview of the PE-level Work Queues.}
    \label{fig:queuing_diagram}
\end{figure}

In this subsection, we investigate the impact of PE-level work queues on the degree of utilization of the PEs in terms of PE idle time measurements and discuss how a queuing mechanism enables integration of advanced schedulers such as Earliest Finish Time (EFT). 

\begin{figure}[h!]
    \centering
    \includegraphics[width=0.7\linewidth]{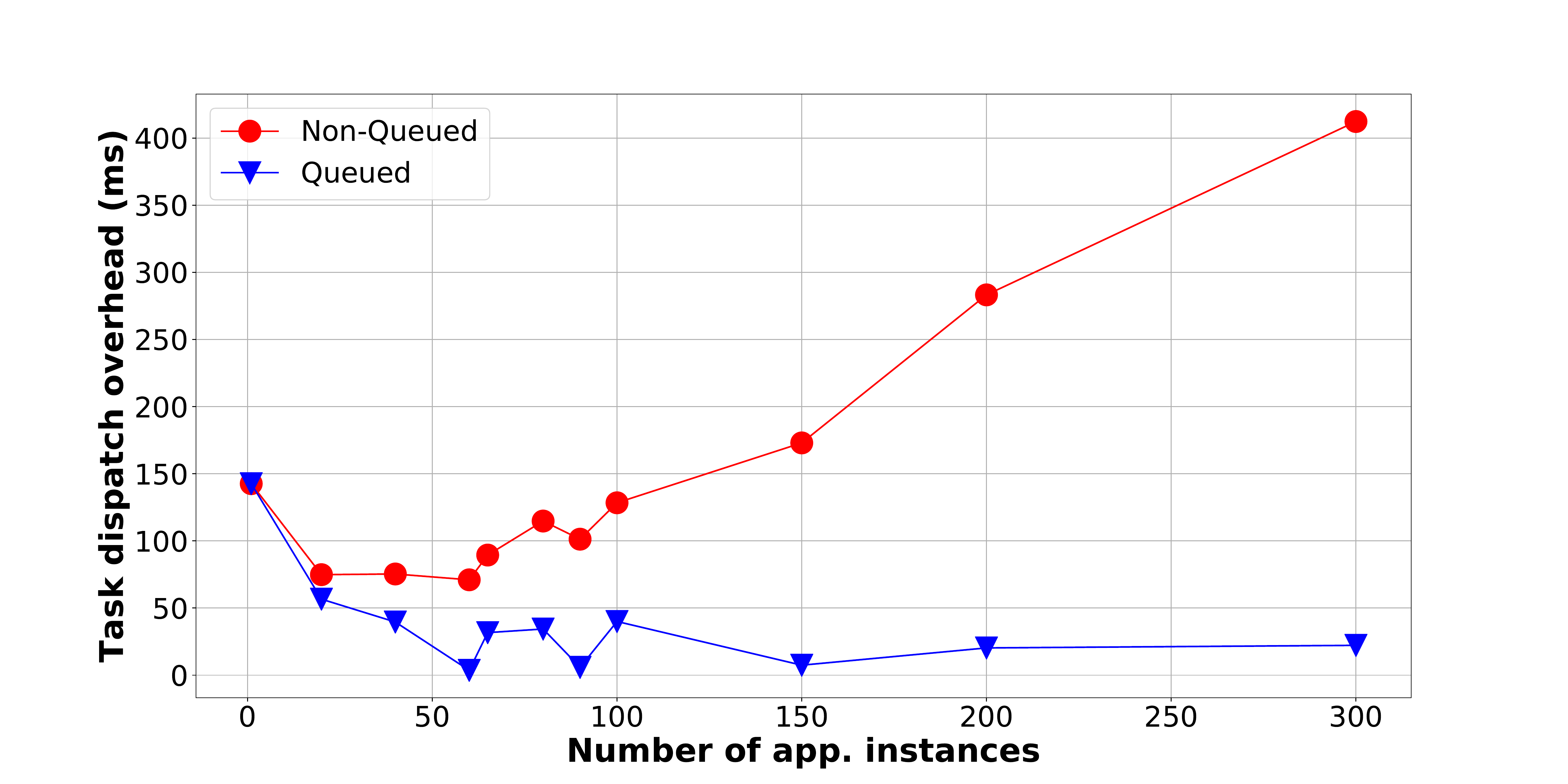}
    \caption{Task dispatch overhead with respect to varying number of Radar Correlator application instances for queuing and non-queuing executions.}
    \label{fig:queuing_results}
\end{figure}

During run time, task to PE mapping event occurs whenever a PE becomes available. 
In our earlier work~\cite{MackUserspace2020}, the emulation framework was limited to single scheduling decision for each mapping event. 
This poses as a key restriction towards development and integration of richer scheduling algorithms. 
For example, the intelligent scheduler proposed by Krishnakumar et al.~\cite{krishnakumar_2020_ILScheduler} requires the ability to schedule multiple tasks on any desired PE without having to wait for that PE to become idle. 
In order to lift this barrier, we incorporate the thread level task queuing mechanism into our framework. 
As illustrated in Figure~\ref{fig:queuing_diagram}, with queuing mechanism, task queues are inserted at the ``input'' and ``output'' interfaces between the CEDR worker threads and the main ``CEDR Management Thread'', and we refer to these queues as the \textit{To-do} and \textit{Completed queues} respectively. 
As the scheduler maps each task, the task is pushed into the \textit{To-do queue} of the corresponding resource.
These tasks are then popped by each worker thread as they become available such that the worker thread can begin computation.
As worker threads finish their tasks, they then signal completion back to the runtime by pushing the tasks back through their \textit{Completed queues}.
This queuing mechanism reduces the task dispatch overhead, which we define as the delay between completion of one task on a worker thread and the beginning of the execution of the next task on the same thread.  
We conduct timing analysis based on the Earliest Finish Time (EFT) scheduler using workload of varying instances of Radar Correlator (1 - 300 instances) with 3 Arm cores. We present our experimental results in Figure~\ref{fig:queuing_results}.
The X-axis of this plot presents the number of instances of Radar Correlator applications, and the Y-axis presents the task dispatch overhead. The blue and red curves capture the trend of task dispatch overhead with respect to the increasing number of applications for queuing-based and non-queuing-based executions respectively. Both queuing and non-queuing approaches show a drop in task dispatch overhead with respect to the increase in number of application instances from 1 to 60. Initially task dispatch overhead is higher due to the fact that there are not sufficient tasks to fully exploit the execution potential of the PEs.
However going beyond this region (application instances $>$ 60), the non-queuing execution shows linear increase in task dispatch overhead with increasing number of application instances, whereas the queuing method tends to approach to a saturation. The linearly increasing trend of non-queuing mechanism is caused by the fact that, under this policy a task is dispatched to each PE upon the PE becoming idle. This implies that with increasing number of application instances (linearly increasing number of tasks), the number of instances where a task is dispatched after waiting for the PE to become idle increases linearly. This in turn increases the task dispatch overhead. On the other hand, in the queuing based approach, while the PEs execute existing tasks assigned to them, the scheduler has the freedom to assign more tasks to the queues of PEs. This approach helps mask the task dispatch overhead behind the execution of tasks by PEs, and enables reducing the dispatch overhead to a saturated value. There are some exceptional points on the curve for the queuing based execution, which makes the curve non-smooth and requires some explanation. These points are for application instances 60, 90, 150- which are all divisible by 3 (the number of CPU cores). Having number of application instances divisible by the core-count enables the scheduler to evenly dispatch and offload the tasks, to the \textit{To-do queues} and from the \textit{Completed queues} of the PEs respectively. This eventually leads to lower number of scheduling rounds, which further reduces the task dispatch overhead.

\subsection{Stream-based DAG execution} \label{subsec:dag_streaming}

\begin{figure}
    \centering
    \subfigure[]{\includegraphics[width=0.48\textwidth]{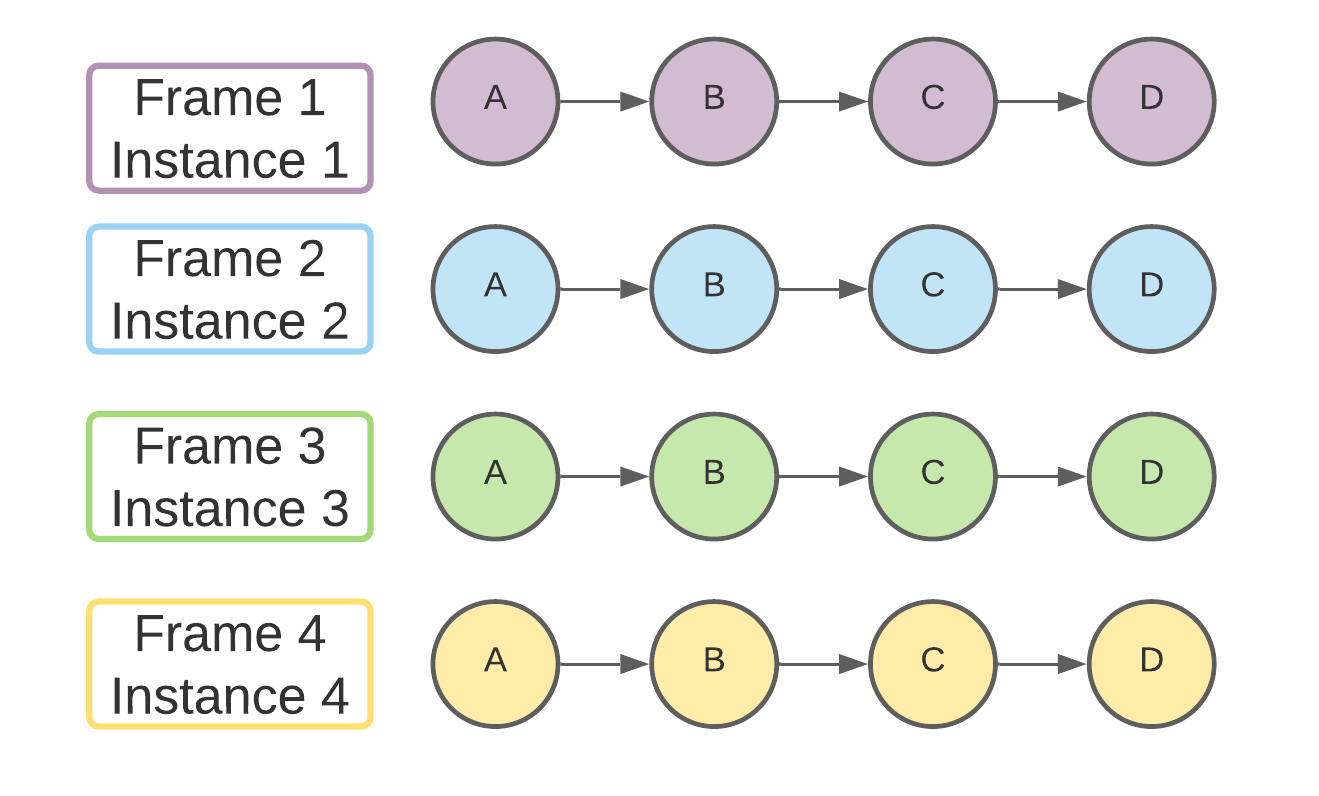}}
    \hspace{0.02\textwidth}
    \subfigure[]{\includegraphics[width=0.48\textwidth]{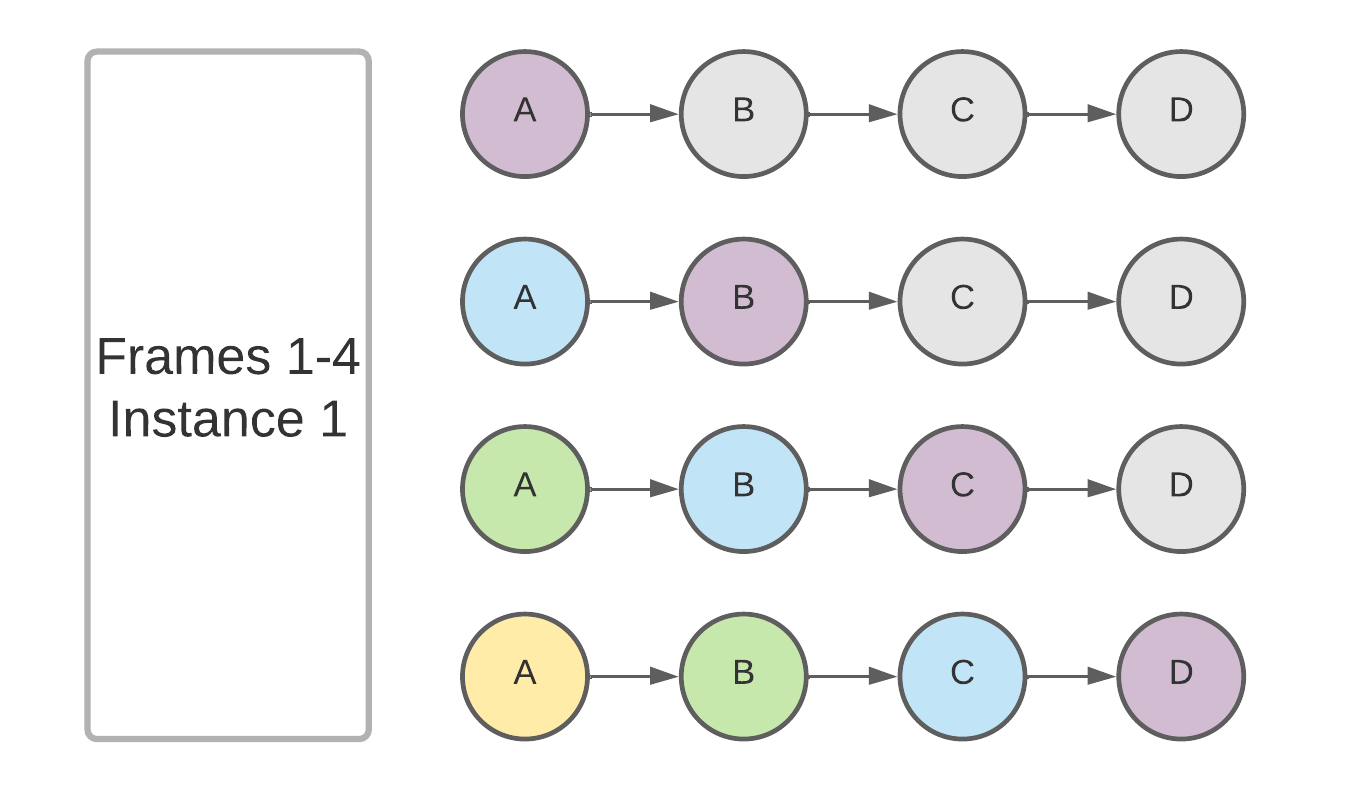}}
    \caption{Non-stream vs Stream-based execution.}
    \label{fig:stream_diagram}
\end{figure}

Until this point, we have demonstrated CEDR's ability to deploy applications on a given hardware configuration under dynamically arriving workload scenarios including dispatching tasks to accelerators. Processing a continuous stream of data is typical for applications in radar based navigation systems, such as the Radar Correlator, that detects the time shift observed in the incoming signal with respect to the reference signal to determine the  distance to other objects. This application could be executed continuously in an autonomous vehicle over frames of streaming, where each frame is composed of  a fixed number of input samples. Supporting this continuous data processing requires invoking a unique instance of Radar Correlator for each incoming frame data. 
This corresponds to invoking a new instance of the entire DAG representing that application and opening the corresponding shared object file for each frame as illustrated in Figure~\ref{fig:stream_diagram}(a) using an example DAG with four nodes (A, B, C, and D). In this execution model, during each application invocation, static memory buffers are allocated between the pairs of dependent DAG nodes, and used by the parent node to write output data to send to the child node. These buffers are freed at the end of each application instance execution.

Each application instance invocation experiences a latency overhead due to the buffer memory allocation and de-allocation. The overall memory demand and latency overhead can grow substantially as the number of frames increases to larger values for a streaming application. In order to address this issue we implemented the support for task-level pipelined execution of streaming applications in CEDR, where we instantiate a single instance of the application, setup the buffers between the dependent DAG nodes once, invoke the application, and execute for each incoming frame data in a pipelined manner as illustrated in Figure~\ref{fig:stream_diagram}(b). This setup avoids replicating the buffers between producers and consumers for each frame and reduces latency associated with setting them up. This pipelined manner of stream-based execution managed by CEDR allows for processing multiple input frames using the same set of memory resources, while ensuring synchronization between parent and child nodes. In the streaming setup, we adopt a double-buffer scheme where each DAG node maintains communication with its adjacent nodes using a pair of buffers, namely even and odd buffers. These pairs are used in an alternating manner by nodes so that at a given time instance, the data-write by a parent node and data-read by a child node use the opposite buffers of the pair. 
Applications implemented for stream-based execution follow the hand-crafted flow of compilation presented in Section~\ref{subsubsec:handcraft_compile}, where buffers between the dependent DAG nodes are defined and allocated by the application developer in the application code constructor and de-allocated in the destructor.

\begin{figure}[h]
\subfigure[]{\includegraphics[width=0.45\linewidth, trim=0cm 0cm 3cm 2.5cm, clip=true]{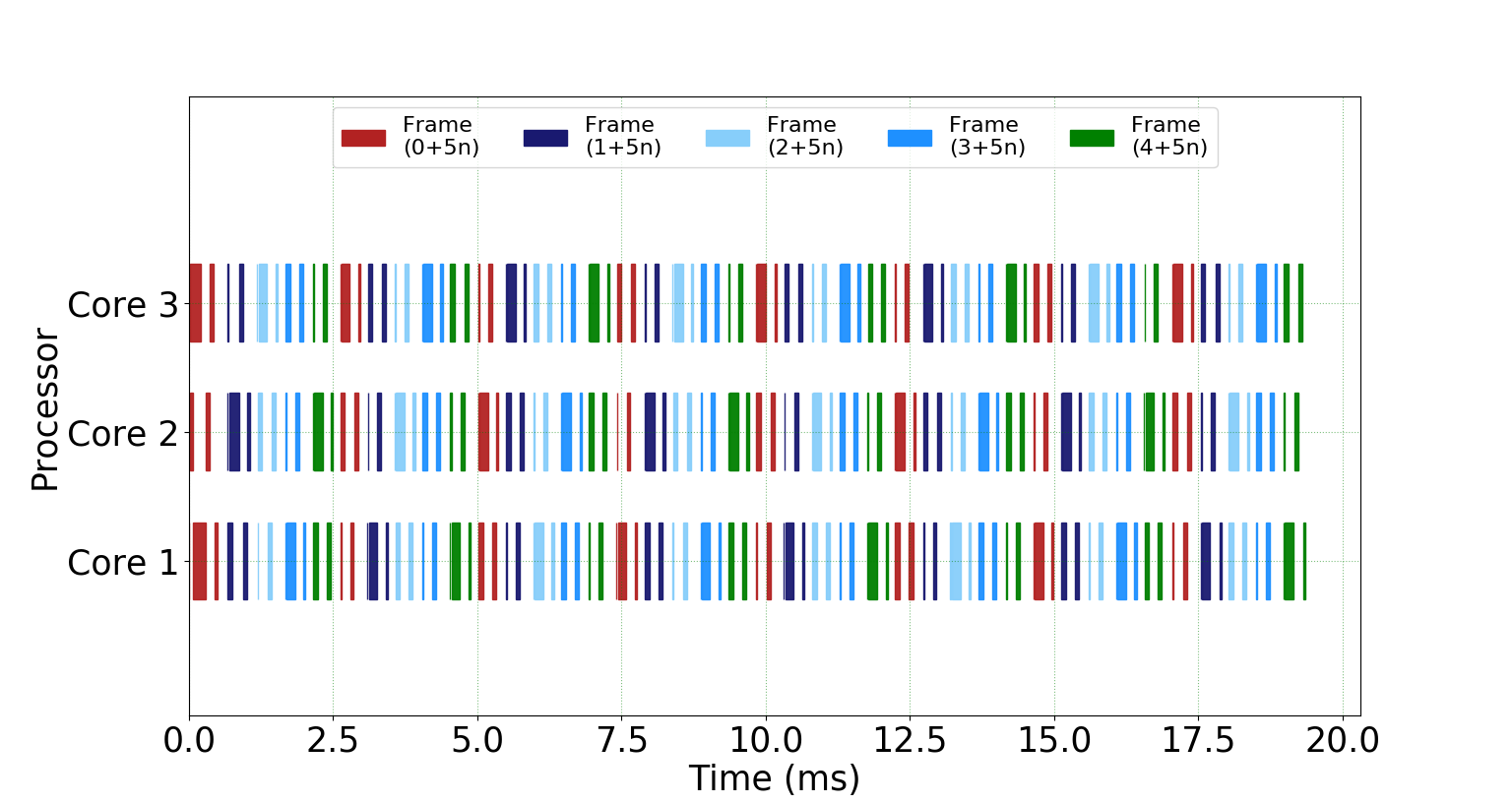}} \hspace{0.2cm}
\subfigure[]{\includegraphics[width=0.45\linewidth, trim=0cm 0cm 3cm 2.5cm, clip=true]{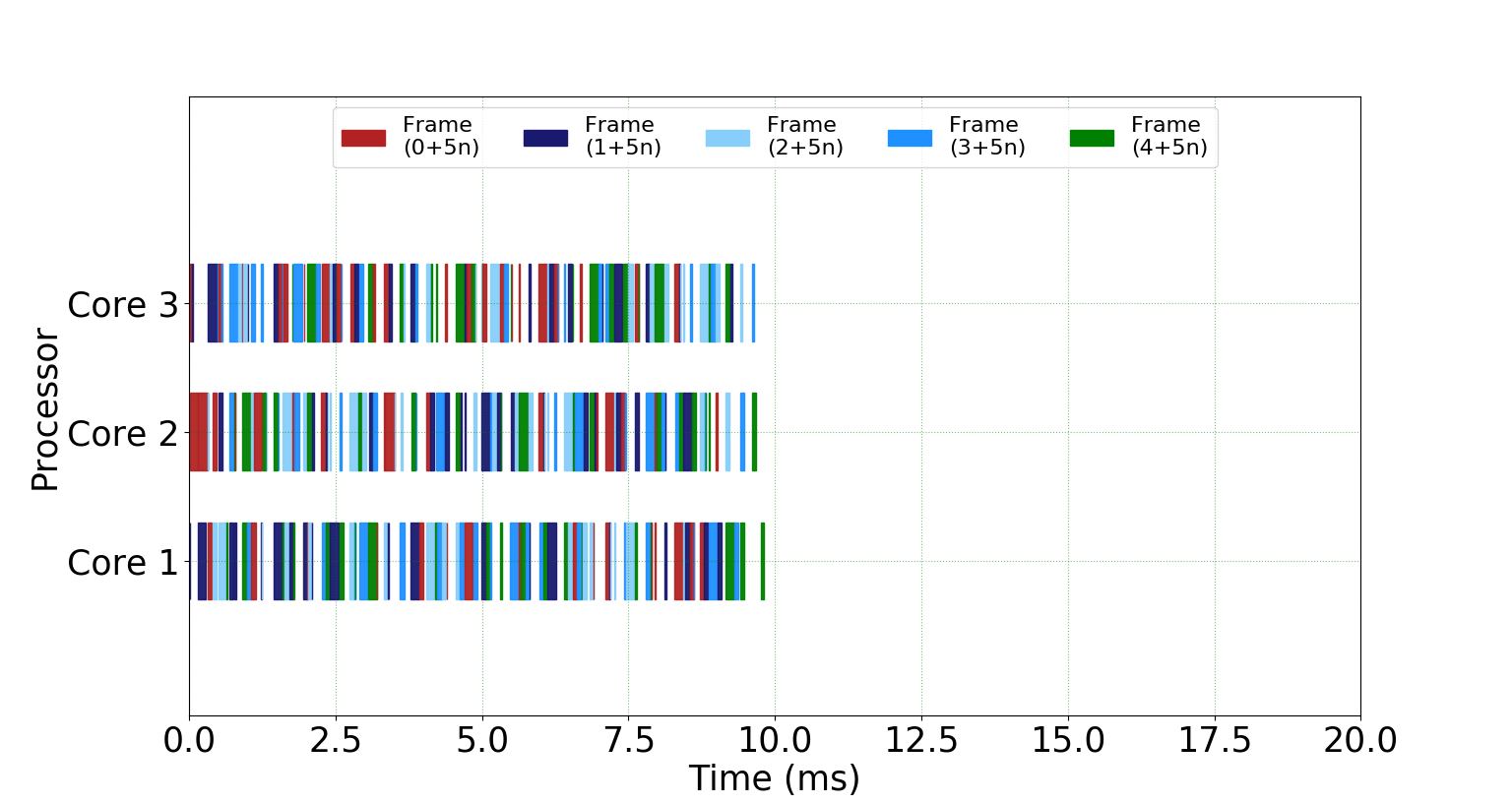}}
\subfigure[]{\includegraphics[width=0.45\linewidth, trim=0cm 0cm 3cm 2.5cm, clip=true]{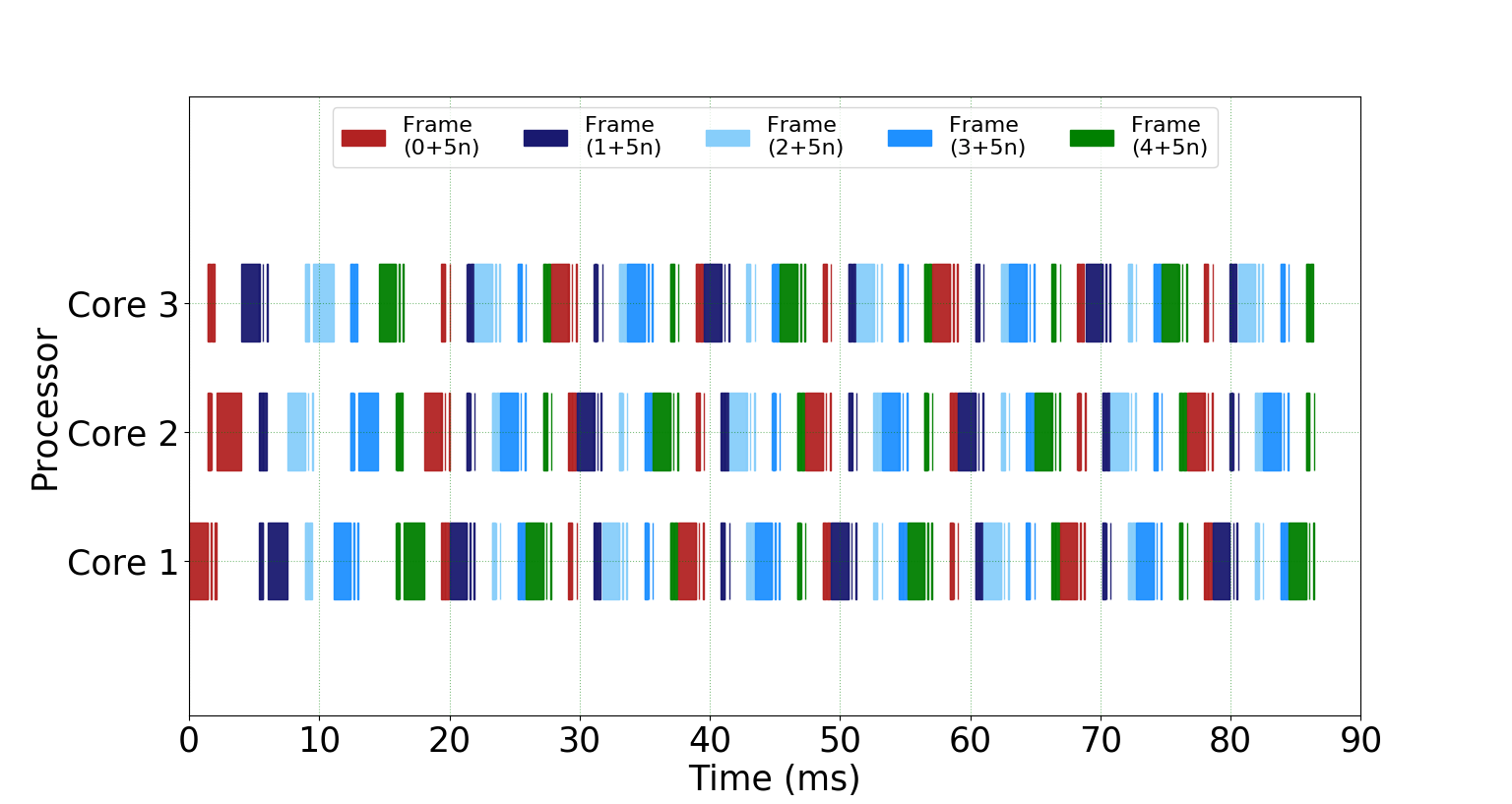}} \hspace{0.2cm}
\subfigure[]{\includegraphics[width=0.45\linewidth, trim=0cm 0cm 3cm 2.5cm, clip=true]{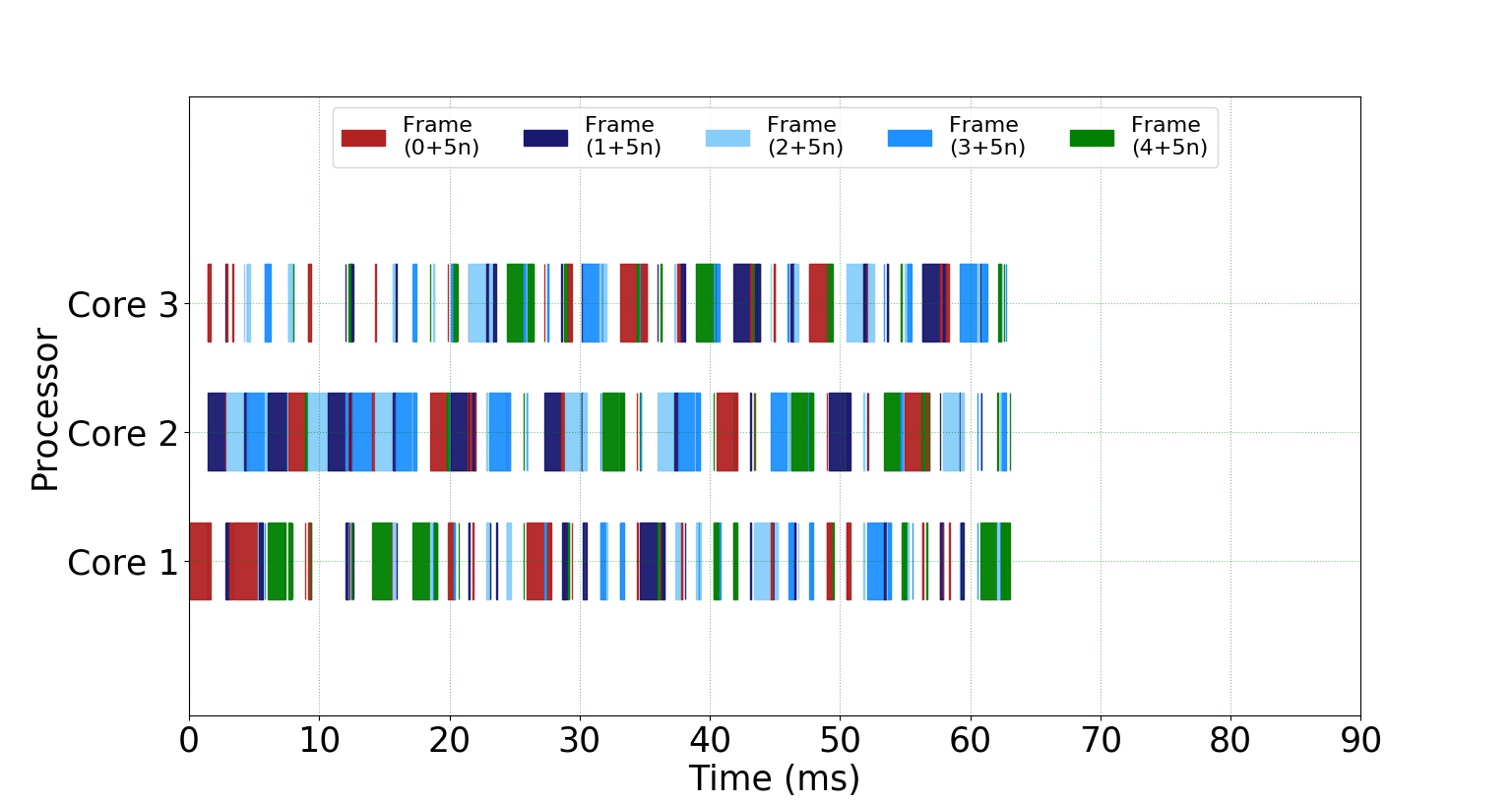}}
\caption{Gantt chart showing execution of 40 frames with Radar Correlator in (a) Non-streaming and (b) Streaming manner, and Temporal Mitigation in (c) Non-streaming and (d) Streaming manner.}
\label{fig:stream_result}
\end{figure}

In order to quantify the benefits of the stream-based execution, we use the streaming adapted versions of Temporal Mitigation and Radar Correlator applications. We use two workloads, one with Radar Correlator and the other with Temporal Mitigation applications, both processing 40 frames to execute on the target MPSoC. We run experiments with both non-stream and stream-based executions using the RR scheduler. In the case of non-stream-based execution, we limit the number of application instances to one in order to make the comparison with stream-based execution fair. The obtained Gantt charts from the experiments are presented in Figure~\ref{fig:stream_result}. The X-axis of these charts indicate execution time and the Y-axis presents the individual PEs. Each colored rectangle indicates the execution event of a task processing a certain frame. These rectangles also present the execution time of each task and the order of different tasks on different PEs. The Gantt charts show that for both Radar Correlator and Temporal mitigation applications, the tasks are scheduled more densely and executed in a pipelined manner, rather than a frame by frame execution. The execution time and average core utilization of these experiments are shown in Table~\ref{tab:stream_vs_nonstream}. This table shows that the stream-based execution reduces execution time by up to 50\% while increasing the average resource utilization by up to 27.6\%.

 \begin{table}[]
    \centering
    \begin{tabular}{|c|c|c|c|c|}
        \hline
        \multirow{2}{*}{Application} & \multicolumn{2}{c|}{Execution time (ms)} & \multicolumn{2}{c|}{Avg. Core Utilization (\%)} \\
        \cline{2-5}
        & Non-stream & Stream & Non-stream & Stream \\
        \hline
        Radar Correlator & 19.35 & 9.80 & 28.74 & 56.32\\
        \hline
        Temporal Mitigation & 86.41 & 63.03 & 36.68 & 50.27 \\
        \hline
        \end{tabular}
    \caption{Execution time and Average Core Utilization percentage for 40 instances of Radar Correlator and Temporal mitigation- with and without stream-based execution.}
    \label{tab:stream_vs_nonstream}
\end{table}

\subsection{Portability Across Platforms} \label{subsec:cedr_portability}

\begin{figure}
    \centering
    \subfigure[]{\includegraphics[width=0.48\textwidth, trim=9cm 1.5cm 8cm 3.2cm, clip=true]{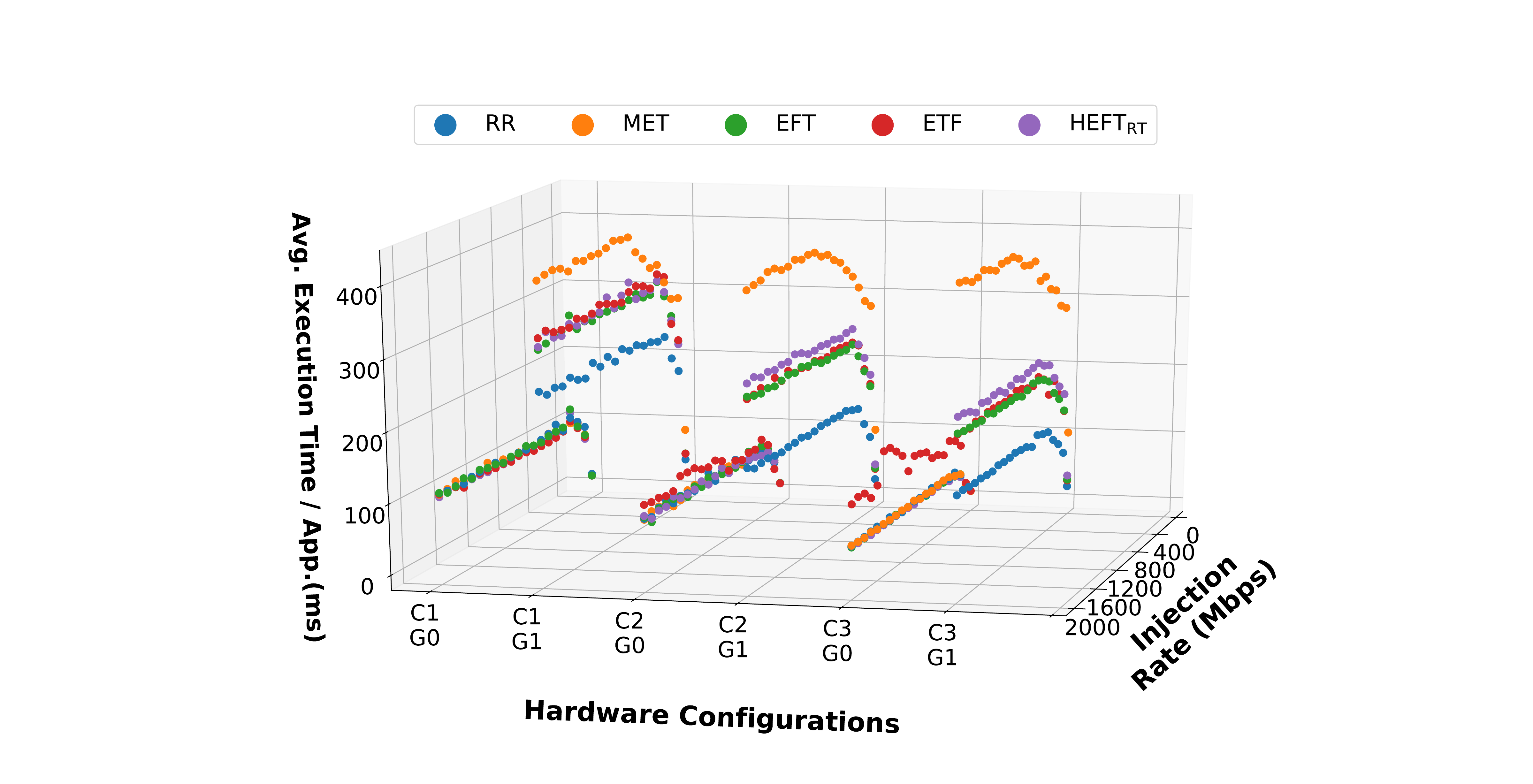}}
    \hspace{0.02\textwidth}
    \subfigure[]{\includegraphics[width=0.48\textwidth, trim=0cm 0cm 0cm 0cm, clip=true]{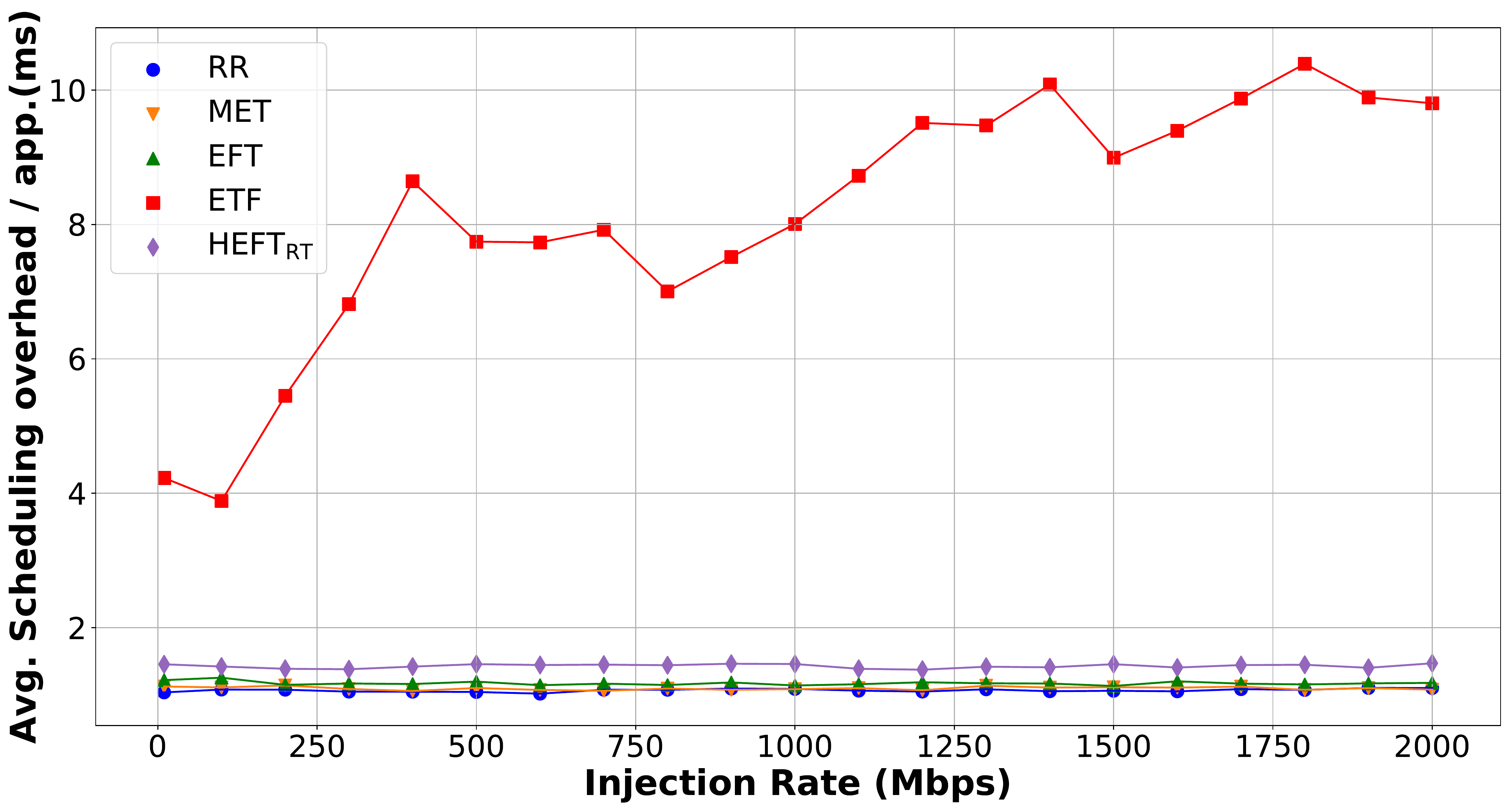}}
    \caption{(a) Average Execution time per application for \textit{high} workload and (b) Average scheduling overhead per application for \textit{high} workload.}
    \label{fig:exec_3D_2D_gpu}
\end{figure}

To illustrate the portability of CEDR, in this case study, we seek to validate experimental trends observed in Section~\ref{sec:experimental_evaluation} on the Nvidia Jetson AGX development board.
As discussed in Section~\ref{sec:experimental_setup}, on this platform, we implemented the FFT and MMULT kernels used there with equivalent sets of CUDA APIs using \texttt{cuFFT} and \texttt{cuBLAS}.
We then adjusted the JSON of each application to support dispatch of its FFT and MMULT kernels to resources of type \texttt{gpu}, and we modified CEDR to support spawning worker threads that are ``tagged'' to support tasks of type \texttt{gpu}.
With this setup, we then performed a similar resource sweep as in Section~\ref{subsec:cedr_verification}.
In this sweep, we tested hardware configurations ranging from 1 to 3 CPU cores and 0 or 1 ``GPU Resources'' available for CEDR to dispatch to, and we swept across all 5 schedulers, using both \textit{low} latency (injection rate range 1-1000 Mbps) and \textit{high} latency (injection rate range 10-2000 Mbps) workloads. We validated the outputs for these experiments by comparing against the serial based execution of the original implementations and observed a one to one match.

As another validation point, among the 3D plots shown in Section~\ref{subsec:runtime_config_sweep}, we present only the average execution time performance plot generated based on the Jetson AGX over the \textit{high} latency workload as shown in Figure~\ref{fig:exec_3D_2D_gpu}(a). 
Here we observe a saturation trend similar to what we observed in Figure~\ref{fig:exec}(b) for all the schedulers with respect to increase in injection rate across all the hardware configurations. Even though not shown, this similar trend behavior holds for average cumulative execution time and average scheduling overhead metrics for both \textit{low} and \textit{high} latency workloads. 

Unlike the ETF that performed worst on the ZCU102, we observe that on the Jetson AGX, MET scheduler performs the worst for the hardware configurations that use a GPU accelerator. For example in the case of most PE rich configuration of 3 CPUs and 1 GPU (C3-G1), the MET scheduler results with the highest execution time per application. Taking a close look at the scheduling overhead for this hardware configuration as shown in Figure~\ref{fig:exec_3D_2D_gpu}(b), interestingly we observe that, as another validation point, it is still the ETF scheduler that has the highest scheduling overhead by significant margin, which is in agreement with our findings on the ZCU102 from Figure~\ref{fig:sched_2d}(b). 

Although the actual FFT computation on GPU is faster than CPU, it suffers from data transfer overhead between CPU and GPU. As the \textit{high} workload contains large number of FFTs and MET schedules all of them on the GPU accelerator, while ETF favors CPU or GPU depending on the state of the PEs. As a result, with the data transfer overhead included, execution time becomes higher with the MET compared to the other schedulers. The diverging behavior observed on Jetson AGX compared to ZCU102 suggests that during DSSoC development, besides the runtime configurations, the platform specific constraints such as the PE to PE data transfer mechanisms and the interconnect structure are needed to be taken into account as well.

With this cross-platform verification in place, we are confident that CEDR can be deployed quickly across a wide variety of heterogeneous architectures quite easily, and because of the portable nature by which its schedulers and applications are defined, little to no work needs to be done to enable this cross-platform compatibility.
In future GPU-related work, we will explore the ability to specify CUDA Streams as distinct resources to better allow tasks to fully utilize the compute elements of the underlying hardware.

\section{Related Work} \label{sec:related_work}

In this section, we present a thorough overview of other works in this space and discuss how CEDR provides a unique set of capabilities within this context.
In the broader scope of modeling and evaluation of heterogeneous systems, a large amount of work has been done in simulators ranging from high level simulators~\cite{beltrame_2009_ReSPNonintrusive, carlson_2011_SniperExploring, fu_2014_PriMEParallel, arda_2020_DS3, matthews_2020_MosaicSim, vega_2020_STOMPTool} to cycle-accurate simulators~\cite{power_2015_Gem5gpuHeterogeneous, cong_2015_PARADECycleaccurate, shao_2016_CodesigningAccelerators, rogers_2019_ScalableLLVMBased, xiao_2019_SelfOptimizing}.
Compared to these approaches, CEDR takes a much more runtime-oriented approach with the view that, despite its increased setup cost, well-developed hardware runtimes coupled with direct architectural experimentation can avoid many of the downsides that can arise with these two approaches.
With high level simulators, these downsides typically take the form of divergence between the fast models utilized and the underlying hardware platforms, while with cycle-accurate simulators, the primary downsides are instead typically related to the long simulation time and complexities associated with modeling advanced hardware platforms that may not fit into the simulator's set of assumptions.
With this in mind, for the remainder of this review, we focus instead on works that seek to provide application-level runtimes.

In the domain of high performance computing (HPC), a large body of previous work exists in creating runtime systems with various capabilities. 
Within each runtime, support for hardware capabilities primarily varies from CPU-only execution~\cite{blumofe_1995_Cilk,papadopoulos_2015_STAPL,mattson_2016_ocr} to CPU/GPU execution~\cite{sabne_2013_cosp, krieder_2014_gemtc, kaiser_2014_hpx} to CPU/GPU/FPGA execution and beyond~\cite{gioiosa_2020_mcl,kim_2012_snucl}, with FPGA support in the latter category primarily enabled through variations of OpenCL. 
However, with the exception of~\cite{gioiosa_2020_mcl}, these environments focus primarily on providing the user with the technical capability to dispatch computations across all of these resources without also providing scheduling heuristics to help them do so in a dynamic, utilization-aware manner. 
Meanwhile, a large body of work exists on HPC-scale job scheduling algorithms~\cite{kumbhare_2020_parco, patki_2015_practical, fan_2019_beyondcpus}, but these scheduling algorithms are detached from any particular runtime system. 
In comparison, CEDR provides an environment that couples all aspects of application programming, scheduling, and execution into a single extensible environment. 
One of the closest environments to CEDR from the HPC domain is likely StarPU~\cite{augonnet2011starpu}.
StarPU is a well known platform for enabling heterogeneous task scheduling and execution on HPC-scale systems, and it has even been extended to enable features like FPGA support~\cite{christodoulis_2018_FPGATarget}.
However, to the best of our knowledge, it has not been applied to SoC-scale systems, and as such the workloads it excels at executing have characteristics that are quite different from those seen in frameworks like CEDR.
Additionally, while it does provide a C-based programming API and rich support for scheduling policies, application programmers are primarily required to rewrite their applications to target the StarPU runtime rather than leverage compilation tooling that can convert off-the-shelf source code to be compatible.
Taken together, while there is a large body of work in the realm of HPC computing that designers of DSSoC runtimes should certainly leverage and seek to learn from where possible, to the best of our knowledge, CEDR fills a unique niche when compared to this body of work. 
For the remaining literature discussions, we will focus primarily on works that target SoC-scale resource scheduling and dispatch.

Within the scope of SoC-scale application-level runtimes, we can further segment these works based on those that target accelerator-rich heterogeneous platforms and those that do not.
Starting first with those that do not, there are a number of works on application-level runtimes that target either homogeneous or single-ISA heterogeneous platforms (such as Arm big.LITTLE architectures).
This category includes works such as SPARTA~\cite{donyanavard_sparta_2016}, SOSA~\cite{donyanavard_sosa_2019}, SEAMS~\cite{maity_2021_SEAMSSelfOptimizing}, and the work of Martins et al.~\cite{martins_hierarchical_2019}.
While each of these works explores highly interesting avenues in the area of application runtime design in their own right (such as the control theory-based approaches leveraged by SOSA~\cite{donyanavard_sosa_2019} and SEAMS~\cite{maity_2021_SEAMSSelfOptimizing}), here, we prioritize discussion of works that address runtimes for multi-ISA or accelerator-rich architectures.
There are a number of studies in the area of developing environments for accelerator-rich platforms.
In HESSLE-FREE~\cite{moazzemi_2019_HESSLEFREEHeterogeneous}, a fuzzy control-based heterogeneous runtime management layer is introduced that allows the user to hook ``sensors'' into standard Linux applications targeting heterogeneous systems and provide the means by which a listening runtime can receive these sensor values and ``actuate'' the underlying hardware platform to modify that application's execution in order to achieve, for instance, improvements in energy consumption.
However, despite the runtime management layer in HESSLE-FREE coordinating aspects such as operating frequency or active cores, it does not appear that this framework itself has support for managing on which resource a given task launches -- for instance, tasks that launch on the GPU look to be statically mapped at compile time to launch on the GPU, with the primary ``actuation'' policies adjusting constraints such as the operating frequency of the GPU.
Bolchini et al.~\cite{bolchini_opencl_2018} present a runtime controller for OpenCL-based applications on heterogeneous architectures.
This OpenCL controller can integrate with Linux, perform mapping decisions via cluster-level mapping, and monitor power and execution metrics on the underlying platform.
However, the authors do not discuss the ability of this runtime to map multiple simultaneous applications, and they do not discuss the ease by which users can adjust their scheduling policy for a new one.

Across the runtimes mentioned thus far, little-to-no support is provided (or, at the very least, discussed) with regards to standalone compilation tooling that can be utilized to easily map applications into their respective frameworks.
As such, we conclude our discussions here by comparing directly to frameworks that present themselves, in some form, as unified compilation \& runtime environments for enabling execution on heterogeneous systems.
Picos++~\cite{tan_picos++_2019} is one such work in this area that proposes a hardware-based runtime that is coupled with support for applications written with OpenMP or OmpSs.
These applications are then mapped to Nanos++ API calls~\cite{nanos++} using the Mercurium Source-to-Source compiler (with current capabilities and status of this compiler described in~\cite{deharo_2021_OmpSsFPGA}).
While the compilation tooling in this ecosystem is quite substantial, by the nature of their hardware design, they are unable to support platform-independent, easily-interchangeable scheduling policies.
Boutellier et al. present PRUNE~\cite{boutellier_prune_2018}, a framework built to enable efficient, heterogeneous execution of signal processing workflows on SoC systems.
As one distinction relative to other works presented, PRUNE explicitly includes ``dataflow rates'' in its underlying model of computation, which allow it to include FIFO buffers between nodes as a part of their model similar to the streaming discussed in Section~\ref{subsec:dag_streaming}.
This modeling assumption, along with related ``design rules'' for dynamic workflows, allow them to propose a compilation framework that receives XML-based, platform-independent application DAGs, generates a set of C code to represent that DAG, and then map that via separate OpenCL and standard C compilation paths to support execution on CPU and GPU hardware.
In this work, the authors determine the actual computational mapping of each node in a primarily static process, with little flexibility for adjusting the mapping at a later point based on the current mixture of workloads.
Additionally, it would appear that the proposed runtime framework cannot itself handle the presence of multiple independent applications.
Auerbach et al.~\cite{auerbach_2012_liquidmetal} present a unified compilation and runtime environment that introduces a Java Virtual Machine (JVM)-based language called Lime that supports generation of OpenCL or Verilog code for GPU or FPGA backends.
Finally, Hsieh et al. present SURF~\cite{hsieh_surf_2019}, a runtime built around enabling efficient execution on heterogeneous SoCs.
In SURF, applications are represented as chains of tasks, where each task has a number of implementations in the form of, for instance, OpenMP, OpenCL, or Hexagon DSP kernels.
In addition, SURF couples this application representation with a dynamic, profiling-driven resource management layer that enables it to sense and adjust its management strategy dynamically at runtime based on system priorities and goals.
While they do provide APIs by which users can specify SURF applications, they do not appear to provide a framework by which users can map novel applications to utilize these calls. 
Additionally, it would appear that SURF applications can only consist of linear chains of kernels given their construction as a sequence of \texttt{surf\_task\_enqueue} API calls.

\begin{table*}
    \centering
    \caption{Feature comparison of CEDR against other state-of-the-art SoC-scale multi-ISA heterogeneous runtime managers with integrated compilation tooling.}
    \label{tab:literature_review}
    \begin{tabular}
    {
    |>{\centering\arraybackslash}m{0.425\linewidth}
    *{4}{|>{\centering\arraybackslash}m{0.05\linewidth}}
    |>{\centering\arraybackslash}m{0.07\linewidth}|
    }
        \toprule
        \textbf{Features} & \textbf{\cite{tan_picos++_2019}} & \textbf{\cite{boutellier_prune_2018}} & \textbf{\cite{hsieh_surf_2019}} & \textbf{\cite{auerbach_2012_liquidmetal}} & \textbf{CEDR}\\
        \midrule
        Portable across Linux-based systems & & \checkmark & \checkmark & \checkmark & \checkmark \\
        \hline
        Supports arbitrary mixtures of dynamically-submitted workloads & & & \checkmark & & \checkmark\\
        \hline
        Flexible support for arbitrary software-based schedulers & & & & & \checkmark\\
        \hline
        Flexible support for arbitrary hardware IP accelerators & \checkmark & \checkmark & & & \checkmark\\
        \hline
        Support for measuring hardware performance counters & \checkmark & & \checkmark & & \checkmark\\
        \hline
        Open source & & \checkmark & & & \checkmark\\
        \hline
        Streaming support & & \checkmark & & & \checkmark\\
        \hline
        DAG based, fine-grained workflow processing & \checkmark & \checkmark & & \checkmark & \checkmark\\
        \hline
        Support for DVFS & \checkmark & \checkmark & & \checkmark & \\
        \hline
        Support for hardware-based scheduling & \checkmark & & & & \\
        \bottomrule
    \end{tabular}
\end{table*}

An overview of CEDR's position against other runtimes that include integrated compilation frameworks is presented in Table~\ref{tab:literature_review}.
We can see that, with its broad support for application and scheduler integration coupled with its portability, support for highly parallel DAG-based workloads, and ability to collect fine-grained hardware performance counters, CEDR maintains a unique set of functionalities and features compared to the rest of the literature.
Namely, to the best of our knowledge, there is no other framework in the literature that provides all of CEDR's features in one unified environment.
For the sake of completeness, we end our discussions here by acknowledging some of the key limitations with CEDR as it currently stands as well as how our goals align with these limitations.
As shown in the last two rows of the table, we do not currently have support for implementing and testing of energy-savings policies that rely on mechanisms such as DVFS or for schedulers that are, themselves, integrated as standalone hardware accelerators.
Support for these features is on the development roadmap for CEDR, however, and despite their exclusions, we believe that even in its current state, CEDR provides a wholly unique set of capabilities relative to those in the rest of the literature.

\section{Conclusion} \label{sec:conclusion}

In this study we present the Compiler Integrated Extensible DSSoC Runtime (CEDR) ecosystem that provides a unified compile time and run time workflow as an abstraction layer over which software can be programmed in a hardware-independent manner and then dynamically mapped and executed over a variety of heterogeneous computation units.
With respect to other runtime environments, CEDR offers unique capabilities with regards to its flexibility at the application, resource management, and hardware integration layers.
CEDR has been implemented and validated on SoC platforms such as Xilinx Zynq UltraScale MPSoC, Odroid XU3, X86 systems, and the Nvidia Jetson Xavier through dynamically arriving workload scenarios. 
Overall we believe that CEDR offers key capabilities to facilitate broader usability and allow multiple players in domain specific architecture research (application developers, hardware architects, scheduler heuristic developers) to experiment with multiple concurrently executing real-life applications over heterogeneous architectures on off-the-shelf SoC and SoC-prototyping platforms.

We see several avenues for enhancing CEDR as a next step in DSSoC research.
First, with its ability to collect incredibly fine-grained timing and performance counter characteristics about the nodes it executes, CEDR is well positioned to be able to leverage this data in helping to train the next generation of intelligent scheduling algorithms that enable scheduling quality to approach that of complex heuristics without incurring the same associated overheads.
Beyond this, there is also a large opportunity for exploration of systems where the scheduler itself is hardware accelerated to better free up the CEDR runtime to instead focus on coordinating task launches and data transfers.
Next, there are a large amount of potential ways in which the trace-based compilation analysis shown here can be generalized to support more complex memory analysis and kernel recognition techniques to enable features such as automatic parallelization of memory-independent DAG nodes or improved heuristics for detecting and enabling automatic heterogeneous hardware execution.
Additionally, with its support for caching of scheduling decisions, algorithmic investigations can be conducted into areas such as determining the optimal eviction polices for such cached decisions (for instance, based on transitions in system load) so that we can ensure that they improve on scheduling overhead without degrading overall system performance too heavily.
Finally, since collecting power consumption information from SoC platforms is a non-trivial issue, several studies in the literature have proposed performance counter based power estimation models covering memory, CPU and interconnect characteristics during application execution.
Along with DVFS policies, we see incorporating power estimation as a valuable enhancement for CEDR to support power aware design decisions and leverage portability of CEDR across different SoC platforms.

\begin{acks}\label{sec:acknowledgement}

This material is based on research sponsored by Air Force Research Laboratory (AFRL) and Defense Advanced Research Projects Agency (DARPA) under agreement number FA8650-18-2-7860. The U.S. Government is authorized to reproduce and distribute reprints for Governmental purposes notwithstanding any copyright notation thereon. The views and conclusion contained herein are those of the authors and should not be interpreted as necessarily representing the official policies or endorsements, either expressed or implied, of Air Force Research Laboratory (AFRL) and Defence Advanced Research Projects Agency (DARPA) or the U.S. Government. \\ \\
The authors would also like to acknowledge Anish Krishnakumar from the Department of Electrical and Computer Engineering at the University of Wisconsin-Madison for his early support in the development and discussions around what eventually became CEDR as well as Conrad Holt from the School of Electrical, Computer, and Energy Engineering at Arizona State University for his work investigating OS-level task schedulers that motivated us to pursue userspace runtime development.

\end{acks}

\bibliographystyle{ACM-Reference-Format}
\bibliography{references}

\end{document}